\documentclass{aa}  

\usepackage{style_paper}

\begin{document} 

   \title{The impact of dynamic pressure bumps on the observational properties of protoplanetary disks}

   \author{Jochen Stadler\inst{1,2,3}
          \and
          Matías Gárate\inst{1}
          \and
          Paola Pinilla\inst{4,1}
          \and
          Christian Lenz\inst{1}
          \and
          Cornelis P. Dullemond\inst{5}
          \and
          \\ Til Birnstiel\inst{6,7}
          \and
          Sebastian M. Stammler\inst{6}
          }

   \institute{$^{1}$Max-Planck-Institut f\"ur Astronomie, K\"onigstuhl 17, 69117 Heidelberg, Germany\\
   $^{2}$Université Côte d’Azur, Observatoire de la Côte d’Azur, CNRS, Laboratoire Lagrange, 06304 Nice, France\\
   $^{3}$Univ. Grenoble Alpes, CNRS, IPAG, 38000 Grenoble, France\\
   $^{4}$Mullard Space Science Laboratory, University College London, Holmbury St Mary, Dorking, Surrey RH5 6NT, UK\\
   $^{5}$Zentrum f\"ur Astronomie, Heidelberg University, Albert-Ueberle-Straße 2, 69117 Heidelberg, Germany\\
   $^{6}$University Observatory, Faculty of Physics, Ludwig-Maximilians-Universti\"at M\"unchen, Scheinerstr. 1, 81679 Munich, Germany\\
   $^{7}$Exzellenscluster ORIGINS, Boltzmannstr. 2, 85748 Garching, Germany\\
   \email{jochen.stadler@oca.eu}
   }

   \date{Received 16 February 2022; accepted 22 August 2022}

  \abstract
   {Over the last years, large (sub-)millimetre surveys of protoplanetary disks in different star forming regions have well constrained the demographics of disks, such as their millimetre luminosities, spectral indices, and disk radii. Additionally, several high-resolution observations have revealed an abundance of substructures in the disk's dust continuum. The most prominent are ring like structures, which are likely caused by pressure bumps trapping dust particles. The origins and characteristics of these pressure bumps, nevertheless, need to be further investigated.}
   {The purpose of this work is to study how dynamic pressure bumps affect observational properties of protoplanetary disks. We further aim to differentiate between the planetary- versus zonal flow-origin of pressure bumps.}
   {We perform one-dimensional gas and dust evolution simulations, setting up models with varying pressure bump features, including their amplitude and location, growth time, and number of bumps. We subsequently run radiative transfer calculations to obtain synthetic images, from which we obtain the different quantities of observations.}
   {We find that the outermost pressure bump determines the disk's dust size across different millimetre wavelengths and confirm that the observed dust masses of disks with optically thick inner bumps ($<40$ au) are underestimated by up to an order of magnitude. Our modelled dust traps need to form early (< 0.1 Myr), fast (on viscous timescales), and must be long lived (> Myr) to obtain the observed high millimetre luminosities and low spectral indices of disks. While the planetary bump models can reproduce these observables irrespectively of the opacity prescription, the highest opacities are needed for the dynamic bump model, which mimics zonal flows in disks, to be in line with observations.}
   {Our findings favour the planetary- over the zonal flow-origin of pressure bumps and support the idea that planet formation already occurs in early class 0-1 stages of circumstellar disks. The determination of the disk's effective size through its outermost pressure bump also delivers a possible answer to why disks in recent low-resolution surveys appear to have the same sizes across different millimetre wavelengths.}


\titlerunning{The bumpy road to planet formation}
\authorrunning{J. Stadler et al.}
\maketitle
%

\section{Introduction}

The discovery of exoplanets shows that planet formation is an efficient process in protoplanetary disks. Observing the circumstellar material around young stars in different star forming regions is crucial to understand the physical processes that regulate the evolution of protoplanetary disks, and hence planet formation. 

Surveys at infrared wavelengths with Spitzer and Herschel provided information about the disk fraction around young stars and provided hints of undergoing disk evolution \citep{Oliveira2010, Fedele2010,Kim2016}. An example of this evolution are the disks that have low infrared emission compared to the averaged disks, suggesting an inner disk depleted of hot dust \citep{Strom1989, Calvet2002,Harvey2007}. Follow-up surveys of protoplanetary disks with state-of-the art telescopes and instruments, such as the Spectro-Polarimetric High-contrast Exoplanet REsearch (SPHERE) and X-Shooter at the Very Large Telescope (VLT) have provided more information about the disk structures of the hot upper layers of the disks \citep[e.g.][]{Avenhaus2018}, the stellar properties and accretion rates \citep[][respectively]{Alcala2014, Alcala2017, Manara2014, Manara2016_b, Manara2017}.

In addition, (sub-)millimetre observations with the Submillimetre Array (SMA) and the Atacama Large Millimetre-submillimetre Array (ALMA) have provided data of more than thousands of protoplanetary disks in different star forming regions \citep[e.g.][]{andrews2013,Pascucci2016,Ansdell2016,Ansdell2017,Ansdell2020,Barenfeld2016,Eisner2018,vanTerwisga2020}, giving demographics of the disk properties and the effect of different physical environments (as for example irradiation from massive stars present in some of the  observed regions). Some of the most studied properties of these disks are their millimetre luminosities (which are used to infer the dust mass, although with a high level of uncertainty), their dust disk sizes, and their spectral indices. 

These observations evidence that the millimetre-flux decreases with age \citep[e.g.][]{Grant2021}, with a couple of exceptions in Corona-Australis and Ophiuchus \citep[][respectively]{Cazzoletti2019, Williams2019}. In addition, a correlation between the size of the disk inferred from millimetre observations and the disk luminosity has been found in disks from different regions \citep{Tripathi2017,Andrews2018_sizelum,Hendler2020}, which seems to flatten in older systems. This correlation may arise from the radial drift of dust particles that is expected to make the disks smaller and less luminous with time, when observed at mm-wavelengths \citep{Rosotti2019}. 
However, \cite{Tazzari2021_sizes} showed that the outer dust disk radii for a sample of disks in Lupus do not change between wavelengths (0.9, 1.3, and 3.1\,mm), which contradicts the expectations of radial drift \citep[see also,][Fig. 7 \& 8]{Long2020}. 

Moreover, most of the high resolution observations of protoplanetary disks show that disks are not smooth, but instead they display substructures \citep[e.g.][]{Andrews2018_DSHARP,Long2018,Cieza2021_ODISEA}. These substructures may be the product of dust trapping in pressure maxima \citep[e.g.][]{Pinilla12,dullemond18} which can reduce or fully suppress the radial drift. Hence, if most of the disks host pressure bumps it is natural that the observed correlations contradict current models where radial drift is very efficient. However, a recent study by \cite{Jiang2021} also questions this link between continuum rings and pressure bumps, as they find no statistically significant evidence for a universal correlation between line emission and continuum substructures.

Pressure maxima can have different origins in the disk and they may have different behaviours and leave different imprints in the disk depending on the origin \citep[][for a review]{Pinilla_Youdin2017}. These origins can be divided in planetary origin, (magneto-) hydro-dynamical \citep{Flock2015}, or dust evolution processes \citep[e.g.][for variations of the dust properties and evolution near the ice lines]{Okuzumi2015}. Currently, it is debatable if most of the observed structures are due to planets, partially because planets have been only confirmed in PDS\,70 \citep{Keppler2018,Mueller2018} and very recently suggested in AB\,Aur \citep{Currie2022,Zhou2022}. 
One of the main differences that are expected from planetary pressure bumps versus (magneto-) hydro-dynamical effects is how long these bumps live and how dynamic they are. On one hand, once a planet is formed in the disk and it is massive enough to perturb the pressure profile of the disk or even the gas azimuth velocities to create traffic jams, the expected structures are quite stable in time (they do not appear and disappear). Still, the bump might jointly migrate (inwards) with the planet \citep[e.g.][]{Meru2019,Perez2019,Weber2019}, as long as it does not undergo rapid runaway migration, for which multiple dust traps can be formed \citep[e.g.][]{Wafflard2020}. On the other hand, pressure bumps that are from (magneto-)hydro-dynamical origin, also known as zonal flows, may appear and disappear in different disk locations \citep[e.g.][]{Uribe2011,Simon2013,Flock2015} with different properties (e.g. amplitudes, widths).

In this paper, we aim to understand if current observations enable us to distinguish between planetary- versus zonal flows- origin of the pressure bumps. We investigate the effect that different pressure bumps properties have on observational quantities such as the disk luminosity (including the estimated dust mass), the spectral index, the disk dust outer radius, and the optical depth. So far, theoretical comparison of observed relations, such as the correlation between the disk dust size the disk luminosity has mostly been studied in detail in the absence of pressure bumps \citep{Powell2019,Toci2021}. Only recently, \cite{Zormpas2022} find that the inclusion of strong pressure bumps (planets) flatten the slope of the size-luminosity relation and that the location of the outermost planet defines the dust size of the disk.
In this work, we focus our analysis of dust evolution on different properties of the pressure bumps, including: their amplitude and location, the number of bumps, and the growth timescale of the bumps. Additionally, we also perform simulations with variable pressure bumps, in order to mimic zonal-flow type bumps.

This paper is structured as follows:
In the upcoming section, the models for the disk and dust evolution and pressure bumps are described. The third section introduces the models' \hyperref[sec:SimSetup]{simulation setups} with the disk's initial conditions, variable pressure bump parameters and the assumptions for the radiative transfer calculations. In Section 4, the results are presented. We start with an extensive \hyperref[sec:parameterstudy]{parameter study} of the pressure bump variables, where in each subsection we present the parameter’s influence on the disks’ observables. The following subsection will focus on the results of the \hyperref[sec:ResultDynamic]{dynamic bumps model}, which tries to mimic zonal flows in disks. In the Section 5, we \hyperref[sec:discussion]{discuss} our results in the context of the various disk observables, such as the dust disk radii, millimetre fluxes, spectral indices, and dust masses. Additionally, we investigate the influence of the dust opacity prescription and planetesimal formation on our results. Finally, in Section 6 we draw our main \hyperref[sec:conclusions]{conclusions}.

\section{Models}
\label{sec:models}
\subsection{Disk evolution}
\label{sec:disk_evol}
In our model we consider a 1D axi-symmetric circumstellar disk whose viscous evolution can be described by the following continuity equation \citep[][]{Lust1952, lbp74, Pringle1981}:
\begin{equation} \label{eq:GasEvolution}
    \partialdiff{t}\Sigma_\textrm{g} = - \frac{1}{r}\partialdiff{r}\left(\Sigma_\textrm{g} r v_\mathrm{g} \right),
\end{equation}
where $\Sigma_\textrm{g}$ is the gas surface density, $r$ is the radial distance to the star and $v_\mathrm{g}$ is the gas radial velocity due to the viscosity which is given by \citep[][]{lbp74}
\begin{equation}
\label{eq:visc_velo}
    v_\mathrm{g}=-\frac{3}{\Sigma_\mathrm{g} \sqrt{r}}\frac{\partial\left(\Sigma_\mathrm{g} \nu \sqrt{r}\right)}{\partial r}.
\end{equation}
For the kinematic viscosity $\nu$, we use the so called $\alpha$-prescription for viscous accretion disks from \cite{Shakura73} that states
\begin{equation}
\label{eq:visc}
    \nu = \alpha c_\textrm{s} h_\textrm{g}.
\end{equation}
Here $\alpha$ describes a dimensionless quantity that measures the efficiency of the angular momentum transport in the disk, ${h_\textrm{g}=c_\textrm{s}/\Omega_\textrm{k}}$ is the gas pressure scale height with $\Omega_\mathrm{k} = \sqrt{GM_\ast/r^3}$ being the Kepler frequency. The isothermal sound speed can be written as  $c_s = \sqrt{ k_\mathrm{B} T/\mu m_\mathrm{p}}$, where $k_\mathrm{B}$ is the Boltzmann constant, $T$ the disk's temperature, $m_\mathrm{p}$ the proton mass and $\mu=2.3$ the mean molecular weight.

\subsection{Pressure bump model}
\label{sec:press_bump}
The focus of this work is on the properties of pressure bumps and how these affect observational properties of disks. Hence, we want to create bumps in the pressure profile of the gas, where the disk pressure in our vertically isothermal assumption is given by
\begin{equation} \label{eq:pressure}
    P = \rho_\textrm{g}\, c_\textrm{s}^2 = \frac{\Sigma_\textrm{g}\,c_\textrm{s}^2}{\sqrt{2\pi} h_\mathrm{g}} 
\end{equation}
with the gas midplane density $\rho_\mathrm{g} = \Sigma_\textrm{g}/\sqrt{2\pi} h_\mathrm{g}$, where most of the mass can be expected. 

Physically, a planetary gap forms when a planet is massive enough for its wakes to create shocks that progressively expel gas out of its horseshoe region. In our model, we implement this by locally increasing the turbulent viscosity, hence the gas radial velocity $v_\mathrm{g}$ has a strong increase in the gap, so the gap remains even if the disk is viscously evolved. The gap in the gas surface density then naturally produces a local pressure maximum or bump at the outer gap edge, since $P(r)$ declines with increasing $r$. This prescription has similarly been used in several previous studies \citep[e.g.][]{dullemond18,Stammler2019,pinilla21}.

To this end, we modify the turbulent viscosity parameter by imposing a Gaussian factor on its constant background value $\alpha_0$ as
\begin{equation} \label{eq:alpha_bump}
    \alpha (t,r) = \alpha_0 \cdot \left(1 + \sum\limits_{i} F_i(t,r)\right),
\end{equation}
where $F_i (t,r)$ for each bump $i$ is given by
\begin{equation} \label{eq:bump_factor}
    F_i(t,r) = A(t)\cdot  \exp{\left(\frac{-(r - r_{\textrm{gap,}i})^2}{2\,w_{\mathrm{gap,}i}^2}\right) },
\end{equation}
with $r_\mathrm{gap}$ the radial gap location, $w_\mathrm{gap}$ the Gaussian width and $A$ the time dependent amplitude of the bump. We parameterize the latter as
\begin{equation}\label{eq:amplitude}
A(t) =  \begin{cases}
            0   &    t < T_\textrm{ini}\\
            A_\mathrm{max}   &    t \geq T_\textrm{fin}\\
            A_\mathrm{max} \cdot \sin{\left(\frac{\pi}{2}\,\left( \frac{t - T_\textrm{ini}}{T_\textrm{fin}-T_\textrm{ini}}\right)\right)} & \rm{otherwise},
      \end{cases}
\end{equation}
with $A_\mathrm{max}$ the maximum amplitude, $T_\textrm{ini}$ the initial time where the bump starts to appear in the disk and $T_\textrm{fin}$ the final time of the growth phase, after which the bump reached its full amplitude $A_\mathrm{max}$. In contrast to these long-lived `planetary' bumps, we will also explore models, in which the bump will disappear again. For those, Eq. \eqref{eq:amplitude} will be adjusted by $A_\mathrm{max}=0$ for $t \geq T_\mathrm{fin}$ and the sin-function is normalised by $\pi$ instead of $\pi/2$.

We note that the above prescribed bumps have fixed radial locations throughout the simulation, therewith neglecting any planetary migration, a simplification which will be discussed in Sect.~\ref{sec:discussion}. Furthermore, the bumpy $\alpha$-profile gets only imposed on the kinematic viscosity, while keeping the global constant value of $\alpha_0$ for the turbulent velocities of the dust particles, as well as for the radial and vertical diffusion of the dust, which will be introduced in the upcoming sections. Physically, this implies that what sets the gas viscous evolution may be detached from the above mentioned processes of the dust evolution, which has been recently studied in detail by \cite{pinilla21}. \\

\subsubsection{Dynamic bumps} \label{sec:dynamic_model}
A special case of the above described disappearing bumps is our  \textit{Dynamic Bumps} model, which mimics zonal flows. First, we divide the disk up into several zones. The zones centres are located in the disk where $r/(5\times h_\mathrm{g})$ are integers and their width is such that, this ratio spans $\pm0.5$ around the integers, for example the boundaries of the first zone are at 2.5 \& 7.5 $h_\mathrm{g}$, respectively. Within each of these zones, pressure bumps can appear and disappear at random positions. However, only one bump can form within each zone and the bumps need to have at least a distance of one pressure scale height $h_\mathrm{g}$ to each other. As in the above `planetary' model, the gaps in the gas surface density get created through a Gaussian bump, imposed only on the turbulent $\alpha$-viscosity profile (Eq. \eqref{eq:bump_factor}, affecting only the gas kinematics but not those of the dust.

Before the bumps actually form and get imposed on the $\alpha$-profile a certain formation time $T_\mathrm{form}$ needs to pass. This is similar to MHD-simulations of the Rossby wave instability (RWI) that creates vortices, where it also takes some time till over-densities accumulate in the disk, which are then able to trap dust \citep[e.g.][]{Meheut2012,Flock2015,Cui_Bai2021}. After the formation time passed, the bumps have a given lifetime $T_\mathrm{life}$, in which their amplitude evolves, like the sinusoidal function defined in Equation \eqref{eq:amplitude}, and after $T_\mathrm{life}$ elapsed they disappear again. We set both $T_\mathrm{form}$ and $T_\mathrm{life}$ proportional to the local orbit time $t_\mathrm{orb}=2\pi/\Omega_k$ at the corresponding radial location of the bump. Once a bump has lived for its given lifetime, it disappears and a new random position within the zone will be drawn for a new bump and the cycle begins anew. 

It is important to state, that this model does not create bumps in the gas pressure based on physical conditions, but is rather forced to act in a similar way as seen in numerical simulations \citep[e.g.][]{Dittrich2013, Lenz2019}. As zonal flows are strictly speaking over-densities in the gas density, we also ran a comparison model inducing gaps rather then bumps in $\alpha$, leading to bumps in $\Sigma_\mathrm{g}$. However, we found no major differences between the two prescriptions, so we choose to implement the pressure bumps as bumps in $\alpha$ similar to the `planetary' bump model.

\subsection{Dust evolution}
\label{sec:dust_evolution}
The evolution of the dust is of particular importance, since its distribution determines the disk's appearance in the mm-continuum that are observed with interferometers like ALMA and which we want to model in this work. It's important to emphasise that the dust dynamics differs from that of the gas, as dust particles drift towards the star or concentrate at the midplane, which will be explained in the following paragraphs. This varying behaviour of dust and gas was predicted decades ago by theory \citep{Whipple1972, Weidenschilling1977} and has been repeatedly seen in observations \citep[e.g. HD 163296,][]{Isella2018_HD16,Varga2021,Oberg2021_maps,Teague2021}.

The most important quantity to track the dust dynamical behaviour is the \textit{Stokes number} $\textrm{St} = \tau_\textrm{stop} \Omega_\textrm{k}$, where the stopping time $\tau_\textrm{stop}$ describes the timescale on which dust particles couple to the gas motion. The strength of the (de)coupling, in turn, depends on the difference between gas and dust velocities ($v_\mathrm{g}$ and $v_d$, respectively), as well as on the particle's Stokes number through the drag force, which gets exerted onto the dust by the gas (per unit mass) as:
\begin{equation}\label{eq:drag_force}
f_D = - \frac{v_d-v_\mathrm{g}}{\tau_\mathrm{stop}}.
\end{equation}
As the global pressure maximum lies towards the star, the gas usually flows at sub-Keplerian velocity due to its own pressure support. Small enough dust particles (St$\ll$1) are well coupled to the gas, meaning they are dragged along with it. Larger boulders (St$\gg$1), however, can decouple from the gas motion and are not affected by it. A special case are mid-sized pebbles (St$\approx$1), which centrifugal force that acts against gravity is most insufficient, causing them to drift (inwards) the fastest.

Assuming spherical dust grains with radii $a$ and a material density $\rho_\mathrm{s}$, the two prevalent regimes for the \textit{Stokes number} at the disk's midplane are \citep{Brauer2008,birnstiel10}:
\begin{equation}
\label{eq:Stokes}
 \textrm{St} = \frac{\pi}{2}\frac{a\,\rho_\mathrm{s}}{\Sigma_\mathrm{g}} \begin{cases}
            1   &  a < 9/4 \cdot \lambda_\mathrm{mfp} \quad\quad  \text{Epstein regime}\\
            \frac{4}{9}\frac{a}{\lambda_\mathrm{mfp}}  &  \text{otherwise} \quad\quad\quad\,\,\,\, \text{Stokes regime 1},
      \end{cases}
\end{equation}
where the mean-free path of the gas is given by ${\lambda_\mathrm{mfp} = 1/(\sqrt{2}\,n\,\sigma_\mathrm{H_2})}$,
with the gas' number density $n=\rho_\mathrm{g}/\mu\,m_\mathrm{p} $ and the cross-section for molecular hydrogen ${\sigma_\mathrm{H_2}=2\cdot 10^{-15}\,\mathrm{cm^2}}$. Based on the above equation, we can note that the Stokes number is essentially a `dynamical grain size' that tracks the dust grain dynamics. We further define the dust's radial velocity following \cite{Nakagawa1986} and \cite{Takeuchi2002}:
\begin{equation} \label{eq:dust_radial_velocity}
    v_\textrm{d} = \frac{1}{1 + \mathrm{St}^2} \, v_\textrm{g} -  \frac{2 \,\mathrm{St}}{1 + \mathrm{St}^2} \, \eta \, v_\textrm{k},
\end{equation}
with $\eta = -\, (1/2)\,  (h_g / r)^2\, \partiallogdiff{P}$ and the Keplerian orbital velocity $v_\textrm{k}$. 

Additionally dust particles also diffuse due to the gas' turbulent motion, which counter acts concentration gradients. The dust diffusivity is hereby given through ${D_\mathrm{d}=\nu /(1+\textrm{St}^2)}$ \citep{Youdin2007}. The overall dust evolution can then be described by an advection-diffusion equation following \cite{ birnstiel10}:
\begin{equation} \label{eq:dustevol}
     \frac{\partial\Sigma_\mathrm{d}}{\partial t}\,+\,\frac{1}{r}\frac{\partial(r\Sigma_\mathrm{d} v_\mathrm{d})}{\partial r}\,-\,\frac{1}{r}\frac{\partial}{\partial r}\left(r D_d\Sigma_\mathrm{g}\frac{\partial}{\partial r}\left(\frac{\Sigma_\mathrm{d}}{\Sigma_\mathrm{g}}\right) \right) = 0,
\end{equation}
which is numerically being solved for the surface density $\Sigma_\mathrm{d}$ of each dust species, i.e. size bin.

We note that we do not consider the effects of the dust back-reaction onto the gas in this work, which can safely be done, since this effect becomes negligible for low dust-to-gas ratios ($\epsilon \leq 0.01$) and long timescales ($\sim$Myr), as shown in \cite{Garate2020}.

\subsubsection{Dust vertical settling}
\label{sec:dust_settling}
The vertical motion of dust particles is important when considering collisions of dust particles and, therewith associated, the growth of grains to larger sizes (which will be explained in more detail in the next subsection). The settling velocity of grains well coupled to the gas (St < 1) is given by $v_\mathrm{z, settl}=-z\,\Omega_\mathrm{k}$St \citep{Safronov1969, Dullemond_Dominik2004}. Thus, larger particles with high Stokes numbers (for St < 1) quickly settle towards the midplane, where $v_\mathrm{z, settl}$ then decreases. Dust settling in protoplanetary disks is nowadays confirmed by high resolution observations of highly inclined disks \citep[e.g.][]{Villenave2020}. 

The height above the midplane which influences the relative velocities between grains and thus their growth, is approximated by the dust scale height $h_\mathrm{d}=h_\textrm{g} \sqrt{\alpha_0/(\alpha_0 +\mathrm{St})}$ \citep{Dubrelle1995}. While this work does not treat the vertical evolution of the dust density distribution, we do compute the 3D total volume density of the dust $\rho_\mathrm{d}$, which will later serve as an input for the radiative transfer calculations. Following \cite{pinilla21}, we define it such that
\begin{equation} \label{eq:rho_dust_3D}
    \rho_\mathrm{d}\, (r, \varphi, z, \text{St}) =
    \frac{\Sigma_\mathrm{d}}{\sqrt{2\pi}\, h_\mathrm{d}} \, \exp\left(- \frac{z^2}{2\,h^2_\mathrm{d}} \right),
\end{equation}
also using the prescription of \cite{birnstiel10} for the dust scale height:
\begin{equation}\label{eq:scaleheight_dust_youd}
h_\textrm{d}\,(\textrm{St}) = h_\textrm{g} \cdot \min\left(1, \sqrt{\frac{\alpha_0}{\min(\mathrm{St},1/2) (1+\mathrm{St}^2)  }}\right),
\end{equation}
where it is ensured that $h_\mathrm{d}$ does not become larger than the gas scale height $h_\mathrm{g}$, which would otherwise be unphysical. We note that we found no differences in the disk's observables of our \textit{Fiducial} setup, whether using the \cite{birnstiel10} or \cite{Dubrelle1995} prescription of $h_\mathrm{d}$.

\subsubsection{Dust growth}
\label{sec:dust_growth}
In this work we simultaneously describe the evolution of dust and gas, including the growth and fragmentation of the dust grains \citep{Brauer2008,birnstiel10}.

In general, the dust phase consists of a distribution of different particle sizes, whose dynamics are mainly determined by their Stokes number (Eq. \eqref{eq:Stokes}). The particles interact and collide with each other, where different collisional outcomes are possible, such as sticking, fragmentation or erosion of grains. The Smoluchowski coagulation equation \citep{Smoluchowksi1916} is solved to model the dust growth. Whether particles grow or fragment in this process mainly depends on their respective sizes or masses and their relative velocities with respect to each other, as described in \cite{birnstiel10}.

While the velocities of small particles ($\leq$ 1 $\mu$m) are governed by Brownian motion, the picture changes for larger grains (see Fig.~1 of \cite{Birnstiel2016}). Then, there are different transport mechanisms that become the dominant contributors of the particles' relative velocities before collision, namely turbulent mixing, vertical stirring or settling and azimuthal or radial drift. If grains grow, fragment, or erode upon collision is thus determined by comparing the computed relative velocities with the fragmentation speed $v_\mathrm{frag}$, a parameter that we take to be constant throughout the disk \citep{Birnstiel2009}. In the fragmentation limit, the maximum Stokes number or size a particle can reach is
\begin{equation} \label{eq:St_frag}
    \text{St}_\mathrm{frag} = \frac{1}{3}\frac{v_\mathrm{frag}^2}{\alpha_0 c_\mathrm{s}^2}.
\end{equation}

\begin{figure}
    \centering
    \includegraphics[width=90mm]{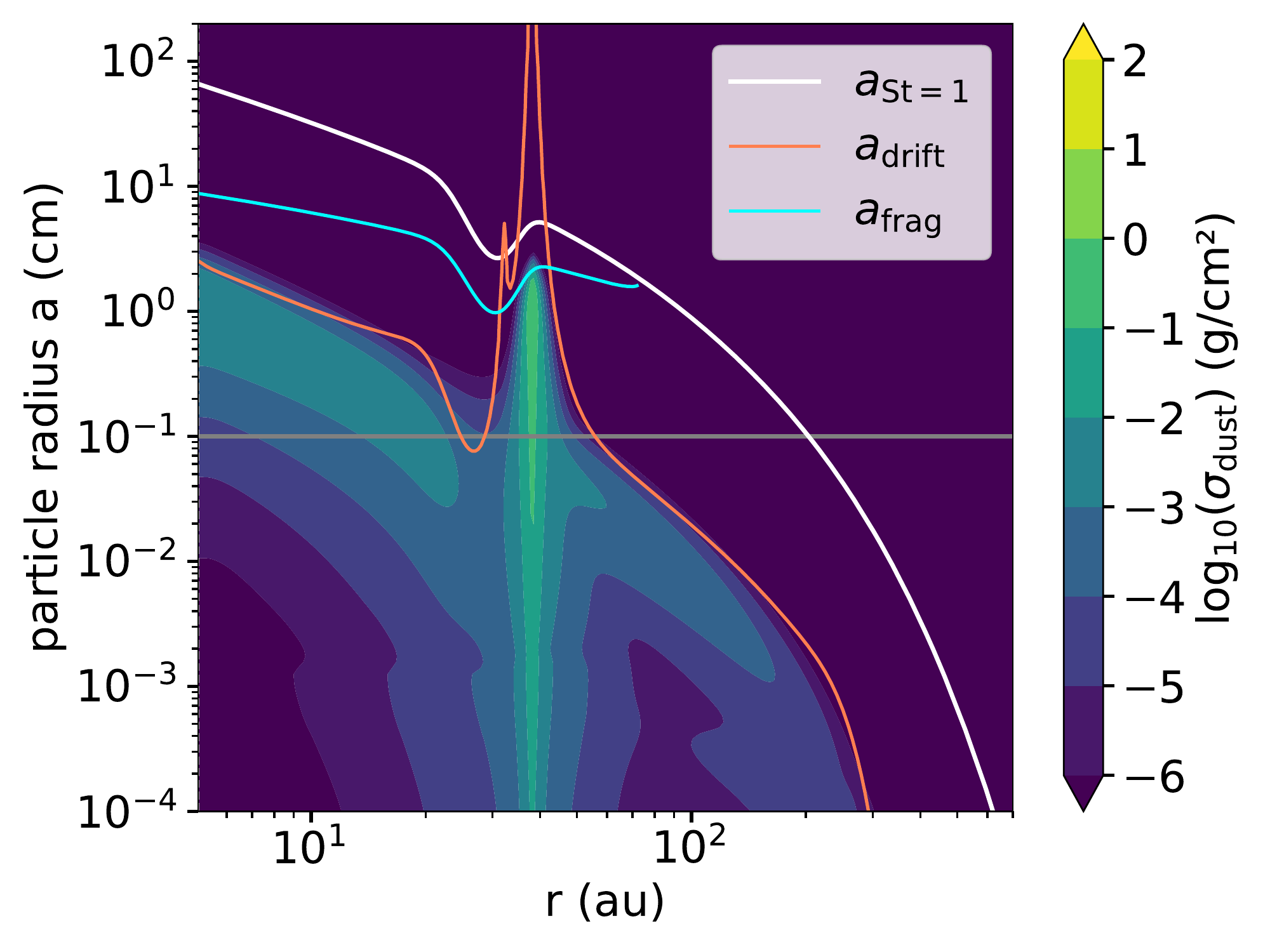}
    \caption{Dust size distribution for the \textit{Fiducial} model at 1 Myr. The contours show the growth limits for drift (orange) and fragmentation (cyan), as well as the particle sizes corresponding to St = 1 (white) and 1 mm (grey).}
    \label{fig:part_dens_distr}
\end{figure}

\indent The other process that locally limits the growth of the particle distribution is radial drift \citep{Brauer2008, birnstiel10}. It dominates when the growth timescale exceeds the drift timescale, so particles drift faster towards the star than they can grow to larger sizes. In this case, the maximum Stokes number is given by
\begin{equation} \label{eq:St_drift}
    \mathrm{St}_{\textrm{drift}} = \left|\frac{\textrm{dln}\, P}{\textrm{dln}\, r }\right|^{-1} \frac{v_k^2}{c_s^2} \,\epsilon,
\end{equation}
with $\epsilon = \Sigma_\mathrm{d}/\Sigma_\mathrm{g}$ being the column density dust-to-gas ratio. The above equations then ultimately set the maximum size a particle can grow to ${\text{St}_\mathrm{max} = \text{min}(\text{St}_\mathrm{frag},\mathrm{St}_\textrm{drift})}$ \citep{Birnstiel2012}, where we neglect dust settling which can also set the maximum grain size \citep{pinilla21}. A display of the both size limits (orange and cyan lines) can be seen in Fig.~\ref{fig:part_dens_distr} that shows the dust size distribution for our \textit{Fiducial} model at 1 Myr of evolution, whose parameters can be found in Table \ref{table:bump_param}. Inside the pressure bumps the maximum grain size is limited by fragmentation because radial drift is reduced at these locations. In the rest of the disk, the maximum grain size is usually limited by drift at Myr timescales. The white solid line in the plot corresponds to the particle size of $\mathrm{St} = 1$, which is proportional to the gas surface density (see Eq. \eqref{eq:Stokes}).

\section{Simulation setup}
\label{sec:SimSetup}
\subsection{Disk setup}
\label{sec:disk_setup}
We carried out our disk simulations with the code \texttt{DustPy}\footnote{For this work version 0.5.3 was used. The newest version of the code package can be found on \url{www.github.com/stammler/DustPy}} by \cite{Stammler2022} which builds on the dust evolution model of \cite{birnstiel10}. This code solves both the gas and dust evolution equations described in the previous section. We assume the disk to be around a Solar mass star and initially set up the gas surface density of the disk following the profile of \cite{lbp74}:
\begin{equation}\label{eq:lbp}
    \Sigma_g(r,t_0) = \Sigma_0 \left(\frac{r}{R_\mathrm{c}}\right)^{-\gamma}\,\exp\left[-\left(\frac{r}{R_\mathrm{c}}\right)^{2-\gamma}\right],
\end{equation}
with $\gamma=1$ and the normalisation factor $\Sigma_0 = M_\mathrm{disk}/(2\pi R_\mathrm{c}^2) $ is defined through the initial disk mass, which we set to be $M_{\mathrm{disk}}=0.05M_\odot$. We take the characteristic radius $R_\mathrm{c}=60\,\mathrm{au}$, from where the profile gets exponentially tapered. The radial grid contains $n_\mathrm{r}=300$ logarithmically spaced cells and spans from 5 to 1000 au, as we are only interested in the outer parts of the disk.

The midplane temperature profile is described by a power law of a flaring passively irradiated disk with 
\begin{equation} \label{eq:temp_profile}
    T(r) = T_\mathrm{eff}\,\left(\frac{0.5 \,\varphi_\mathrm{irr}\, R_*^2}{r^2} \right)^{1/4} \approx 160\,\mathrm{K}\left(\frac{r}{1\,\mathrm{au}}\right)^{-1/2},
\end{equation}
where the stellar effective temperature $T_\mathrm{eff}=3500\,\mathrm{K}$, the stellar radius $R_*=2\,R_\odot$ and the irradiation angle $\varphi_\mathrm{irr}=0.05$. We choose this lower than solar value of $T_\mathrm{eff}$, to shrink the width of our imposed gaps, which are determined by the pressure scale height and therewith the disk's temperature (see next section). Additionally, we assume the dust and gas to have the same temperature and the disk to be vertically isothermal.

For the mass distribution of the dust particles, we also take a logarithmically spaced grid with $n_\mathrm{m}=141$ cells, covering sizes from $0.5 \,\mu\mathrm{m}$ up to $2\,\mathrm{m}$. The initial dust-to-gas ratio is set to be $\epsilon_0=0.01$ and initially the dust particles will have a size distribution $n(a)\propto a^{-3.5}$, following the so-called MRN distribution of interstellar grains by \cite{Mathis1977} with a maximum grain-size of $\mathrm{a_0}=1\,\mathrm{\mu m}$. We assume the grains to be compact spheres with a mass given by $m=4\pi a^3 \rho_\mathrm{s}/3$ and their material density $\rho_\mathrm{s}=1.675\,\mathrm{g/cm^3}$, in line with our opacity composition. Finally, we assume the fragmentation velocity to be $v_\mathrm{frag}=10\,\mathrm{m/s}$ throughout the whole disk \citep{Gundlach2011, Gundlach2015}. An overview of the disk model parameters are listed in Table \ref{table:model_param}.

    \begin{table}[h]
    \centering
    \begin{threeparttable}
     \caption[]{Disk model parameters.}
    \label{table:model_param}
    \begin{tabular}{l l c l}
     \hline \hline
    Symbol & Description & Value & Unit \\
     \hline
   $\alpha_0$ & Viscosity parameter & 0.001 & \\
   $M_*$ & Stellar mass & 1.0 & $M_\odot$ \\
   $R_*$ & Stellar radius & 2.0 & $R_\odot$ \\
   $T_\mathrm{eff}$ & Stellar effective temperature & 3500 & K \\
   $\varphi_\mathrm{irr}$ & Irradiation angle & 0.05 & rad \\
   $M_\mathrm{disk}$ & Initial disk mass & 0.05 & $M_\odot$ \\
   $R_\mathrm{c}$ & Characteristic radius & 60 & au \\
   $\rho_\mathrm{s}$ & Dust material density & 1.675 & g/$\mathrm{cm^3}$ \\
   $\epsilon_0$ & Initial dust-to-gas ratio & 0.01 &  \\
   $v_\mathrm{frag}$ & Fragmentation velocity & 10 & m/s \\

    \hline
    \end{tabular}
    \end{threeparttable}
    \vspace{-3.0mm}
    \end{table}

\subsection{Pressure bump variables}
\label{sec:Pressure_bump_var}

    \begin{table}[t]
    \centering
    \begin{threeparttable}
     \caption[]{Parameter space of the `planetary' pressure bumps.}
    \label{table:bump_param}
    \begin{tabular}{l l c l}
     \hline \hline
    Variable & Description & Value & Unit \\
     \hline
    $r_\mathrm{gap}$ & Gap location & [10, $\mathbf{30}$, 80] & au \\
    $w_\mathrm{gap}$ & Gaussian bump width & [0.10, \textbf{0.12}, 0.15] & $r_\mathrm{gap}$ \\
    $T_\mathrm{ini}$ & Initial formation time & 1000 & yrs \\
    $T_\mathrm{fin}$ & Final formation time & [0.5, $\mathbf{1}$, 5, 10]$\times 10^5$ & yrs \\
    $A_\mathrm{max}$ & Bump amplitude & [1, $\mathbf{2}$, 4, 10] & \\
    $M_\mathrm{planet}$ & Planet mass & [0.2, $\mathbf{0.3}$, 0.6, 1.4] & M$_\mathrm{jup}$ \\
   
    \hline
    \end{tabular}
    \begin{tablenotes}
      \small
      \item NOTE - Values of the reference \textit{Fiducial} setup are written in \textbf{bold}. Planet masses are inferred from the above bump amplitudes at $r_\mathrm{gap}$=30\,au using Equations 10 \& 11 of \cite{Kanagawa2018}.
      \vspace{-2mm}
    \end{tablenotes}
    \end{threeparttable}
    \end{table}

To study the nature of pressure bumps and their observables, we want to explore a variety of possible parameters in our pressure bump models. The key parameters to be varied are the gap location $r_\mathrm{gap}$ and the time-dependent amplitude of the Gaussian bump $A(t)$. The latter is defined through its maximum amplitude $A_\mathrm{max}$ and the growth time $T_\mathrm{grow}=T_\mathrm{fin}-T_\mathrm{ini}$ (see Eq. \eqref{eq:amplitude}). To this end, we vary each of these parameters with respect to a \textit{Fiducial} model, while the other parameters are kept constant. The parameters of the \textit{Fiducial} model, which later will be taken as a reference point, are $r_\mathrm{gap}=30$ au, $A_\mathrm{max}=2$ and $T_\mathrm{fin}=0.10$ Myr. An overview of the parameter space for the `planetary' pressure bump models can be found in Table \ref{table:bump_param}.

Initially, we include the bumps at $T_\mathrm{ini}=1000\,\mathrm{yrs}$ and then let them grow over several thousand years. However, how fast bumps can form is locally governed by the viscous timescale \citep{armitage10}. Inserting $\Delta r = w_\mathrm{gap}$ and simplifying, we obtain:
\begin{equation} \label{eq:t_visc_bump}
    \tau_\mathrm{visc}(r)\approx \frac{(\Delta r)^2}{\nu} =\frac{w_\mathrm{gap}^2}{\nu}=\frac{4}{\alpha_0\, \Omega_\mathrm{k}},
\end{equation}
since we induce them in $\alpha$. For the Gaussian width of the bump, in the above equation, we take $w_\mathrm{gap}=2\,h_\mathrm{g}(r_\mathrm{gap})$ at the gap location and also inserted the kinematic viscosity. The width needs to be larger (or equal) than $h_\mathrm{g}$ to ensure the stability of the bumps \citep[][]{Li2000, Ono2016}. This timescale is inversely proportional to the Keplerian frequency, so outer disk regions have longer viscous time scales, and accordingly bumps in the outer parts will take longer to grow, which naturally sets a lower limit of $T_\mathrm{grow}\geq \tau_\mathrm{visc}(r_\mathrm{gap})$.

In order to actually create pressure bumps, where dust particles can be trapped, it is essential that there is a radius where the pressure gradient changes sign, meaning there exists a local pressure maximum in the disk \citep{Pinilla12}. For radially increasing gap locations, the pressure scale height, thus the Gaussian width $w_\mathrm{gap}$ and the disk's aspect ratio (${h_\mathrm{g}/r \propto r^{1/4}}$) increase, leading to a smaller Gaussian factor in Eq. \eqref{eq:alpha_bump} and a shallower pressure gradient. This puts a constraint on the amplitude of the bump $A_\mathrm{max}$ which needs to be large enough to fulfil the above condition. To this end, we take a higher value $A_\mathrm{max}=3$ at the gap location of $r_\mathrm{gap}=80\,\mathrm{au}$, to ensure to create a particle trap.

\subsection{Dynamic bumps model}
\label{sec:setup_dynamic_model}
For the \textit{Dynamic Bumps} model, we do not vary the parameters for certain bumps in terms of their amplitude and location, but rather want to focus on which effect the bumps' lifetimes have on the disk's observational properties. As described in Sect.~\ref{sec:dynamic_model} the disk will be divided up into several zones. Since we are only interested in the outer parts of the disk, we restrict the regions where bumps can form to be between 8 - 100 au. We set the Gaussian bumps width to be $w_\mathrm{gap}=2h_\mathrm{g}$, in line with the radial extent of vortices found in global MHD-simulations \citep[e.g.][]{Flock2015,Cui_Bai2021}. As explained in the previous section, we take $A_\mathrm{max}=3$, as to create efficient dust traps across the whole disk. However, this leads to the fact that dust trapping will be more efficient in the inner zones, where the pressure gradient will be steeper due to the flared nature of our disk ($h_\mathrm{g}/r\propto r^\mathrm{1/4}$). We assume that all bumps start forming at $T_\mathrm{form}=500 \,t_\mathrm{orb}$ local orbit times. We then run three simulations varying the lifetimes of the dynamic bumps by $T_\mathrm{life}=(200, 400, 1000)\,t_\mathrm{orb}$.

We want to compare these dynamic bumps models to a static (`planetary') one. To accurately do that, we set up another simulation where we induce three bumps at $r_\mathrm{gap}= [10, 30, 80]$ au into the disk. They take $T_\mathrm{form}=500 \,t_\mathrm{orb}$ to appear and then another $T_\mathrm{grow}=200\,t_\mathrm{orb}$ to reach their final amplitude of ${A_\mathrm{max}=3}$, after which they remain unchanged in the disk.

\subsection{Radiative transfer calculations}
\label{sec:Rad_transfer_calc}
To infer the observational appearance of the disk, we perform radiative transfer calculations with the code \texttt{RADMC-3D}\footnote{The freely available code and its manual can be found on: \url{www.ita.uni-heidelberg.de/~dullemond/software/radmc-3d/}, version 2.0} \citep{Dullemond2012}. As input, we take the 3D dust density distribution (see Eq. \eqref{eq:rho_dust_3D}) for each grain size obtained from our \texttt{DustPy} simulations at an evolution time of 1 Myr.

We assume the composition of the grains as stated in the Disk Substructures at High Angular Resolution Project (DSHARP) paper of \cite{birnstiel18}, consisting of water ice \citep{warren&brandt08}, astronomical silicates \citep{draine2006}, as well as troilite and refractory organics \citep{henning1996}. We then calculate the opacities for each grain size using the program \texttt{Optool}\footnote{Freely available on: \url{www.github.com/cdominik/optool}, version 1.7.2} \citep{Dominik2021}, where we further assume the grains to be compact spheres. The scattering phase function for grains that are comparable in size to the modelled wavelength has a strong forward peak. This poses a problem, due to the limited resolution of our radiative transfer model. Therefore, we set the scattering opacity of the grains to an upper limit within the first 5° of the phase function, which thereafter needs to be re-normalised again. This means that within those first degrees forward scattering will effectively be treated as no scattering at all.

In accordance with our disk model, we set up a Solar mass star with an effective temperature of $T_\mathrm{eff}=3500\,\mathrm{K}$ which is treated as a blackbody point source and emits $10^7$ photon packages for our radiative transfer calculations. The spherical grid of the model has $N_\mathrm{r}=300$ radial, $N_\phi=64$ azimuthal and $N_\theta=180$ polar cells with the star at its centre. The latter corresponds to three or six cells per pressure scale height at 10 and 100 au, respectively. The wavelength grid, which is needed to compute the dust temperature, spans from 0.1 $\mu \mathrm{m}$ to 1 cm and contains 150 cells.

First of all, the 3D dust equilibrium temperature gets computed, which is then used to calculate synthetic observations of the disk in the 1.3\,mm and 3.0\,mm dust continuum. Hereby, we use the full treatment of anisotropic-scattering of dust grains with polarisation \citep{Kataoka2015} of the code. Even though, it is computationally expensive, we later want to compare this to models where we do \textit{not} treat scattering and investigate the differences. 

For all calculations, we assume the disk to have an inclination of $i=\mathrm{20^\circ}$ and to be at a distance of $d=140\,\mathrm{pc}$, which is the mean value of the nearest star forming regions. In a final step, the synthetic images get convolved by a Gaussian point-spread-function with a full-width at half-maximum (FWHM) of 0.04" $\times$ 0.04" to mimic the current angular resolutions obtained by interferometers like ALMA \citep[e.g.][]{andrews18} and to make sure to resolve the disk's substructures.

\begin{table*}[t]
    \centering
    \small
    \begin{threeparttable}
    \caption{Observational disk properties of different pressure bump models}
    \label{table:disk_properties}
    \begin{tabular}{l c c c c c c c c c c c c c c}
     \hline \hline
     Name & $r_\mathrm{gap}$ & $A_\mathrm{max}$ &  $T_\mathrm{fin}$ &   $M_\mathrm{d}^\mathrm{1Myr}$  & $M_\mathrm{d}^\mathrm{3Myr}$ &  $M_\mathrm{d,>0.1\mathrm{mm}}^\mathrm{1Myr}$ & $M_\mathrm{d, obs}$&  $F_\mathrm{1.3mm}$ & $F_\mathrm{3.0mm}$ & $\alpha_\mathrm{mm}$ &  $R_\mathrm{d,68\%}$ & $R_\mathrm{d,90\%}$ & $\tau_\mathrm{abs}^\mathrm{peak}$ &
     $\tau_\mathrm{obs}^\mathrm{peak}$ \\
    
      & [au] &  & [Myr]& [$\mathrm{M_\oplus}$] & [$\mathrm{M_\oplus}$] & [$\mathrm{M_\oplus}$] & [$\mathrm{M_\oplus}$] &  [mJy] & [mJy]  & & [au] & [au] & & \\
    
    (1) & (2) & (3)& (4)&(5)&(6)&(7)&(8)&(9)&(10)& (11)&(12)&(13)& (14)& (15) \\
    
     \hline
     Inner Bump	& 10 & 2 & 	0.10 & 115	& 	100 & 102 &	9.9 &  17.9 & 2.0 &	2.6 &	13	&16 & 35 &0.41\\
     \hline
     Fiducial & 30 & 2 &	0.10	& 87.3 & 	77.7 & 71.2 & 26.9	& 48.5 &	5.1	&2.7&	38&	42 & 3.5 & 0.70\\ 
     \hline
     Smaller A & 30 & 1 &	0.10 & 16.5 &  2.4 & 11.4 & 11.8	&  21.3	& 1.2 & 3.4& 36 & 41 & 0.26 & 0.19 \\ 
     \hline
     Bigger A & 30 & 4 & 0.10 & 94.7 & 	92.4 & 77.2 & 24.7	 & 44.6	& 4.8	& 2.7	& 40	& 44 & 3.8 & 0.66\\ 
     \hline
     Strong A & 30 & 10 & 0.10 & 97.0  & 96.3 & 78.7  & 23.1	& 41.7 & 4.6 & 2.6 & 42 & 46 & 3.9  & 0.59\\
     \hline
     Fast T & 30 & 2 &	0.05 & 90.5 & 80.8 & 74.1 & 27.2 & 49.1	&5.2	&2.7	&38 &	42 & 3.6 & 0.71\\ 
     \hline
     Long T & 30 & 2 &	0.50	& 52.5 & 44.9 & 41.6 & 24.0	& 43.2 & 3.9 &	2.9 &	38 & 42 & 1.8 & 0.63\\ 
     \hline
     Longer T & 30 & 2 &	1.0	& 30.6 & 24.9 & 23.0 & 19.4	& 35.0 & 2.4 & 3.2 &	38&	42 & 0.86 & 0.47 \\ 
     \hline
     Outer Bump & 80 & 3 &	0.20 & 38.9 & 37.0 & 24.2 & 19.0 & 34.3 & 1.6 &	3.7 & 106 & 110 & 0.27 & 0.24 \\ 
     \hline
     Two Bumps & 30\&80 & 2\&3 & 0.1\&0.2 & 88.1 & 79.6 & 66.3 & 32.7 & 58.9 & 4.8 & 3.0 & 39 & 100 & 1.9 & 0.60  \\
      \hline
     No Bump & -- & -- & -- & 6.8  & 1.7 & 3.4  & 6.5 & 11.7 & 0.5 & 3.7 & 35 & 55 & 0.02 & 0.02 \\
    \hline
    \end{tabular}
    \begin{tablenotes}
      \small
      \item NOTE - (1) Given name of the pressure bump model. (2),(3),(4) Various model parameters as described in Sect.~\ref{sec:press_bump}. (5),(6) Total dust masses at 1 Myr and 3 Myr of disk evolution. (7) Dust mass of grains larger than 0.1 mm at 1 Myr. (8) Observed disk dust mass (Eq. \eqref{eq:dust_mass_obs}). (9), (10) Synthetic fluxes of the disk at 1.3\,mm and 3.0\,mm. (11) Millimetre spectral index (Eq. \eqref{eq:spec_index}). (12), (13) Dust disk radii that encloses either 68\% or 90\% of the 1.3\,mm flux. (14), (15) Peak absorption and observed optical depth at location of dust ring, both at $\lambda=1.3$ mm (Eq. \eqref{eq:opt_depth}, \eqref{eq:opt_dept_obs}).
    \end{tablenotes}
    \end{threeparttable}
    \end{table*}

\begin{figure*}[t]
    \centering
    \includegraphics[width=60mm]{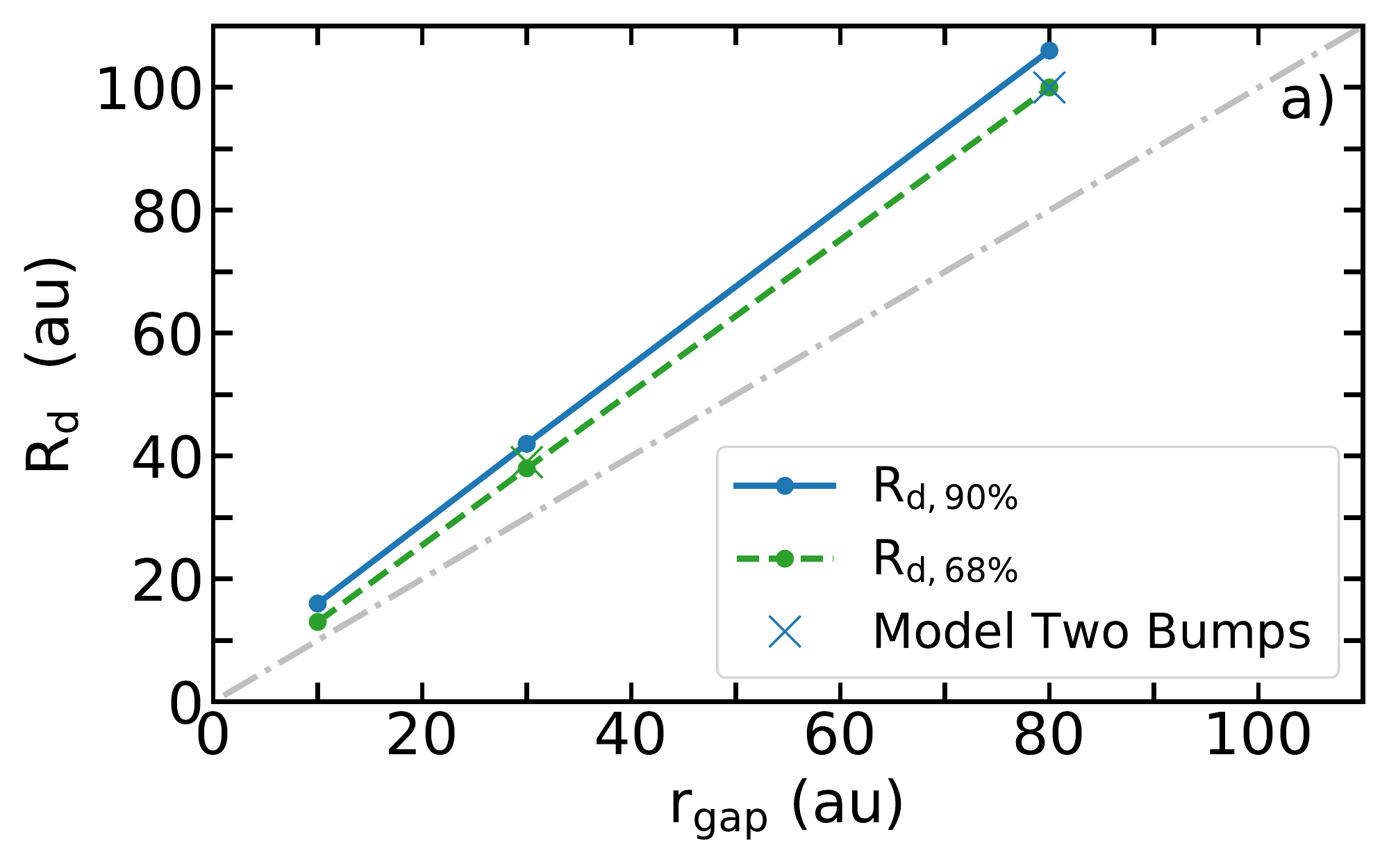}
    \includegraphics[width=60mm]{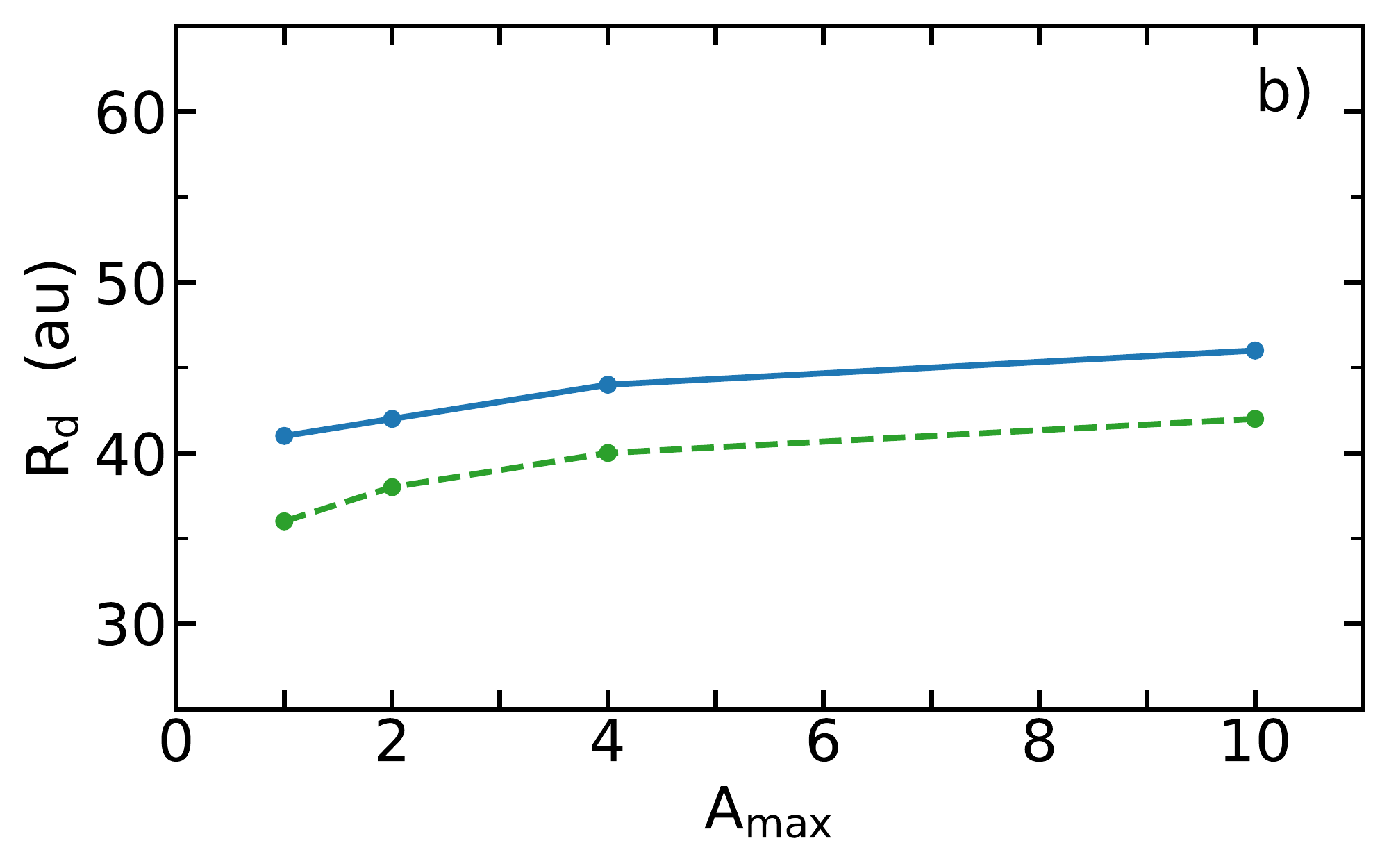}
    \includegraphics[width=60mm]{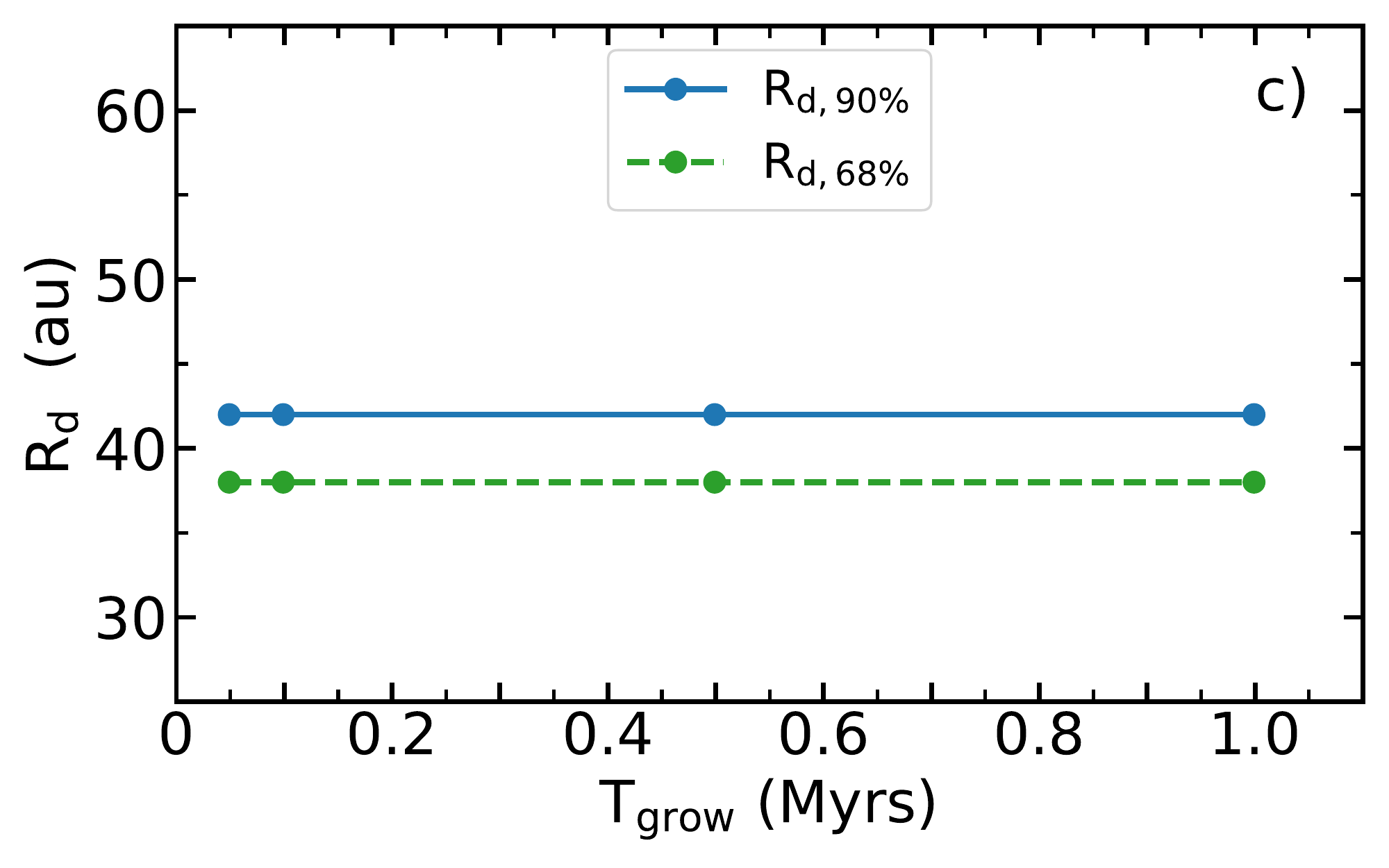}
    \caption{Dust disk radius parameter plots. Dust disk radii enclosing 68\% (green) or 90\% (blue) of the 1.3\,mm cumulative flux plotted over the gap location, bump amplitude and bump growth time (from a) - c)).  In sub-plot a) the cross markers correspond to the \textit{Two Bumps} model and the dashed-dotted line shows the 1:1-line.}
    \label{fig:radius_effect}
\end{figure*}

From the convolved intensity profiles of the disks, we can calculate their millimetre fluxes at the two wavelengths, as well as their dust disk radii. The latter is defined through the radius where 68\% or 90\% of the emission at 1.3\,mm (or additionally 3.0\,mm) is enclosed, by assuming that the sensitivity of the observations has a peak signal-to-noise ratio, with respect to the maximum of the emission, of 1000. We choose this peak SNR, instead of a constant value, following \cite{pinilla21}. The reason behind this is that each observation has a different sensitivity, depending on the total flux of the disk, as well as on the limitations of the integration time. Current observations conducted with ALMA are able to achieve this SNR for the brightest disks, however for fainter disks this might not be achievable without too long integration times.

Furthermore, we calculate the disk's radially integrated millimetre spectral index taking its fluxes at 1.3\,mm and 3.0\,mm as
\begin{equation}
\label{eq:spec_index}
    \alpha_\mathrm{mm} = \frac{\log_\mathrm{10}\left(F_\mathrm{1.3mm}/F_\mathrm{3.0mm}\right)}{\log_\mathrm{10}\left(3.0/1.3\right)}.
\end{equation}

Another quantity, which we want to infer from our synthetic observations is the dust mass, that can be calculated through the optically thin approximation of \cite{hildebrand1983}:
\begin{equation}\label{eq:dust_mass_obs}
    M_\mathrm{d, obs} \simeq \frac{d^2 \, F_\mathrm{1.3mm}}{\kappa_\nu \, B_\nu (T(r))},
\end{equation}
where $d$ is the distance to the source, $F_\mathrm{1.3mm}$ is the flux taken at $\lambda=1.3\,$mm and $B_\nu$ is the Planck function, for which mostly a blackbody temperature of $T=20\,\mathrm{K}$ is assumed. For the opacity, we take a frequency-dependent prescription, that is frequently used in disk surveys and is given by $\kappa_\nu = 2.3\,\mathrm{cm^2\,g^{-1}}\times (\nu/230\,\mathrm{GHz})^{0.4}$ \citep{andrews2013}. It is important to stress that this relation only holds as long as most of the disk has an optical depth of $\tau\leq 1$, otherwise the optically thin approximation breaks down and the dust mass cannot accurately be traced anymore.

\begin{figure*}[t]
    \centering
    \includegraphics[width=60mm]{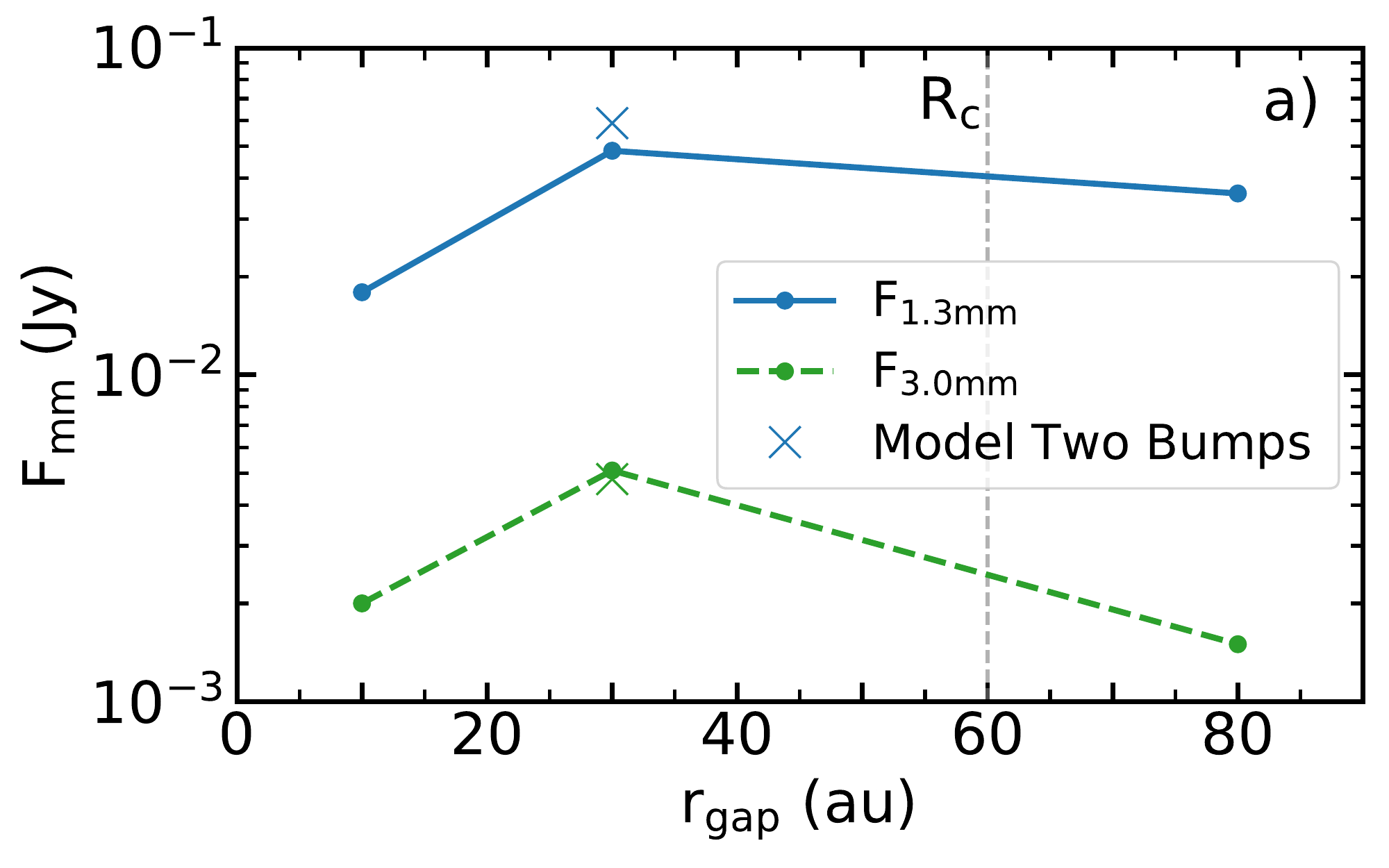}
    \includegraphics[width=60mm]{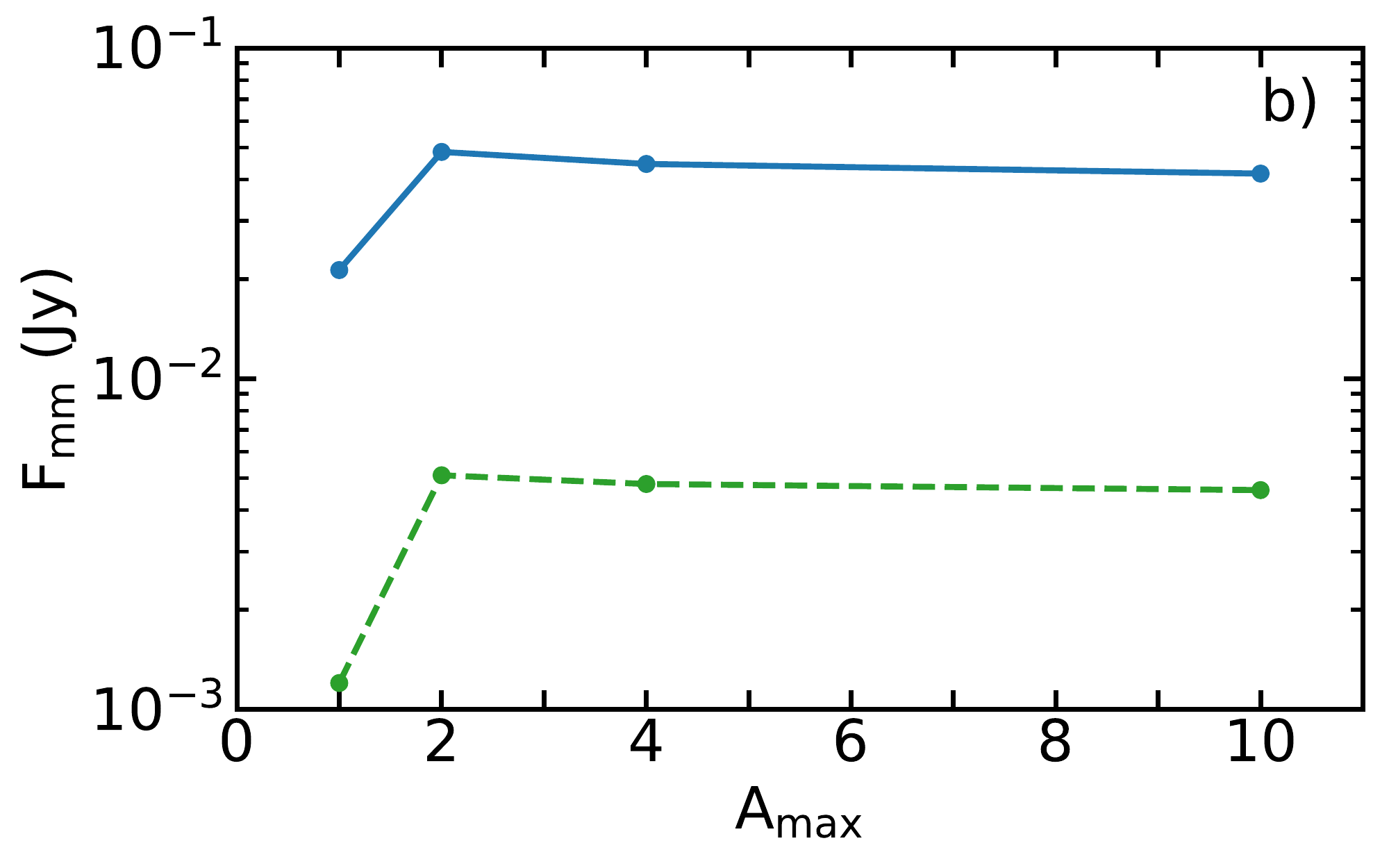}
    \includegraphics[width=60mm]{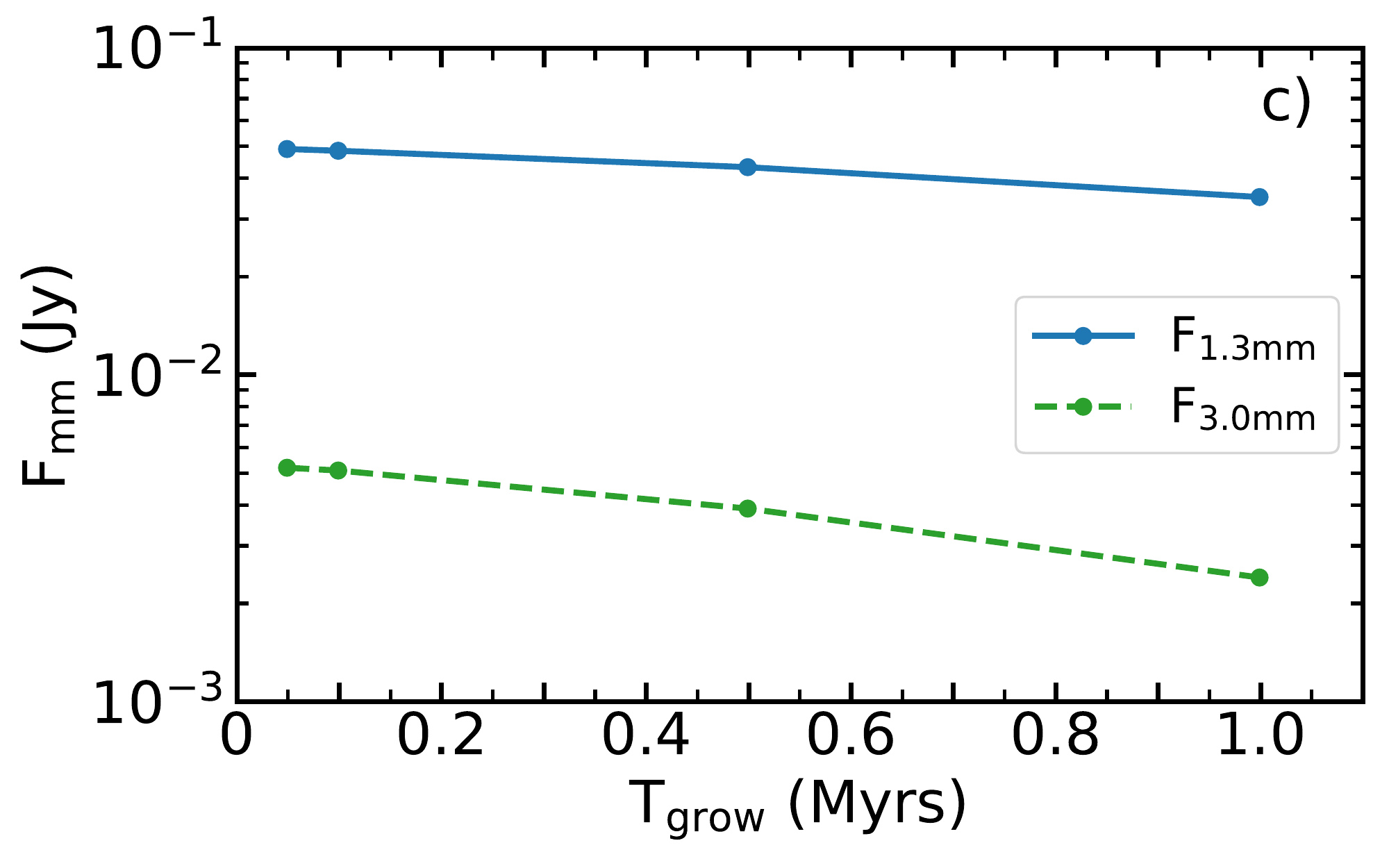}\\
    \includegraphics[width=60mm]{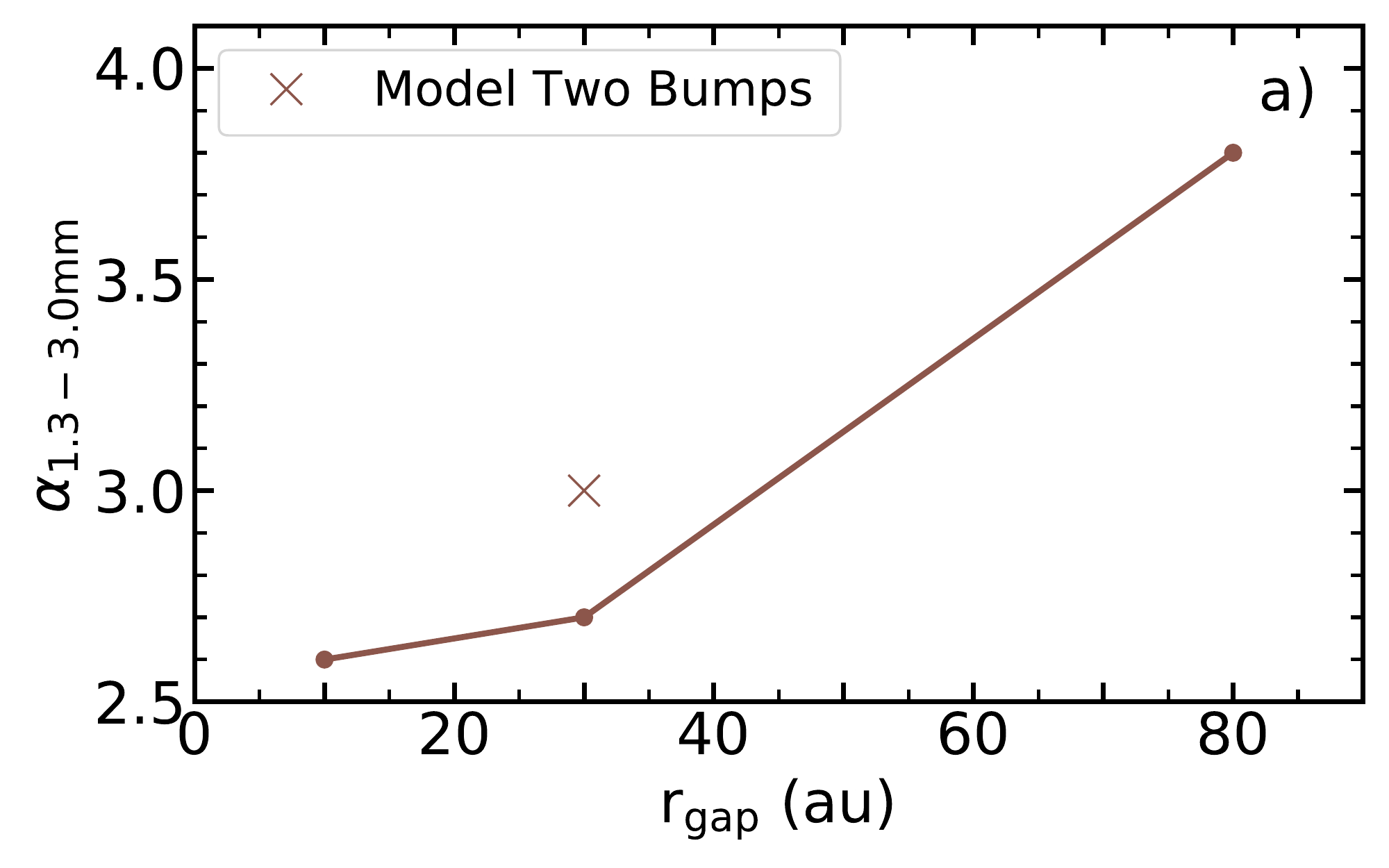}
    \includegraphics[width=60mm]{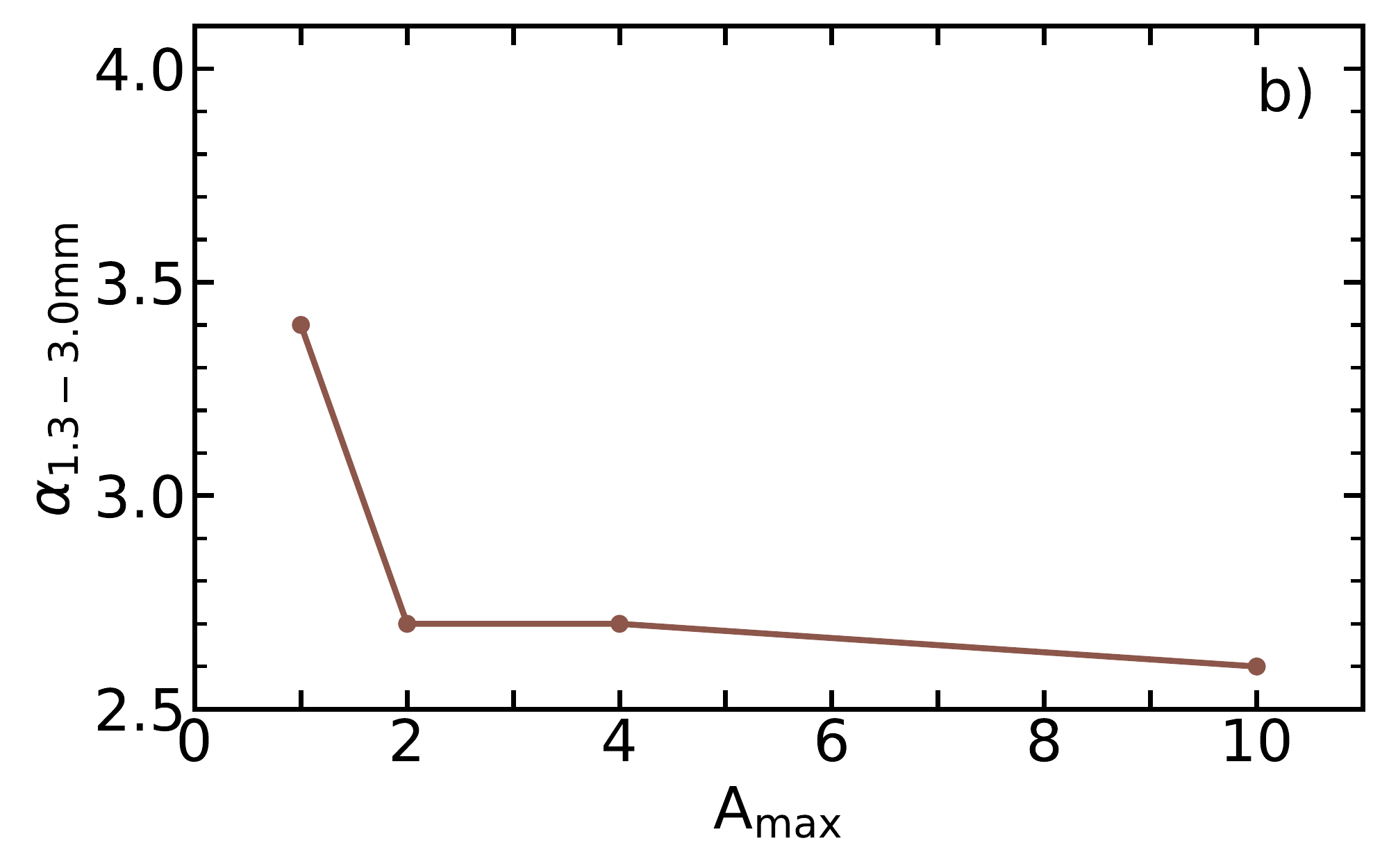}
    \includegraphics[width=60mm]{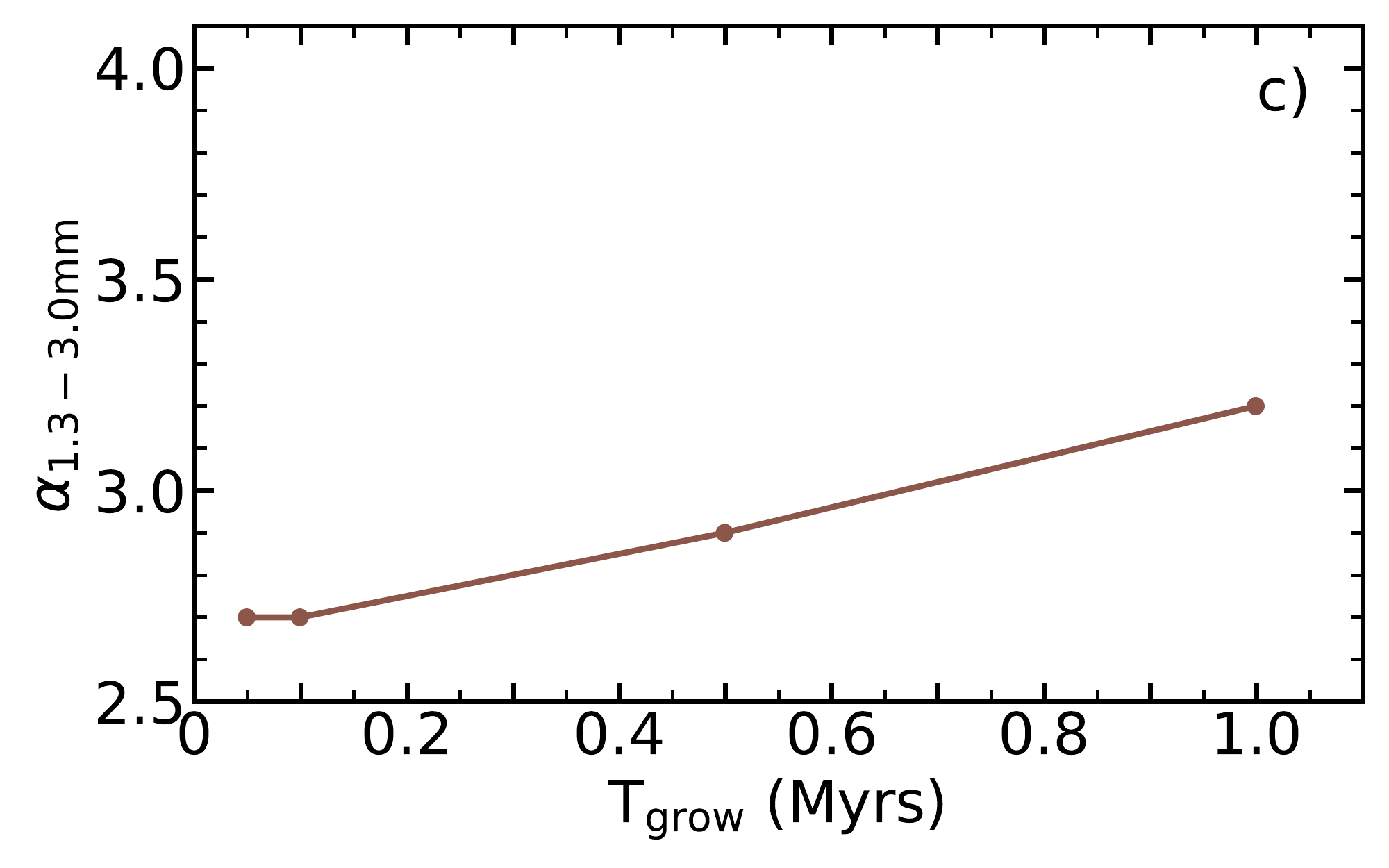}
    \caption{Flux and spectral index parameter plots. \textsl{(Upper)} Disk millimetre fluxes at 1.3\,mm (blue) and 3.0\,mm (green) and according millimetre spectral indices \textsl{(Lower)} plotted over the gap location, bump amplitude and bump growth time (from a) - c)).}
    \label{fig:flux_alpha_effect}
\end{figure*}

Lastly, we actually want to infer the optical depth from our synthetic observations and compare this to the absorption optical depth of our models:
\begin{equation} \label{eq:opt_depth}
    \tau_\mathrm{\nu\,,abs}(r) = \sum^{N}_{i} \Sigma_\mathrm{d}^i(r) \,\kappa_\nu^{i,\mathrm{abs}},
\end{equation}
with the absorption opacity $\kappa_\nu^{i,\mathrm{abs}}$ calculated from the DSHARP opacities \citep{birnstiel18} and where the sum index corresponds to the different dust species. The highest emission of our bump models is naturally measured at the peak location of dust ring $r_\mathrm{peak}$, as can be seen in Fig.~\ref{fig:SynImages_res}. Therefore, we also want to focus on computing the optical depth at this peak location, since there we might expect the emission to become optically thick. Observationally, this is usually calculated by solving the radiative transfer equation for the optical depth \citep[see Eq. 11 of][]{dullemond18}
\begin{equation} \label{eq:opt_dept_obs}
\begin{aligned}
    & \,I_\nu(r)=B_\nu(T_\mathrm{d}(r))(1-\exp{[-\tau_\nu(r)]}) \\
    \Leftrightarrow &\, \tau_\mathrm{obs} = - \ln \left(1-\frac{I_\mathrm{1.3mm} (r_\mathrm{peak})}{B_\mathrm{1.3mm}(T_d(r_\mathrm{peak}))} \right),
\end{aligned}
\end{equation}
where both the intensity $I_\nu$ and the temperature of the blackbody function $B_\nu$ get evaluated for $\lambda = 1.3\,$mm. For the temperature $T_\mathrm{d}$, we use the profile introduced in equation \eqref{eq:temp_profile}.

\section{Results}
\label{sec:results}
\subsection{Parameter effects on observables} \label{sec:parameterstudy}
In this subsection, we will present the parameter study of our `planetary' pressure bump model and how the three main variables, namely the gap location $r_\mathrm{gap}$, the maximum amplitude $A_\mathrm{max}$ and the growth time of the bump $T_\mathrm{grow}$, affect the observational properties of our disk. To this end, we vary each of the parameters with respect to our \textit{Fiducial} model, while the other parameters are kept constant. An overview of the different models with their corresponding parameters and observational properties can be found in Table \ref{table:disk_properties}.

\subsubsection{Dust disk radii} \label{subsec:disk_radii}

The main result of this subsection is that the (farthest out) pressure bump determines the disk dust radius. We show this, investigating how the dust disk radius changes with our different bump models. In Fig.~\ref{fig:radius_effect}, we plotted dust disk radius enclosing 68\% and 90\% of the cumulative flux at 1.3\,mm over each model parameter. An overview plot of all model's millimetre intensity profiles with the location of each emission radius can be found in Fig.~\ref{fig:NormIntenstiy_Profiles} of the Appendix.

In plot \ref{fig:radius_effect} a) we find that the disk radius shows a linear relationship to the gap location of our model. Hence, the (farthest out) pressure bump determines the disk dust radius. Interestingly, the $R_\mathrm{d,68\%}$ and $R_\mathrm{d,90\%}$ (green and blue crosses) of the \textit{Two Bumps} model, plotted in the figure, each trace its corresponding bumps at 30 and 80 au, while the rest of the models have no big difference between their 68\% and 90\% radii. All the data points lie above the 1:1 line, since the cumulative flux traces the emission from the dust rings, which lie exterior to the corresponding gap location. 

The amplitude of the bump has only a minor effect on the disk radius, as long as a particle trap is created ($A_\mathrm{max}\gtrsim 2$). This can be seen in b), where higher amplitudes create deeper gaps that move the local pressure maximum further out, therewith increasing the disk radius by a small factor.

In subplot c), we see that the growth time of the bump has no effect on the dust disk radius at all, which just stays constant with varying $T_\mathrm{grow}$.

\subsubsection{Millimetre fluxes}
Small gap locations ($r_\mathrm{gap}<R_\mathrm{c}$) and short growth times of the bumps are needed to observe high millimetre fluxes. This can be seen in Fig.~\ref{fig:flux_alpha_effect}, where we study the fluxes at $\lambda=1.3\,\mathrm{mm}$ and $\lambda=3.0\,\mathrm{mm}$ dependent on the bump parameters.

\begin{figure*}[t!]
\centering
\includegraphics[width=60mm]{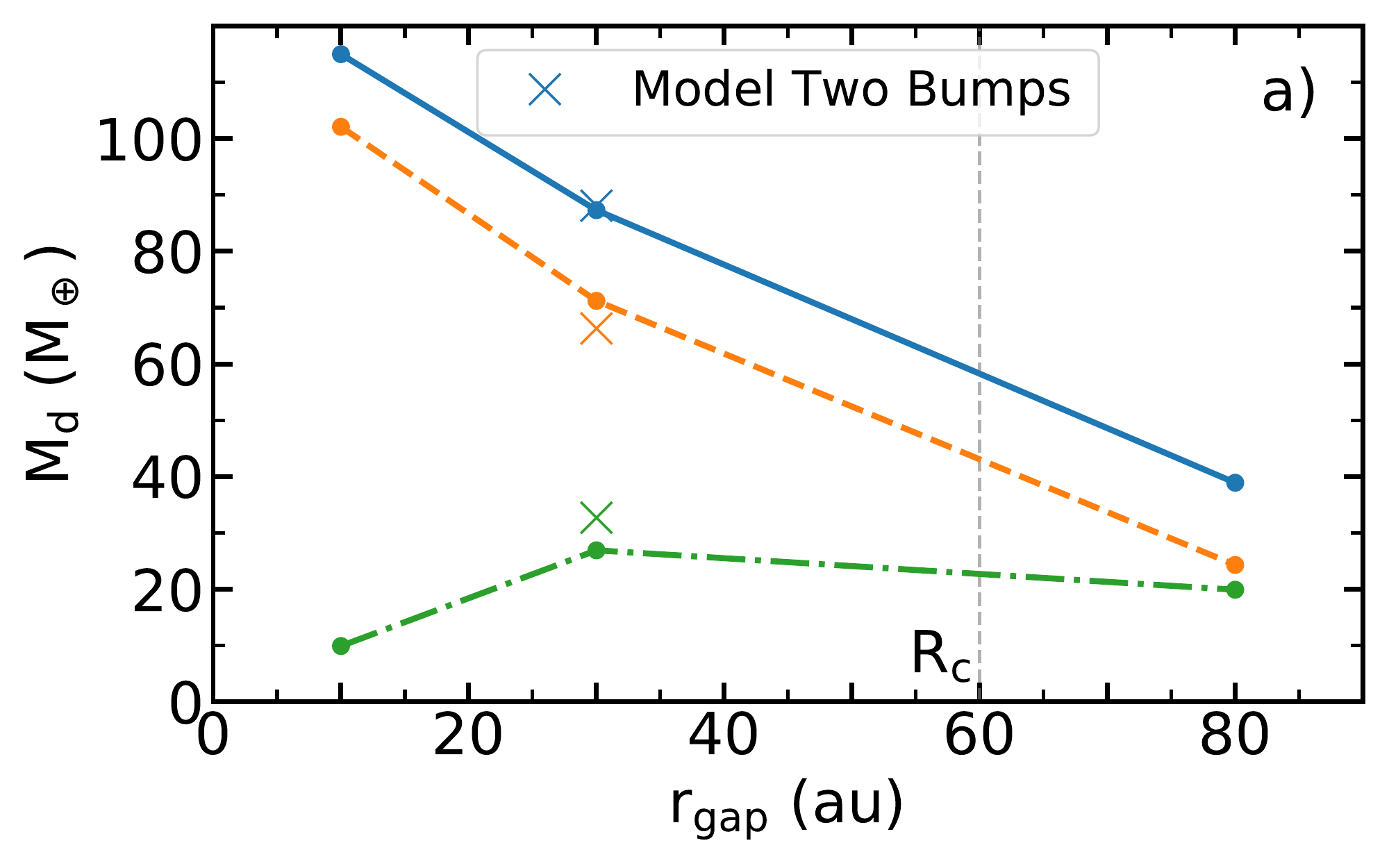}
\includegraphics[width=60mm]{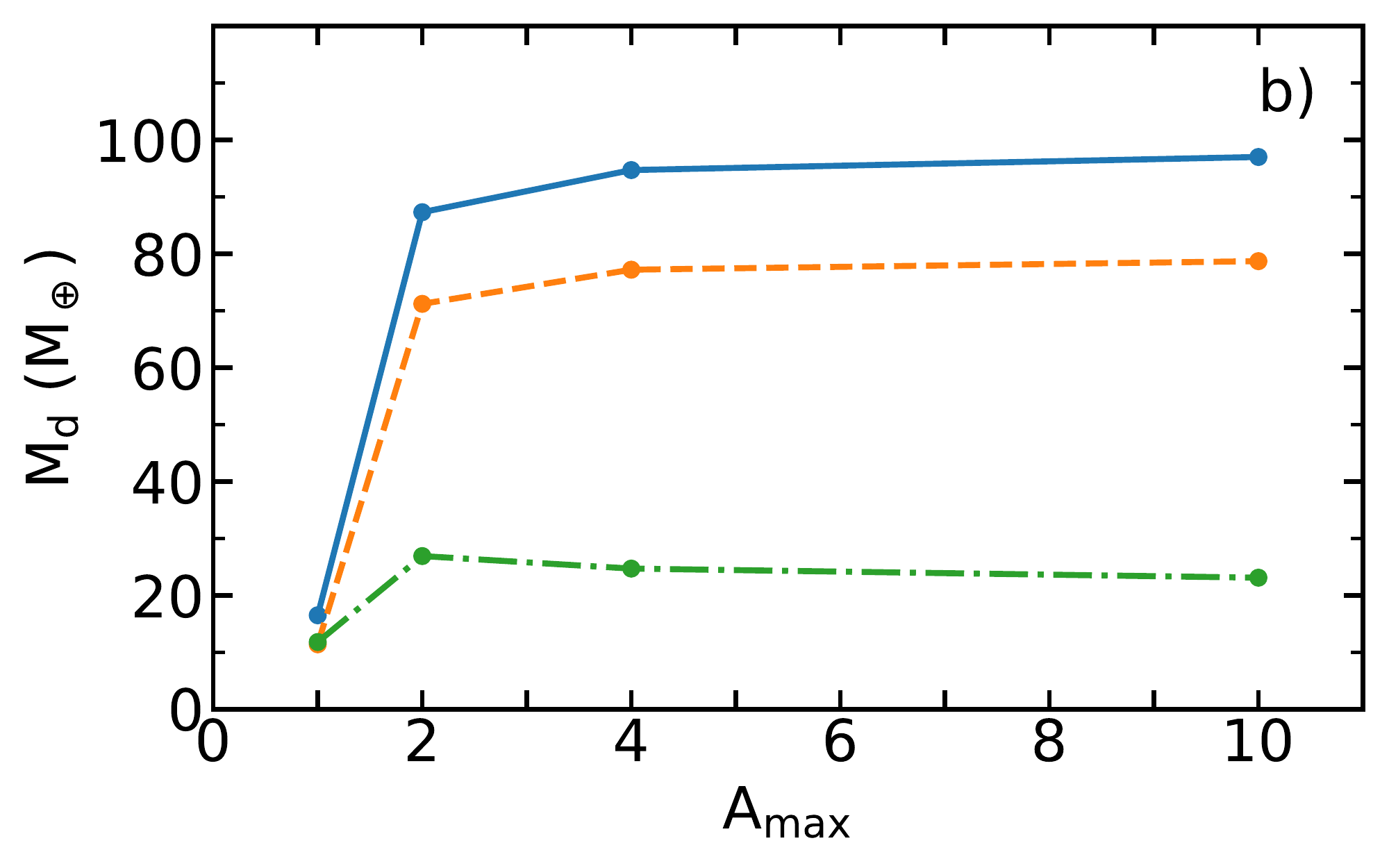}
\includegraphics[width=60mm]{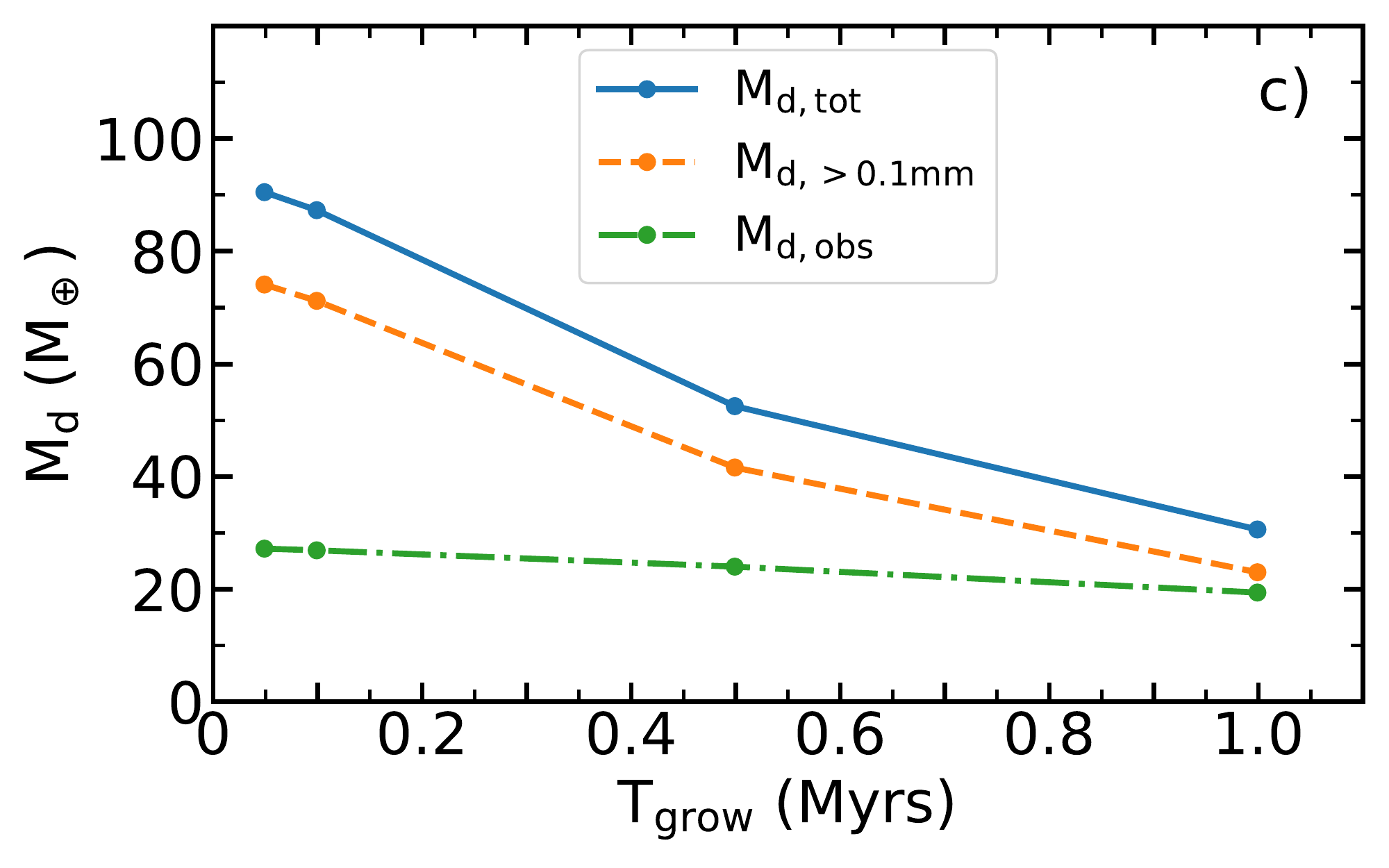}\\
\includegraphics[width=60mm]{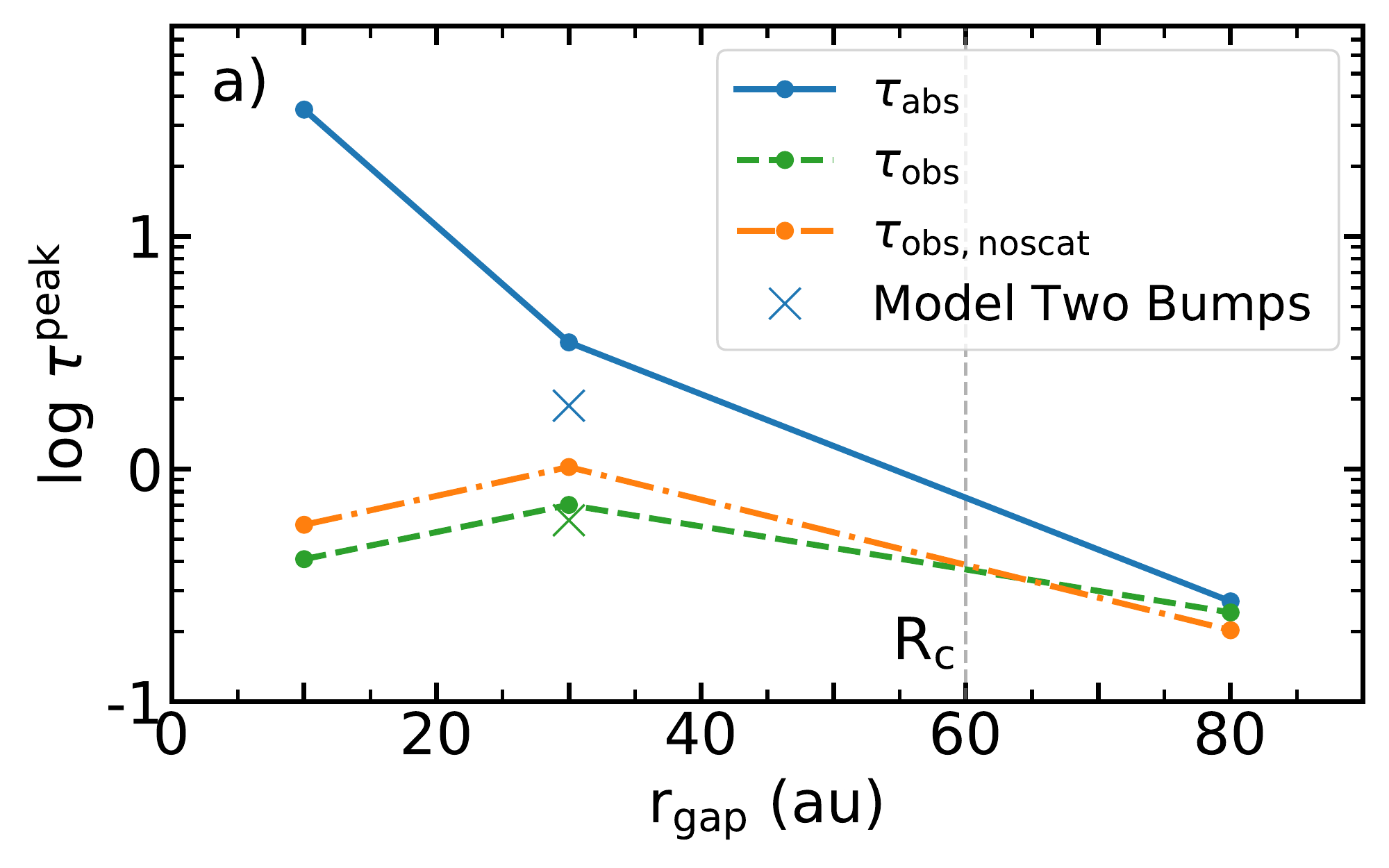}
\includegraphics[width=60mm]{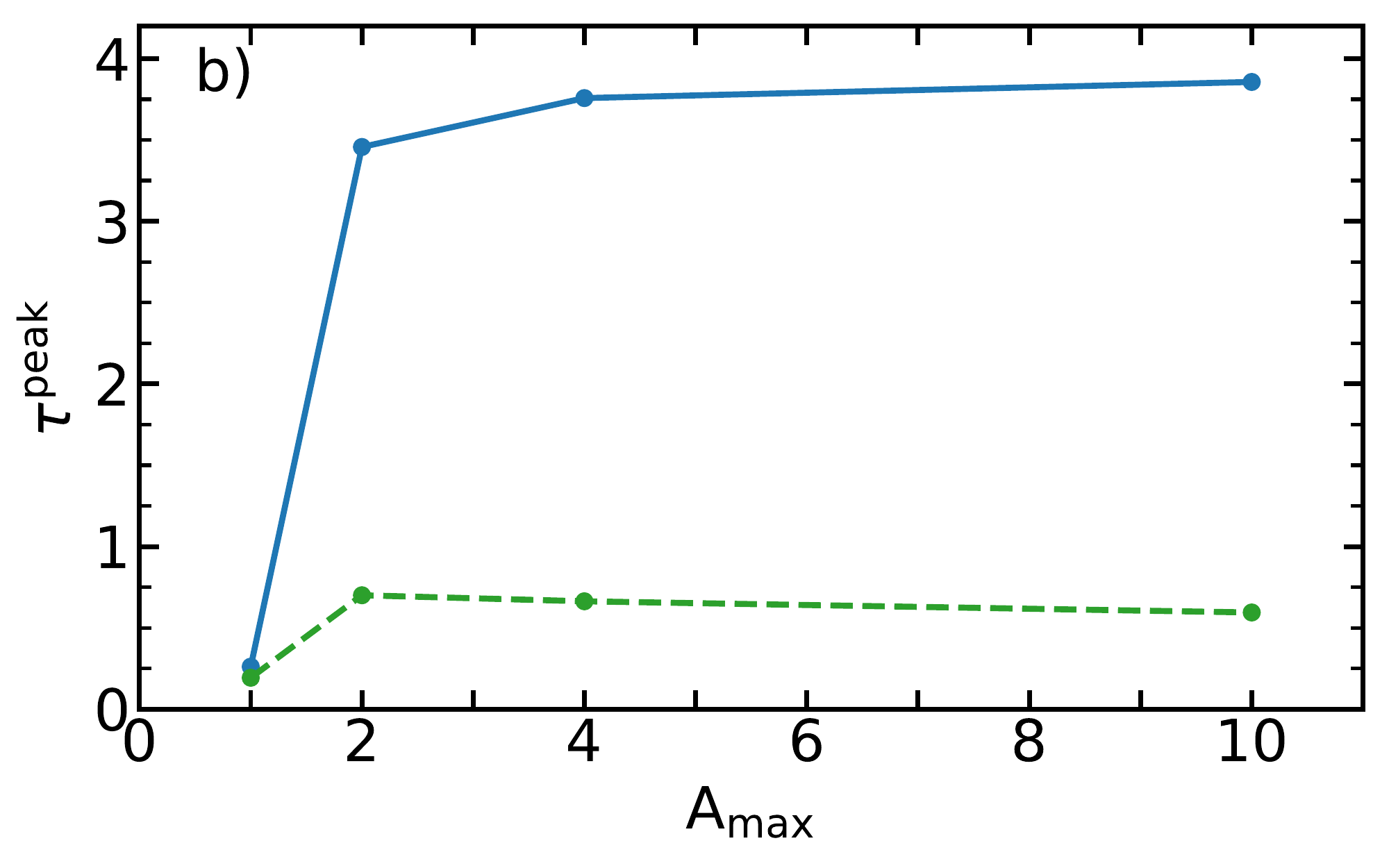}
\includegraphics[width=60mm]{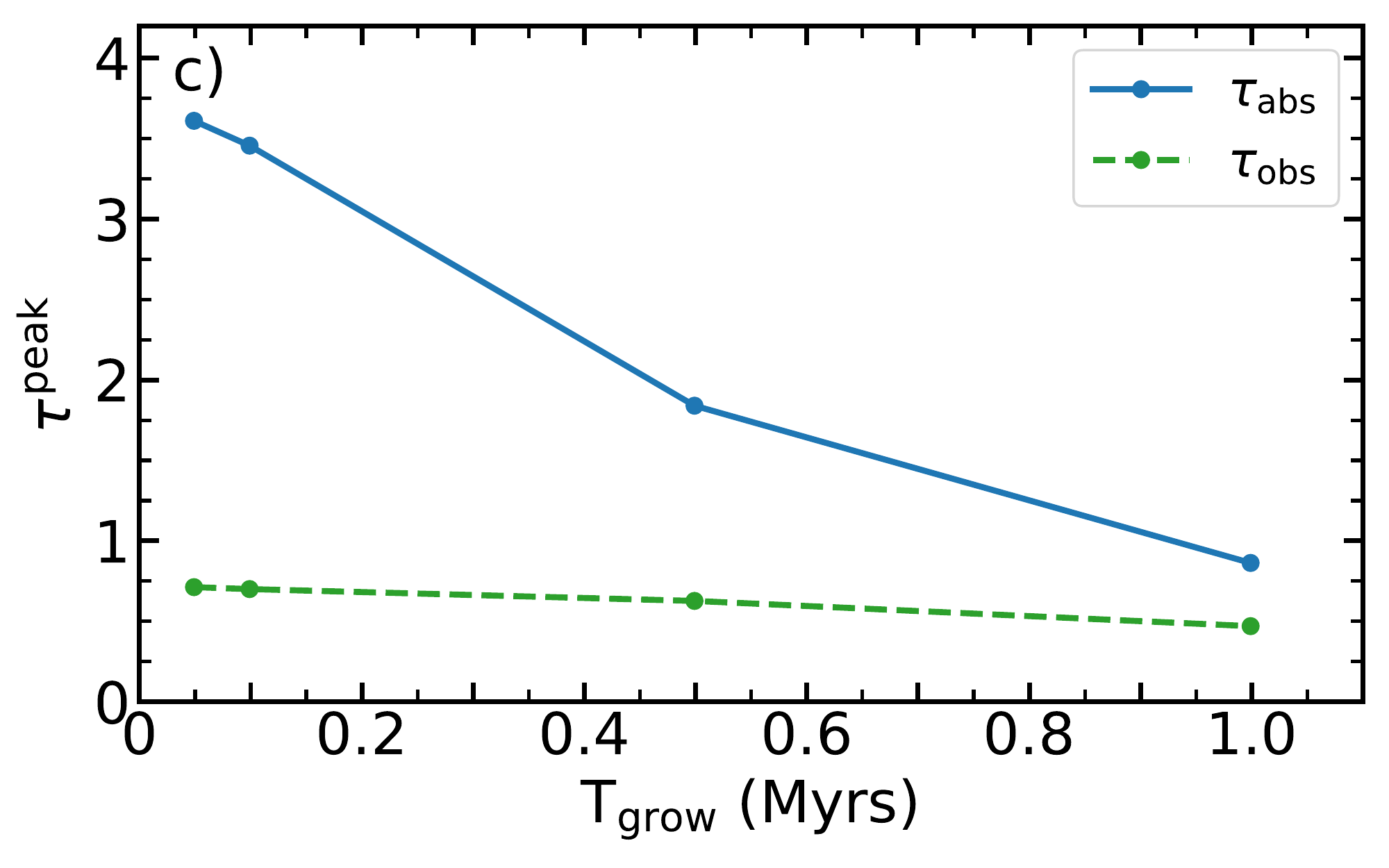}
\caption{Dust mass and optical depth parameter plots. \textsl{(Upper)} Total dust mass (blue), dust mass for grains larger than 0.1 mm (orange) and observed dust mass (green) plotted over the gap location, bump amplitude and bump growth time (from a) - c)). \textsl{(Lower)} Absorption and observed peak optical depth also in dependence on bump model parameters. The \textit{Inner Bump} model reaches $\tau_\mathrm{abs}=35$ which is why we used a logarithmic scale in subplot a)}
\label{fig:Mdust_tau_effect}
\end{figure*}

In the first plot a), we see that the \textit{Fiducial} model, with the gap at 30 au, displays the highest millimetre flux. Although, the model with the gap further in traps more dust, its millimetre flux is lower since the width of the dust ring is narrower ($w_\mathrm{gap}=2\,h_\mathrm{g}(r_\mathrm{gap})$) which leads also to stronger dust trapping around the pressure maximum. This increases the optical depth of the \textit{Inner Bump} model, thereby shielding more dust from our view and reducing its emission. We will explain the influence of the optical depth on our results in more detail in a coming subsection. 

For the \textit{Outer Bump} model, with its gap further out, the flux is lower as it traps less dust, because $r_\mathrm{gap} > R_\mathrm{c}$ and most of the dust mass is located within the characteristic radius. Hence, the bump cannot trap enough dust mass from drifting inwards, in order to grow sufficient millimetre pebbles for high millimetre fluxes. This also explains, why the $F_\mathrm{3.0mm}$ drops more than the $F_\mathrm{1.3mm}$, as longer wavelengths trace larger particles.

A last thing to point out in Fig.~\ref{fig:flux_alpha_effect} a) is that the fluxes for \textit{Two Bumps} model (cross markers) very well coincide with the \textit{Fiducial} one, even though it has a second bump at 80 au. The model's  additional bump only increases $F_\mathrm{1.3mm}$ by 20\%, while $F_\mathrm{3.0mm}$ decreases compared to the \textit{Fiducial} setup.

In subplot b), we investigate the influence of the bump's amplitude on the millimetre fluxes. Here, the only major change in fluxes happens when the amplitude $A_\mathrm{max}$ increases from 1 to 2, because an amplitude factor of 1 is insufficient to actually trap particles. This bump only slows down the particle drift and is not able to halt a large amount of dust mass from accreting onto the star (see also Fig.~\ref{fig:Mdust_tau_effect} c)). The minor decrease in the fluxes for $A_\mathrm{max}>2$ is due to the fact, that for larger bump amplitudes, we increase the pressure gradient. Thus, the grains accumulate closer around the local pressure maximum, leading again to higher optical depths.

The third subplot c) displays the flux dependency on the growth time of the pressure bumps. We find the general trend that both millimetre fluxes decrease for longer growth times. This is straightforward to understand, since the earlier the pressure bump is in the disk, the more dust it can trap from drifting inwards and also the more time dust particles have to grow to larger mm-sizes. We further note, that $F_\mathrm{3.0mm}$ decreases faster for larger growth times than $F_\mathrm{1.3mm}$ which will have a direct effect on the spectral indices, as we will discuss in the following subsection.

\subsubsection{Spectral indices}
Strong and fast-forming dust traps in the disk interior are needed to produce small spectral indices of disks. They are determined through the logarithmic ratio of the mm-fluxes of our disk models and we show their dependence on the bump parameters in Fig.~\ref{fig:flux_alpha_effect}.

The lower left subplot shows, that the spectral index increases for bumps located further out in the disk, which is a consequence of less mm-dust-grains prevalent in the bumps, leading to lower 3.0\,mm-fluxes, as well as to lower optical depths at the dust ring location. Curiously, the \textit{Two Bumps} model has a higher spectral index than the \textit{Fiducial} model, due to its higher $F_\mathrm{1.3mm}$ (see Fig.~\ref{fig:flux_alpha_effect}a). This happens as some of the dust grains that were trapped in the inner, optically thick regions in the \textit{Fiducial} model, are now retained in the optically thin outer regions.

In plot b), we see the spectral index' dependency on the bump amplitude which decreases for higher values of $A_\mathrm{max}$. Since the \textit{Smaller A} model does not trap particles sufficiently, it has a quite high spectral index. For larger amplitudes, $\alpha_\mathrm{mm}$ then only decreases a little which is due to the stronger trapping around the pressure maxima, as explained in the earlier section.

The lower plot c) in Fig.~\ref{fig:flux_alpha_effect} shows that the spectral index increases for longer growth times of the bump. As already mentioned in the last subsection and which can be seen in plot c) above, $F_\mathrm{3.0mm}$ decreases faster for long growth times than $F_\mathrm{1.3mm}$. Thus, for higher $T_\mathrm{grow}$ less dust mass is trapped to grow sufficiently large mm-sized pebbles with high opacities that particularly contribute to $F_\mathrm{3.0mm}$.

As already mentioned in the last subsection, it understandably needs longer times and higher accumulations of dust masses to grow sufficiently large  mm-sized pebbles with high opacities that especially contribute to a higher $F_\mathrm{3.0mm}$ compared to $F_\mathrm{1.3mm}$, whose ratio in the end determines $\alpha_\mathrm{mm}$.

\subsubsection{Dust masses}
\label{sec:result_dust_mass}
The observed dust masses of models with optically thick inner bumps are underestimated by up to an order of magnitude in comparison with the actual dust mass. This can be seen by the bump model's influence on the various inferred dust masses, at an evolution time of our disk of 1 Myr. In Fig.~\ref{fig:Mdust_tau_effect}, we plot the total dust mass of our simulation $M_\mathrm{d,\,tot}$, the dust mass contained in grains larger than 0.1 mm $M_\mathrm{d,\,>0.1mm}$ and the observed dust mass $M_\mathrm{d,\,obs}$ over each model parameter. The latter is computed through the optically thin approximation and therefore solely mirrors the 1.3\,mm flux of the models. We investigate $M_\mathrm{d,\,>0.1mm}$, since at an observing wavelength of a few millimetre we do not trace $\mu \mathrm{m}$-sized grains, so it is more insightful to compare $M_\mathrm{d,\,obs}$ to  $M_\mathrm{d,\,>0.1mm}$, instead of $M_\mathrm{d,\,tot}$.

In subplot a), we find our simulation dust masses decreasing for larger gap radii, since the closer in the bump is, the more dust it can halt from accreting onto the star. Comparing the observed dust mass with the actual ones, we find that the largest discrepancy occurs for the \textit{Inner Bump} model, where $M_\mathrm{d,\,obs}$ is an order of magnitude smaller than $M_\mathrm{d,\,tot}$. This is not surprising, since this model has the highest of all dust masses, mainly contained within the dust trap, therefore also having the highest optical depth at its peak location. Consequently, with the optically thin approximation we can only trace dust down to optical depths of roughly $\tau\approx1$, meaning the largest fraction of the dust mass is hidden from our view. The \textit{Fiducial} model and the \textit{Outer Bump} model have both similar observed dust masses, though the former underestimates the real mass, while the latter traces its millimetre mass pretty well.

Looking at the \textit{Two Bumps} model, we see that it has a higher \textit{observed} dust mass than the \textit{Fiducial} one, despite having a lower mm-sized dust mass. This is due to the second outer bump of the model. Instead of trapping more particles in the inner bump, therewith just increasing its optical depth, we now have an otherwise hidden fraction of the (mm) dust mass trapped in the outer bump which contributes to its 1.3\,mm flux.

Moving on to plot \ref{fig:Mdust_tau_effect} b), we find the best match between the observed and actual dust masses for the \textit{Small A} bump model. The models with an increased amplitude, on the one hand, increase their model dust masses. On the other hand, their observed dust masses decrease a small amount, which can be traced back to stronger trapping around the pressure maxima, leading to higher optical depths of the dust ring.

The last subplot c), shows the dust mass dependency on the growth time of the bumps. As elaborated in previous sections, the model dust masses decrease for longer growth times, since the trapping is less efficient, if the bumps need longer times before they fully develop their local pressure maximum. This leads to a convergence of the two lines with $M_\mathrm{d,\,obs}$, which does not drop as steeply. 

\subsubsection{Optical Depths}
\label{sec:result_opt_depth}
The main result of this last subsection is that the observed optical depths of all dust traps in the inner regions of the disk are underestimated. We find this by comparing the absorption optical depth $\tau_\mathrm{abs}^\mathrm{peak}$, obtained through the dust surface density at the dust ring location of our \texttt{DustPy} models (Eq. \eqref{eq:opt_depth}), to the observed optical depth $\tau_\mathrm{obs}^\mathrm{peak}$, calculated through the ring peak intensities of our synthetic observations (Eq. \eqref{eq:opt_dept_obs}). The corresponding plots are shown in Fig.~\ref{fig:Mdust_tau_effect}. 

First of all, we find all values of the observed optical depth to lie between ${\tau_\mathrm{obs} \approx 0.2-0.7}$ for the models with bumps. On the other hand, the absorption optical depth usually spans values up to ${\tau_\mathrm{abs} \leq 4}$ (lower panels). However, in the case of the \textit{Inner Bump} it reaches up to $\tau_\mathrm{abs}=35$. The absorption optical depth generally increases for larger amplitudes, shorter growth times and smaller gap radii.

The general trend of the curves in subplots a) - c) show the same dependency on the bump parameters for the absorption and observed optical depths, as already seen in the upper panel of Fig.~\ref{fig:Mdust_tau_effect}, between the total and observed dust masses, respectively. This is not surprising since they trace the same quantities: $\tau_\mathrm{abs}$ and $M_\mathrm{d, tot}$ trace the dust surface density and $\tau_\mathrm{obs}$ and $M_\mathrm{d, obs}$ the 1.3\,mm flux density, both at the corresponding dust ring location. 

In the lower left subplot a), we additionally plotted the observed optical depths $\tau_\mathrm{obs, noscat}$ (orange) for our models where we did not treat scattering in the radiative transfer calculations. As can be seen, this increases the peak flux and therewith the optical depth of our rings in the inner parts of the disk by roughly ${40 \%}$, and slightly decreases it for the outer bump by ${15 \%}$ (see also Fig.~\ref{fig:RadFlux_Scat}). In the inner rings, we find a large amount of mm- to cm-sized dust pebbles which have a high scattering opacity (albedo). Furthermore, the scattering phase function of such grains has a strong forward peak, as already mentioned in Sect.~\ref{sec:Rad_transfer_calc}. So, if we treat scattering in these optically thick models, a considerable amount of radiation gets scattered out of the line-of-sight leading to a decreased emission and optical depth compared to the models without scattering. In other words, scattering reduces the (optical) depth in our rings from where photons can escape \citep{Zhu2019}. This effect does not play a role in the optically thin ($\tau_\mathrm{abs}<1$) outer parts of the disk, where the non-treatment of scattering even leads to a small increase in the observed optical depth.
To sum up, we find that for nearly all models the observed optical depth is underestimated. Only for models without dust trapping (\textit{No Bump} \& \textit{Small A}) and for the \textit{Outer Bump} model, can $\tau_\mathrm{abs}$ be well traced by $\tau_\mathrm{obs}$.

A final remark on the parameter effects: We also ran models where `planetary' bumps form and then disappear after a finite lifetime of 0.5 - 1 Myr. However, those disappearing bumps display no lasting effect on the disk's dust evolution in terms of observability. After a few hundred thousand years they cannot be differentiated anymore from the model without any bump, regardless of their lifetime and bump location. 

\subsection{Dynamic bumps} \label{sec:ResultDynamic}
\begin{figure}[t]
\centering
\includegraphics[width=\linewidth]{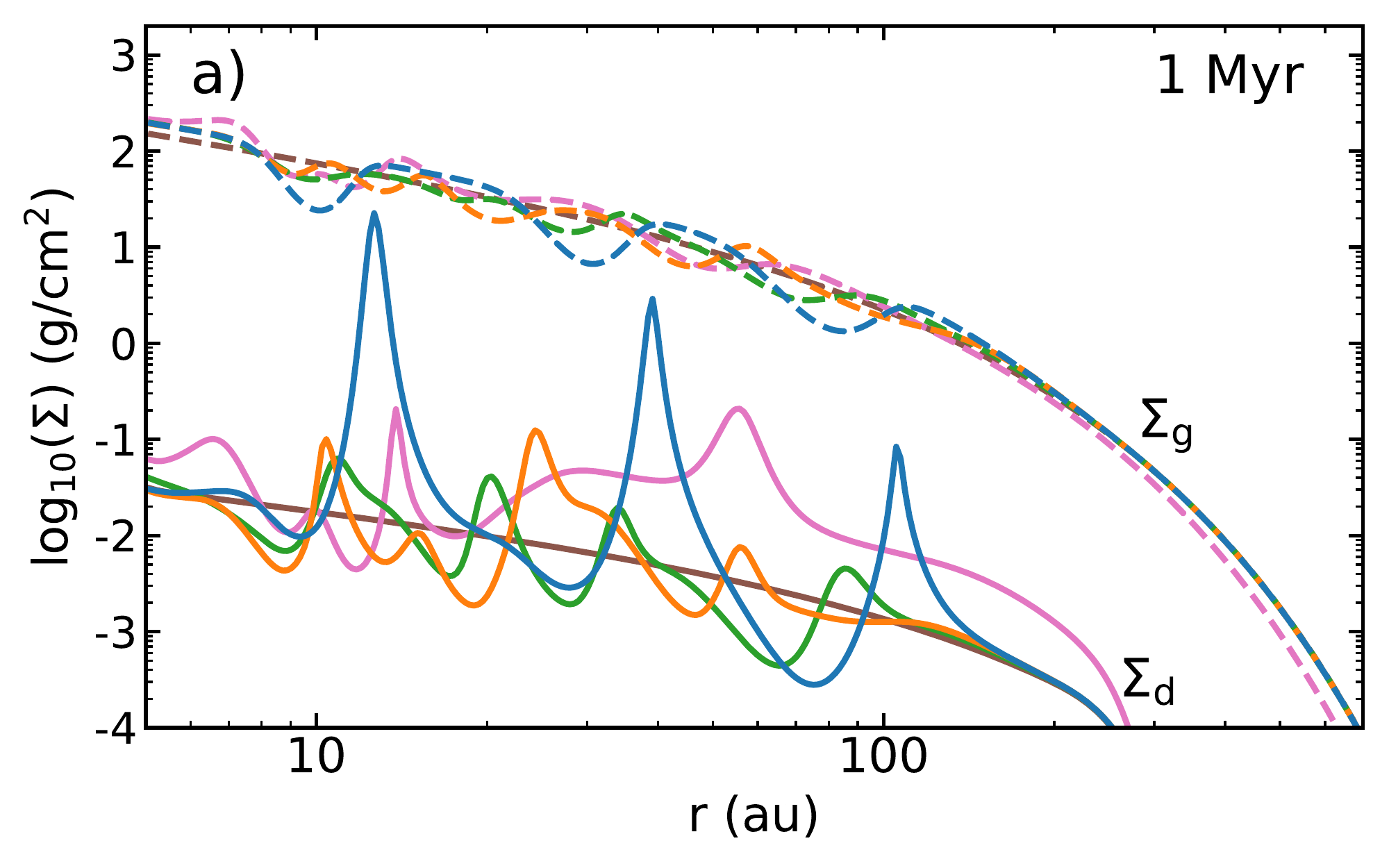} \\
\includegraphics[width=\linewidth]{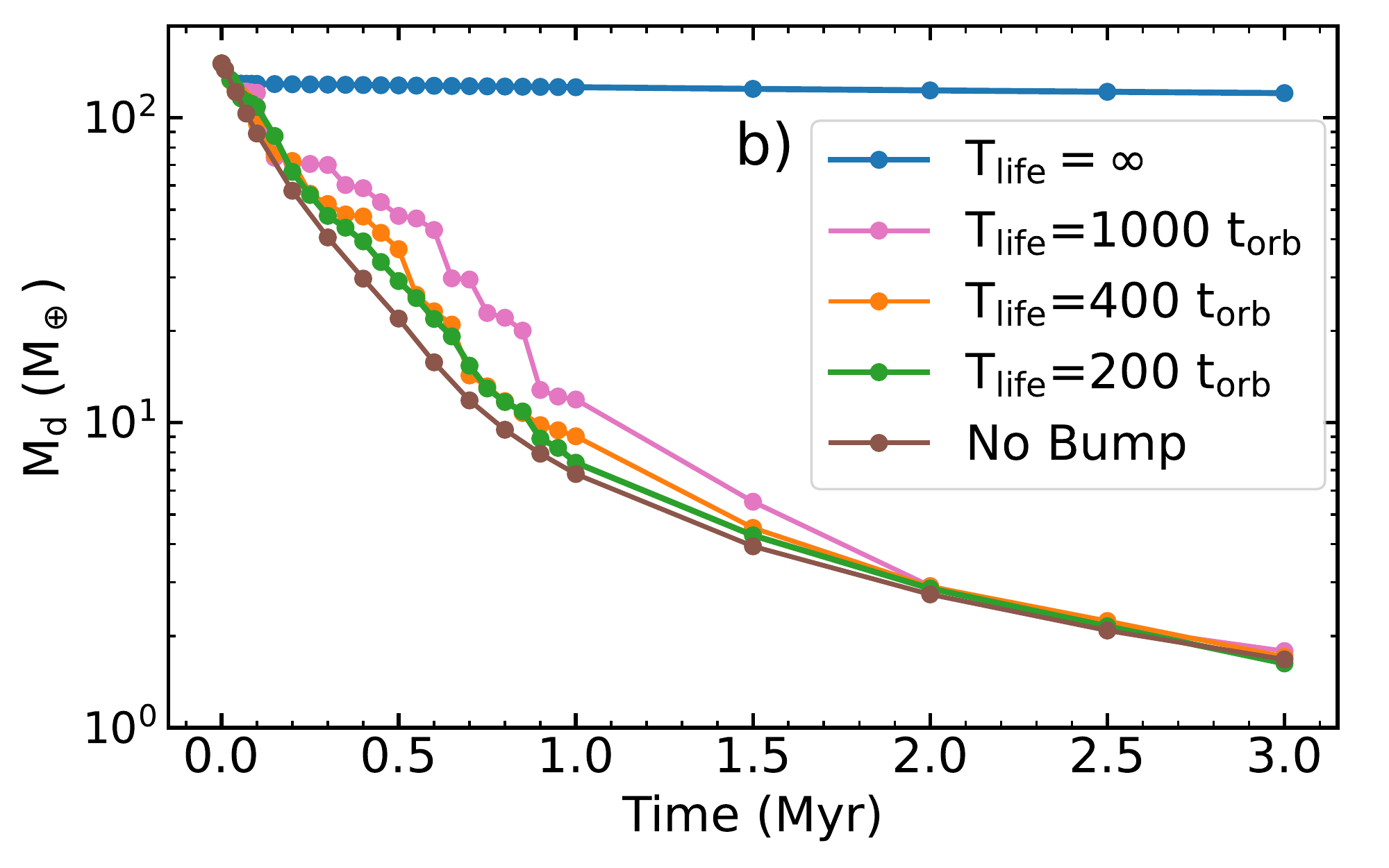}
\caption{Surface density and dust masses plots of dynamic bumps models. \textbf{a)} Gas (dashed line) and dust (solid line) surface densities at 1 Myr of evolution for the different dynamic bumps models and the model without any bumps for comparison. \textbf{b)} Total dust mass of each model over the disk lifetime.}
\label{fig:dynamic_bumps}
\end{figure}

The main result of this section is that dynamic bumps, in contrast to `planetary' ones, cannot efficiently halt the drift of pebbles in order to display high millimetre fluxes and low spectral indices observed in protoplanetary disks.  We present the observational properties of our \textit{Dynamic Bumps} models, which are set to mimic zonal flows in protoplanetary disks, in Table \ref{table:dynamic_obs_proper}. In Fig.~\ref{fig:dynamic_bumps}, we show the gas and dust surface densities at the evaluation time of the observations at 1 Myr, as well as the total dust masses over the models' disk lifetime. The detailed dust size distribution of the models at 1 Myr can be found in Fig.~\ref{fig:DustSizeDens_Dynamic} of the Appendix.

The dynamic models, shown in plot \ref{fig:dynamic_bumps} a) \& b), with bump lifetimes of 200 (\textit{Std}), 400 (\textit{Long}) and 1000 local orbital times, show quite low dust surface densities and nearly lost all of their initial dust mass already after 1 Myr of the disk's evolution. After 3 Myr, less than 2  $\mathrm{M_\oplus}$ worth of dust are retained in the disk. In b), it is worth noting that the doubled lifetime of the \textit{Long} model does not slow down the exponential loss of the dust mass, compared to the \textit{Std} setup. So it comes as no surprise that these models also show very low millimetre fluxes, in fact, they even have the lowest of all models in this work, even lower than the model with no bump, shown as a comparison in the plots. Thus, their respective millimetre spectral index of $\alpha_\mathrm{mm}=3.6$ lies only slightly below the value of $\alpha_\mathrm{ISM}\simeq3.7$, which is typically found in the optically thin emission of interstellar medium (ISM) grains \citep{Tazzari2021_3mm}, where no dust growth is expected to happen.

On the contrary, the \textit{Static} bumps model displays the highest dust mass and millimetre luminosity of all models in this work. The reason for this is its strong dust trap (inner bump) at 10 au, which is already fully developed after $\simeq20$ kyrs (700 orbits), due to the short orbital timescales here. Once the trap has formed, it efficiently halts the dust drift and therefore the disk's dust mass does not significantly decrease anymore, 
as can be seen in the blue line of plot \ref{fig:dynamic_bumps} b). This is the important difference between the dynamic and the static model. If the dynamic ones would also have efficient dust traps in the inner parts of their disk, they would retain their dust mass and thereby reaching higher luminosities.

A last thing to point out about the \textit{Static} model is that it has the highest total dust mass of all models, however, its observed dust mass is underestimated by a factor of five. Here, most of the disk's dust mass is contained in the inner bump thus making it highly optically thick ($\tau^\mathrm{peak}_\mathrm{abs}=15$). Hence, the optically thin approximation for measuring the dust mass is not applicable anymore.

To conclude this section, we shortly want to touch on the measured disk radii of the three different models. Looking at the Table \ref{table:dynamic_obs_proper} above, we find the $R_\mathrm{d,68\%}$ of the \textit{Static} model to be the largest, but not significantly larger than for the the dynamic ones. Interestingly however, is the fact that $R_\mathrm{d,90\%}$ is the lowest for the \textit{Static} model, even though it displays a dust ring in $\Sigma_\mathrm{d}$ at $\sim100$ au of Fig.~\ref{fig:dynamic_bumps} a). This ring indeed, does not substantially contribute to overall flux anymore which is why we do not accurately trace it within the 90 \% of the emission and as a result, cannot trace the outer edge of the disk anymore.

 \begin{table}[h!]
    \centering
    \begin{threeparttable}
    \caption{\centering Observational disk properties of the \textit{Dynamic Bumps} model.}
    \label{table:dynamic_obs_proper}
    \begin{tabular}{c |c |c c c}
     \hline \hline
    Property & Unit & Std & Long & Static \\
    \hline
    $T_\mathrm{life}$&$\mathrm{t_{orb}}$ & 200 & 400 & $\infty$ \\
    \hline
    $M_\mathrm{d, 1Myr}$ & $\mathrm{M_\oplus}$  & 7.4 & 9.0 & 126 \\
    $M_\mathrm{d, 3Myr}$ & $\mathrm{M_\oplus}$ &  1.6 & 1.7 & 121 \\
    $M_\mathrm{d,>0.1\mathrm{mm}}^\mathrm{1Myr}$&$\mathrm{M_\oplus}$ &  3.7 & 5.0 & 106 \\
    $M_\mathrm{d, obs}$&$\mathrm{M_\oplus}$ & 3.6 & 5.0 & 24 \\
    $F_\mathrm{1.3mm}$ & mJy & 6.0 & 9.0 & 44 \\
    $F_\mathrm{3.0mm}$ & mJy & 0.3 & 0.5 & 4.4 \\
    $\alpha_\mathrm{mm}$ &  & 3.6 & 3.6 & 2.7 \\
    $R_\mathrm{d,68\%}$ & au & 29 & 27 & 34 \\
    $R_\mathrm{d,90\%}$ & au & 48 & 45 & 41 \\
    \hline
    \end{tabular}
    \begin{tablenotes}
      \small
      \item NOTE - All observational properties have rows similarly labelled as columns in Table \ref{table:disk_properties}. The first row shows the parameter of the model's bump life time.
    \end{tablenotes}
    \end{threeparttable}
\end{table}

\section{Discussion}
\label{sec:discussion}
\subsection{Disk sizes}
\label{sec:disc_disk_sizes}
We find nearly all of the measured disk sizes ($R_\mathrm{d,68\%}$) coincide with the location of the farthest out pressure bump, as was already proposed by earlier studies \citep{Pinilla2020,Zormpas2022}.
The exception are the models with multiple bumps, where the measured sizes ($R_\mathrm{d,68\%,90\%}$) may indicate the location of the inner dust traps (see Fig.~\ref{fig:NormIntenstiy_Profiles}, \textit{Two Bumps} and \textit{Static} models).

In a recent study, \cite{Franceschi2021} shows that in the presence of efficient dust traps, the dust disk's outer edge (or `dust line') only traces the trap location across different wavelengths. Therefore, the `dust line' cannot be used as a tracer of radial drift anymore to determine the disk's mass, as proposed by earlier studies of \cite{Powell2017,Powell2019}. The results of \cite{Franceschi2021} are in agreement with the findings of our dust trapping models, when we calculate the emission radii for our intensity maps at 1.3\,mm and compare them to the disk sizes at 3.0\,mm (see Fig.~\ref{fig:NormIntenstiy_Profiles} in the Appendix). The maximal deviation for $R_\mathrm{d,68\%}^\mathrm{1.3mm}$/$R_\mathrm{d,68\%}^\mathrm{3.0mm}$ is 2\% and for $R_\mathrm{d,90\%}$ less than 5\%. Again, one exception is the \textit{Two Bumps} model, where $R_\mathrm{d,90\%}^\mathrm{1.3mm}=100$ au traces the second bump, while $R_\mathrm{d,90\%}^\mathrm{3.0mm}=41$ au coincides with the first.

A recent multi-wavelength survey of the Lupus star-forming region by \cite{Tazzari2021_sizes} finds that their observed disk sizes are nearly the same across different millimetre wavelengths. 
The authors find a ratio between their sample's mean millimetre dust disk radii of $R_\mathrm{68,3.0mm}$/$R_\mathrm{68,1.3mm}=0.94\pm0.23$ which matches our results well. This finding is however in  contrast to dust evolution models \citep[e.g.][]{Birnstiel2012,Long2020} of smooth disks which predict much smaller radii at 1.3\,mm and 3.0\,mm, than were observed (see their Fig.~4). As their survey was conducted at low resolution (~0.3''), it is not possible to resolve hidden substructures within their disks. However, some of their (larger) sources overlap with the DSHARP sample, that resolved them and found substructures, which could explain why $R_\mathrm{68\%}$ is constant with wavelength at least for some disks.

\begin{figure*}[t]
\centering
\includegraphics[width=150mm]{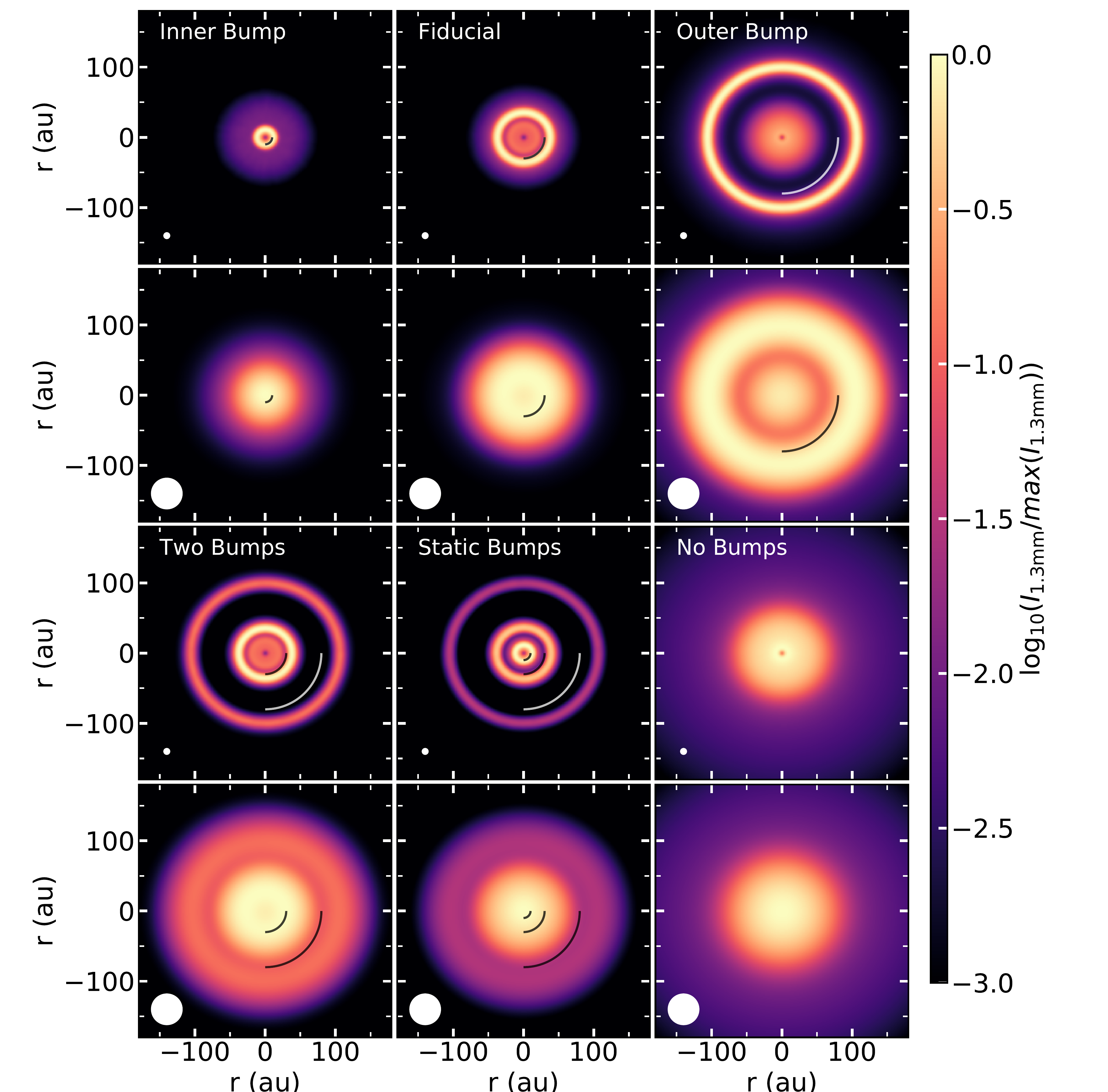}
\caption[Synthetic observations of models at low and high angular resolution]{Synthetic observations of models at low and high angular resolution. Images for $\lambda=1.3$ mm, observed with high (0.04'') and low (0.30'') angular resolution, the according beams are indicated in the lower left corner of the plots. The black and white arcs in the images indicate the gap locations of the models.}
\label{fig:SynImages_res}
\end{figure*}

To compare our results to the survey of \cite{Tazzari2021_sizes}, we cautiously note the caveat that we only consider a solar-type star with a fixed disk setup. Nevertheless to be as accurate as possible, we convolve our synthetic images with the survey's larger beam size of 0.30'' (versus our 0.04'') and further assume a lower SNR of 200 (versus our 1000). Fig.~\ref{fig:SynImages_res} displays an overview of the synthetic images of our models at high and low angular resolution.

\begin{figure}[t]
\centering
\includegraphics[width=\linewidth]{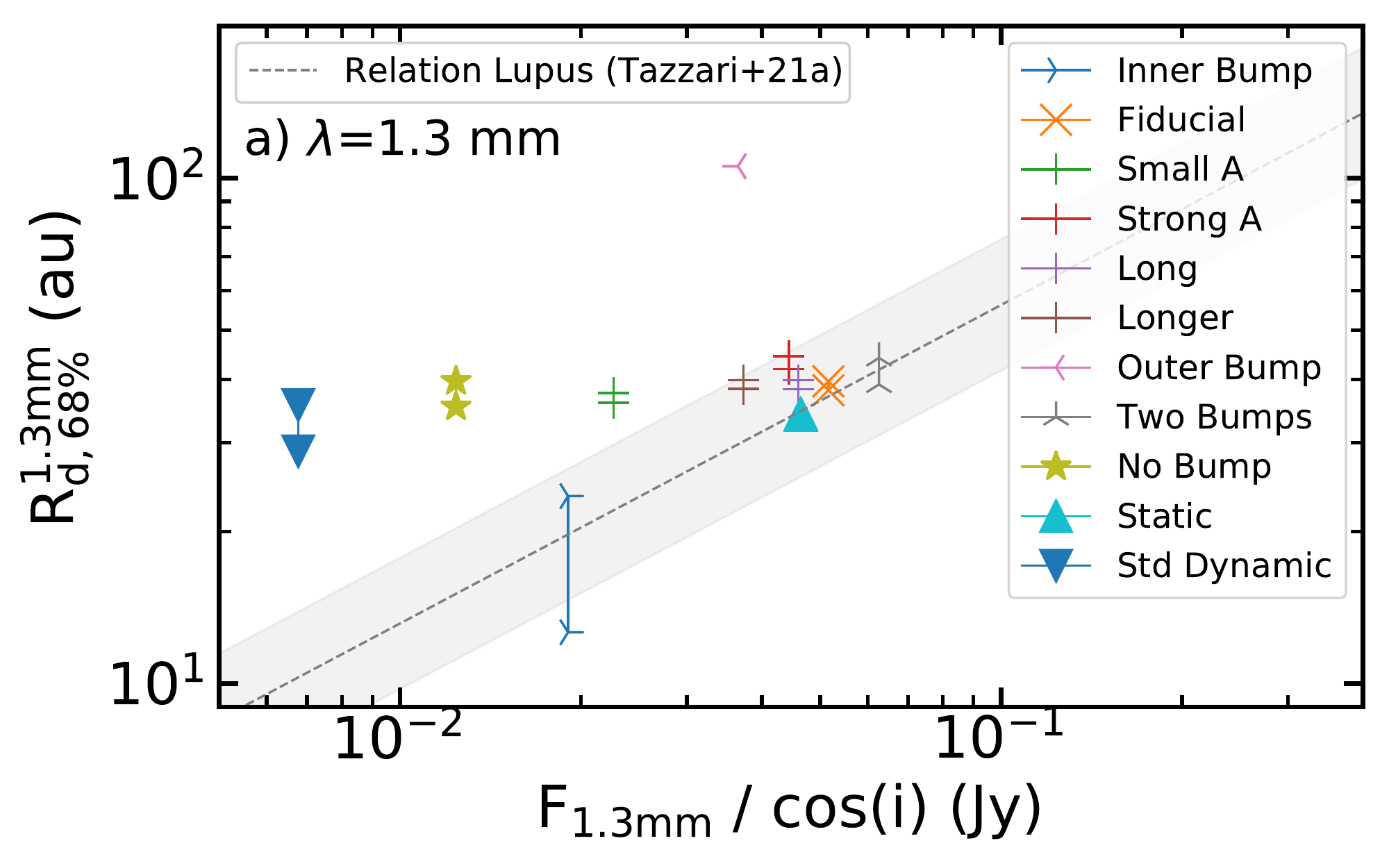}
\includegraphics[width=\linewidth]{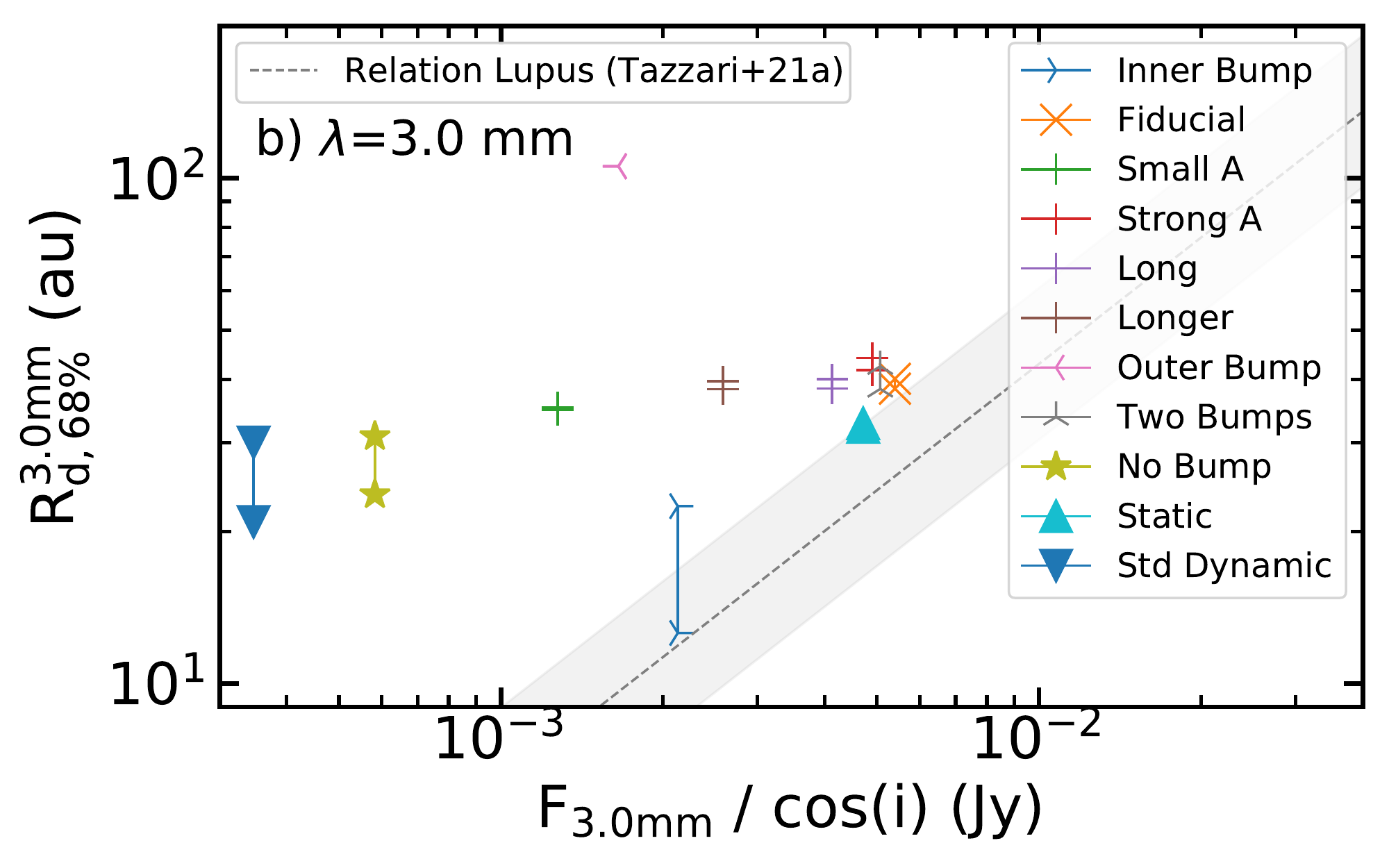}
\includegraphics[width=\linewidth]{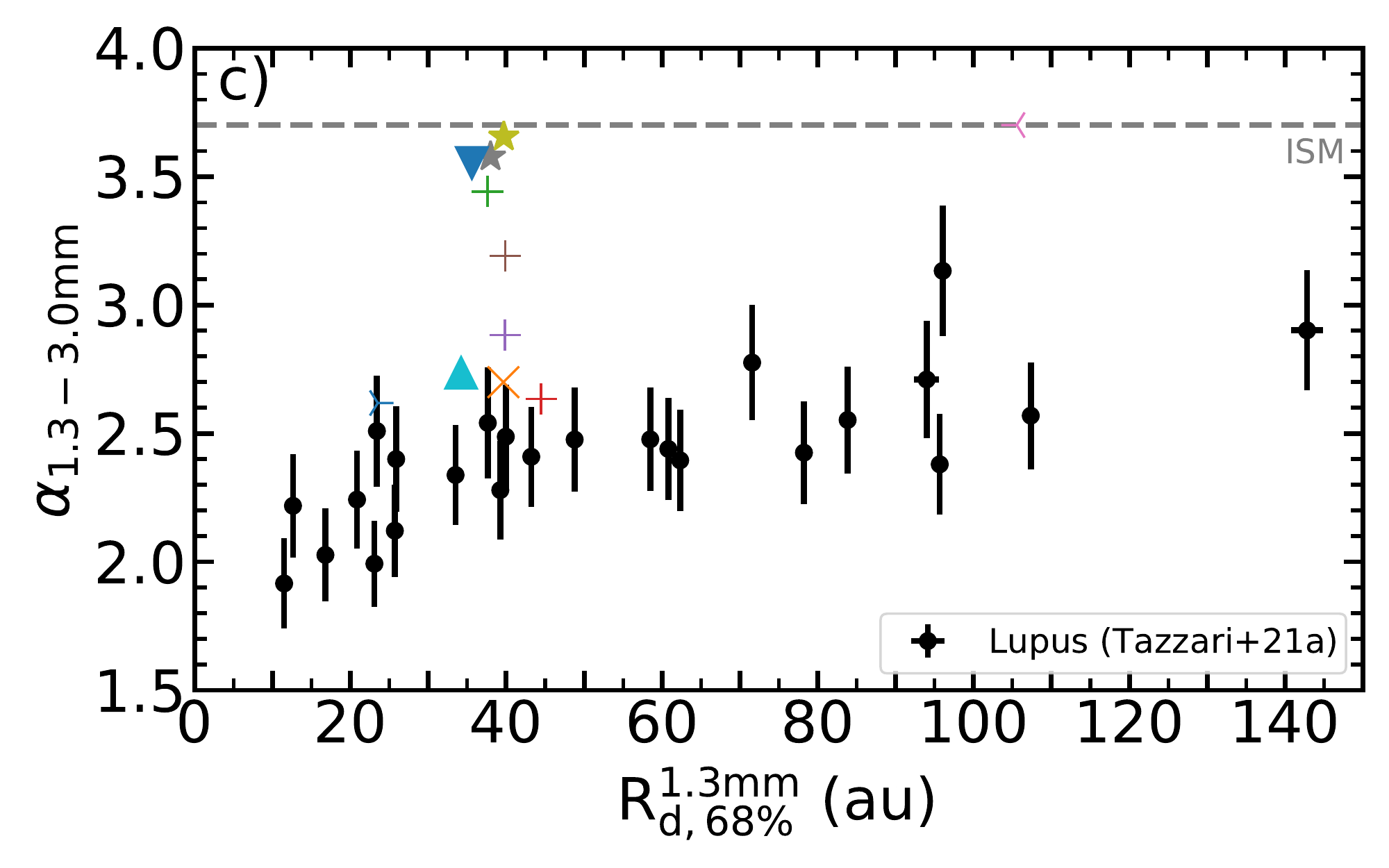}
\caption[Size-luminosity relations and spectral indices dependent on disk sizes]{Size-luminosity relations and spectral indices dependent on disk sizes. In \textbf{a)\,+\,b)} the fluxes are divided by the cosine of the inclination (i = 20°).  The plotted 68\% emission disk radii are found to decrease from low (0.30'') to high (0.04'') resolution for all models irrespective of wavelength which is highlighted by lines connecting both radii. The grey dashed lines show the relations for Lupus at $\lambda=$1.3\,mm and $\lambda=$3.0\,mm found by \cite{Tazzari2021_sizes}, where the grey area represents the 68\% confidence interval around the median. \textbf{c)} Spectral index ($\alpha_\mathrm{1.3-3.0mm}$) over disk effective size (R$_\mathrm{d, 68\%}$) at $\lambda=$1.3\,mm with 0.30'' resolution, black dots show Lupus data also from \cite{Tazzari2021_sizes}.}
\vspace{-4mm}
\label{fig:Tazzari_relations}
\vspace{-0.5mm}
\end{figure}

We show our derived disk sizes at low and high angular resolution, in dependence of their 1.3 and 3.0 millimetre flux in the upper plots of Fig.~\ref{fig:Tazzari_relations} and then compare our models to the size-luminosity relation(s) found in Lupus by \cite{Tazzari2021_sizes}. In the lower plot c) of Fig.~\ref{fig:Tazzari_relations}, we compare our (low resolution) disk sizes to each of the Lupus sample and find them to span the same range of values. The discussion about the spectral indices and the luminosities is in the next subsection. 

The decrease in resolution used to match the sample of \cite{Tazzari2021_sizes} increases the measured disk sizes by a small factor of a few percent for most models.
Only the \textit{Inner Bump} model shows a significant increase in its measured disk size at the lower resolution (doubling its 68\% emission radius at 1.3\,mm, from 13 au to 24 au), which occurs because both the pressure bump located at 10 au and the disk's physical extent are unresolved, leading to a spread of the emission over a wider area for the larger beam size.

Nevertheless, both measured radii of the \textit{Inner Bump} model still lie within the correlation found in the Lupus sample, as shown in the plots of Fig.~\ref{fig:Tazzari_relations}, both for the 1.3\,mm radius and for the matching low spectral index. If, the smallest disks of the sample, that are comparable in size to the beam ($R_\mathrm{d}<40$ au), exhibit substructures, they cannot be resolved at this low angular resolution (see Fig.~\ref{fig:SynImages_res}, \textit{Inner Bump} and \textit{Fiducial} models). Indeed, it might even be favourable that small disks do contain unresolved dust traps, as they also show the lowest spectral indices of the Lupus sample. The presence of pressure bumps can help to obtain such low values of $\alpha_\mathrm{mm}$, as has been shown by \cite{Pinilla12} and is in agreement with our findings.

Building on the assumption of unresolved dust traps, we might find the smallest (Lupus) disks to be even smaller, when observed at higher angular resolution. This correlation has also been observed for some disks in Taurus by \cite{Hendler2020} (see Fig.~12), supporting our results. These points lead us to an important conclusion regarding the size-luminosity relation of disks. Since the measured luminosities and the sizes of larger disks, do not change or strongly decrease with higher angular resolution, we expect the slope of the relation to become steeper as the smallest disks are resolved. However, high angular resolution observations with  high SNR are especially challenging to obtain for small disks, as most of them appear to be faint  \citep[e.g.][]{Long2019, Kurtovic2021}. Still, they do find their sample's size-luminosity slope to steepen by including the resolved disks of DSHARP into their sample.

To conclude this section, we want to discuss the effect of neglecting planetary migration in our models, i.e. that we assume a fixed radial location of the bumps throughout the simulation. 
Looking at the corresponding planetary masses of our bump amplitudes in Table \ref{table:bump_param} and taking into account our disk setup, our planets might be subject to either type II or type III migration \citep[e.g.][]{Masset2003}. In the former case, we would expect the bump to jointly migrate inwards with the planet, thus just decreasing the disk's effective size. In the case of rapid runaway (type III) migration, however, the planet is able to form multiple pressure bumps on its discontinuous inward migration \citep[see Fig.~5 of][]{Wafflard2020}. The lifetime of such pressure bumps can be between several thousand years to $10^{5}$ years depending on the disk's turbulent viscosity. For the case when the disk viscosity is high ($10^{-3}$), models of type III migration are similar to the models with disappearing bumps that we have investigated. These cases show no lasting effect on the disk's dust evolution in terms of observability, except one observes the system at a `lucky” time, while the structures are still present. However, when the viscosity is very low ($10^{-5}$), the pressure bumps formed by planets in type III migration can be long-lived and reproduce several observable properties of disks as expected from static bumps.

We lastly note that the other bump model parameters, such as the growth times and amplitudes, do not have a major influence on the measured dust disk radii. The latter only if the amplitude factor is not large enough to act as an efficient dust trap, which usually sets the disks outer edge.

\subsection{Millimetre fluxes and spectral indices}
\label{sec:disc_flux_specindex}
In this section, we discuss which model parameters influence the millimetre fluxes of our disks and their according spectral index. 

First of all, we compare how our measured millimetre fluxes fit into the size-luminosity relations shown in Fig.~\ref{fig:Tazzari_relations}. We find all models with efficient dust traps in the inner regions of the disk ($<40$ au) have high enough intensities (and sizes) to fit within the confidence interval of the 1.3\,mm relation. Yet, the models with outer bumps display too low 1.3\,mm fluxes, compared to their large sizes, since their bumps lie too far out ($r_\mathrm{gap}>R_\mathrm{c}$) to trap enough dust. Similarly, the model without any dust trap, is also too faint for its medium disk size. Our models with more than one bump fit the size-luminosity relation well. Having more than one pressure bump in their disks, preferably in the inner parts, reduces their measured disk radii while simultaneously increasing their millimetre intensity. 

In subplot b) of Fig.~\ref{fig:Tazzari_relations} we also compared our models' 3.0\,mm flux (and radii) to the size-luminosity relation found at this wavelength by \cite{Tazzari2021_sizes}. They fit the correlation worse than for 1.3\,mm in plot a). While the disk sizes, at least for the models with bumps are constant with wavelength (as can be seen comparing the two plots), all of our models show too insufficient emission at 3.0\,mm. Only the \textit{Static} model, with its three bumps, still lies within the confidence interval of the correlation-line. The discrepancy of too low $F_\mathrm{3.0mm}$ is again especially pronounced for the models without efficient dust traps and also for the outer bump model.

\begin{figure*}[t]
\centering
\includegraphics[width=0.33\linewidth]{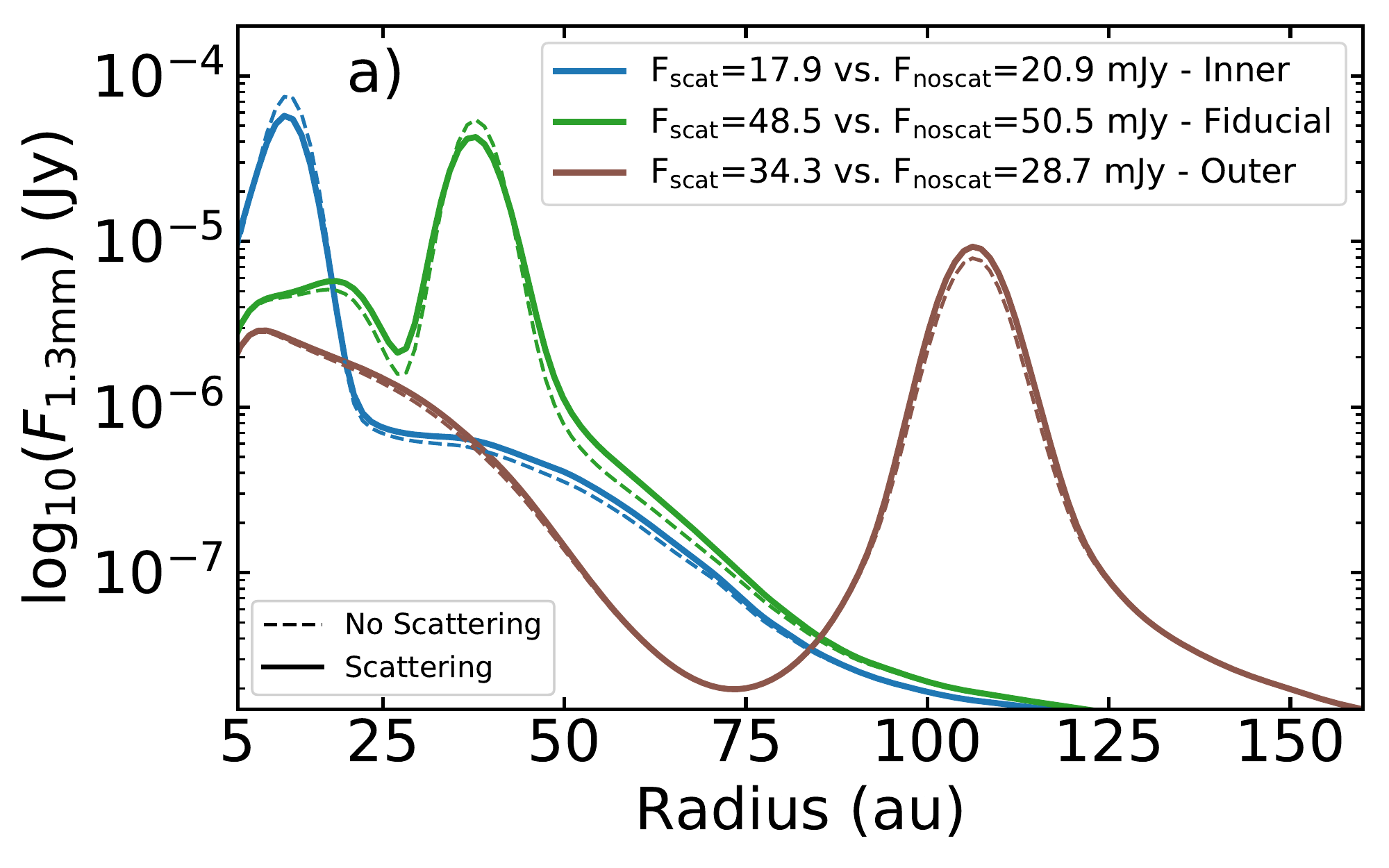} 
\includegraphics[width=0.33\linewidth]{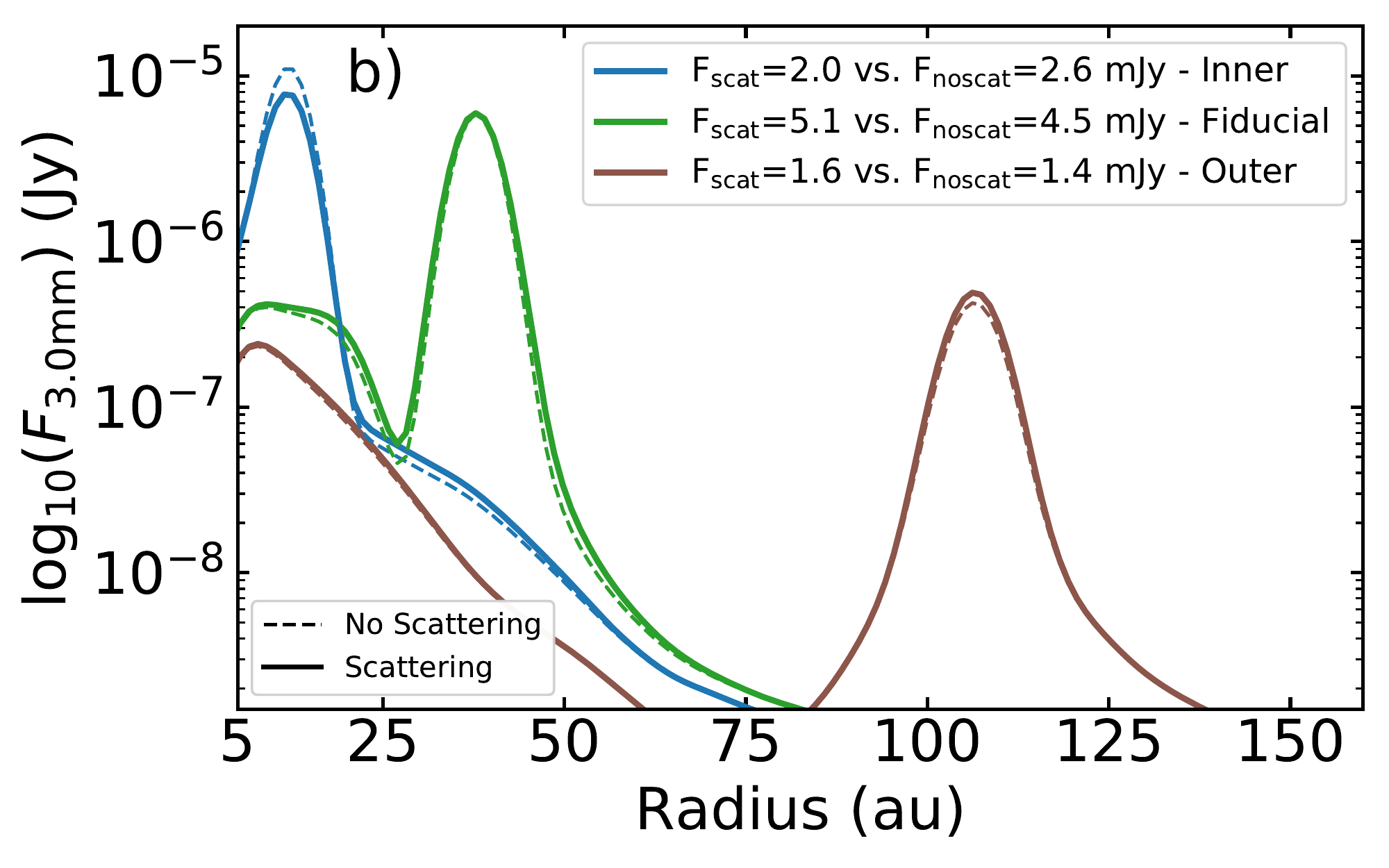}
\includegraphics[width=0.33\linewidth]{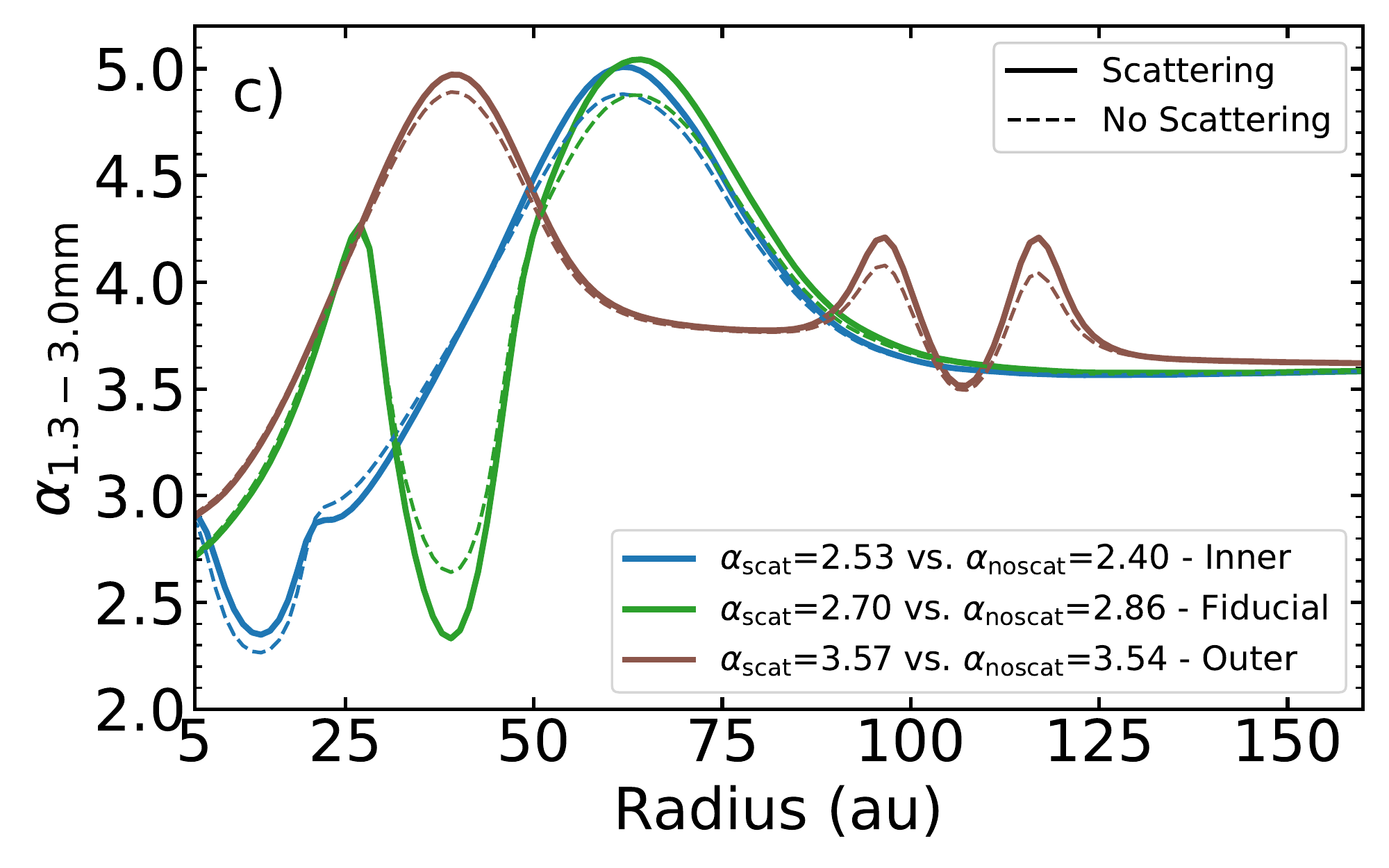}
\caption[Radial flux and spectral index profiles for the three models with varying gap location with and without scattering.]{Radial flux and spectral index profiles for the three models with varying gap location, with and without scattering. \textbf{a)\,\&\,b)} Radial flux profiles at $\lambda=1.3$ \& 3.0\,mm. \textbf{c)} The models' radial profile of the spectral index. In all plots, the solid lines treat scattering in the radiative transfer, the dashed lines not.}
\label{fig:RadFlux_Scat}
\end{figure*}

The low $F_\mathrm{3.0mm}$ fluxes of our disks translate directly into high millimetre spectral indices, as shown in the lower plot of Fig.~\ref{fig:Tazzari_relations} c). Here, we compare $\alpha_\mathrm{mm}$ of our models to the values found in the Lupus sample by \cite{Tazzari2021_sizes}, which distribution is statistically indistinguishable form the star forming regions of Taurus and Ophiuchus \citep{Tazzari2021_3mm}. We first note that all of our derived spectral indices are higher than what was observed for disks with similar sizes in Taurus. The only models that still fit within the given errors are the ones with fast-forming efficient dust traps at 10 and 30 au, respectively. Models with longer bump growth times, no or less efficient dust traps, and also the ones without a close-in bump, show too high spectral indices of $\alpha_\mathrm{mm} \geq 3.0$. Some of their computed $\alpha_\mathrm{mm}$ are even close to the value of 3.7, found in the interstellar medium (dashed line) \citep{Goldsmith1997}.  Nevertheless, we still observe the same general trend in subplot \ref{fig:Tazzari_relations} c), namely of an increased $\alpha_\mathrm{mm}$ with larger disk size of our models.

These results suggest that in order to obtain the low spectral indices observed, we need disks to include long-lived and efficient dust traps, especially in their inner disk parts to halt the dust drift. This finding, together with low millimetre luminosities of our \textit{Dynamic Bumps} model, makes us disfavour the zonal-flow origin of pressure bumps, in which the bumps induced by over-densities in the disk have a short lifetime. On the other hand, this then supports the planetary-origin of the dust traps, in which case the pressure bumps are indeed long-lived and thus can trap the large amounts of dust that we observe with high resolution observations of disks \citep{dullemond18}.

About the effect of scattering on our results, Fig.~\ref{fig:RadFlux_Scat} show the radial flux profiles of our three models with varying gap location, either treating scattering in the radiative transfer or not. In the left subplot a), we find that the inclusion of scattering into our calculation decreases the 1.3\,mm flux of our models, at the location of the optically thick inner dust rings. However, in the optically thin regions scattering increases the flux, as it is the case for the whole disk of the \textit{Outer Bump} model. The same holds true for the 3.0\,mm flux profile in subplot b). The only differences here is that the dust ring of the \textit{Fiducial} model is not optically thick anymore at this wavelengths, hence its peak emission flux does not decrease when scattering is included. 

The implications of the (non-)treatment of dust scattering in models with optically thick regions or substructures has been studied in detail by \cite{Zhu2019}. They find that dust scattering can significantly reduce the emission of optically thick disk regions, in line with our findings. They further state that the spectral index can be used to determine the optical thickness of disks. Using the DSHARP-opacities, they expect a jump of $\alpha_\mathrm{mm}$ in the radial direction from optically thick ($\alpha_\mathrm{mm}<2.5$) to optically thin regions ($\alpha_\mathrm{mm}>2.5$) of the disks. This is exactly what we find, looking at the radial profile of the spectral index in subplot c) of Fig.~\ref{fig:RadFlux_Scat}, where especially the \textit{Inner Bump} model lies below 2.5 at its location of the optically thick dust ring. \cite{Zhu2019} also state that in the optically thick regime, the spectral index depends on the albedo rather than the opacity, as in the optically thin regime. Therefore, we include scattering in our models to be sure to correctly determine the emission of the optically thick dust rings.

Even though our optically thick substructures display regions of low $\alpha_\mathrm{mm}$ in the radial profile, the overall values are still too high compared to profiles obtained by multi-wavelengths observations (e.g. see \cite{Tazzari2021_sizes}, \cite{Huang2018_TW}), where $\alpha_\mathrm{mm}\leq3.5$ in most cases, even in the outer disk regions. These generally unobserved high spectral indices of our models arise from the low flux values at 3.0\,mm, since they overall contain too few mm-sized dust grains with high opacities. There are multiple issues regarding the low $F_\mathrm{3.0mm}$. Firstly, most of the dust mass in the disk is trapped in the dust rings. Due to optical depth effects, it can only contribute a certain portion to the emission. Large grains that contribute most to $F_\mathrm{3.0mm}$ are trapped even closer around the pressure maximum, compared to grains of lower opacity. This leads to higher optical depths and lower emission.

Secondly, the bulk of the dust mass is stored in the largest centimetre-sized grains for all of our models, also for the one(s) with a bump at 80\,au. Within the dust traps, the particle size is determined through the fragmentation limit, as displayed in Fig.~\ref{fig:part_dens_distr}. Within this regime, the maximum size a particle can grow to through collisions, is proportional to the square of the fragmentation velocity (see Eq. \eqref{eq:St_frag}). Moderately decreasing $v_\mathrm{frag}$ to values lower than $10$\,m/s helps to grow a larger amount of mm-sized pebbles, whose drift is reduced and which usually have opacities an order of magnitude higher than grains of cm sizes (see Fig.~\ref{fig:Opacities}), therewith obtaining higher disk fluxes. This is indeed the case for a simulation we ran with $v_\mathrm{frag}=$ 3\,m/s which displays 62\,\% and 24\,\% higher fluxes at 1.3\,mm and 3.0\,mm, respectively, compared to our \textit{Fiducial} setup. However, the lower $v_\mathrm{frag}$ also leads to an increase of $\alpha_\mathrm{mm}=3.0$ versus the original 2.7.

\begin{figure}[t!]
\centering
\includegraphics[width=\linewidth]{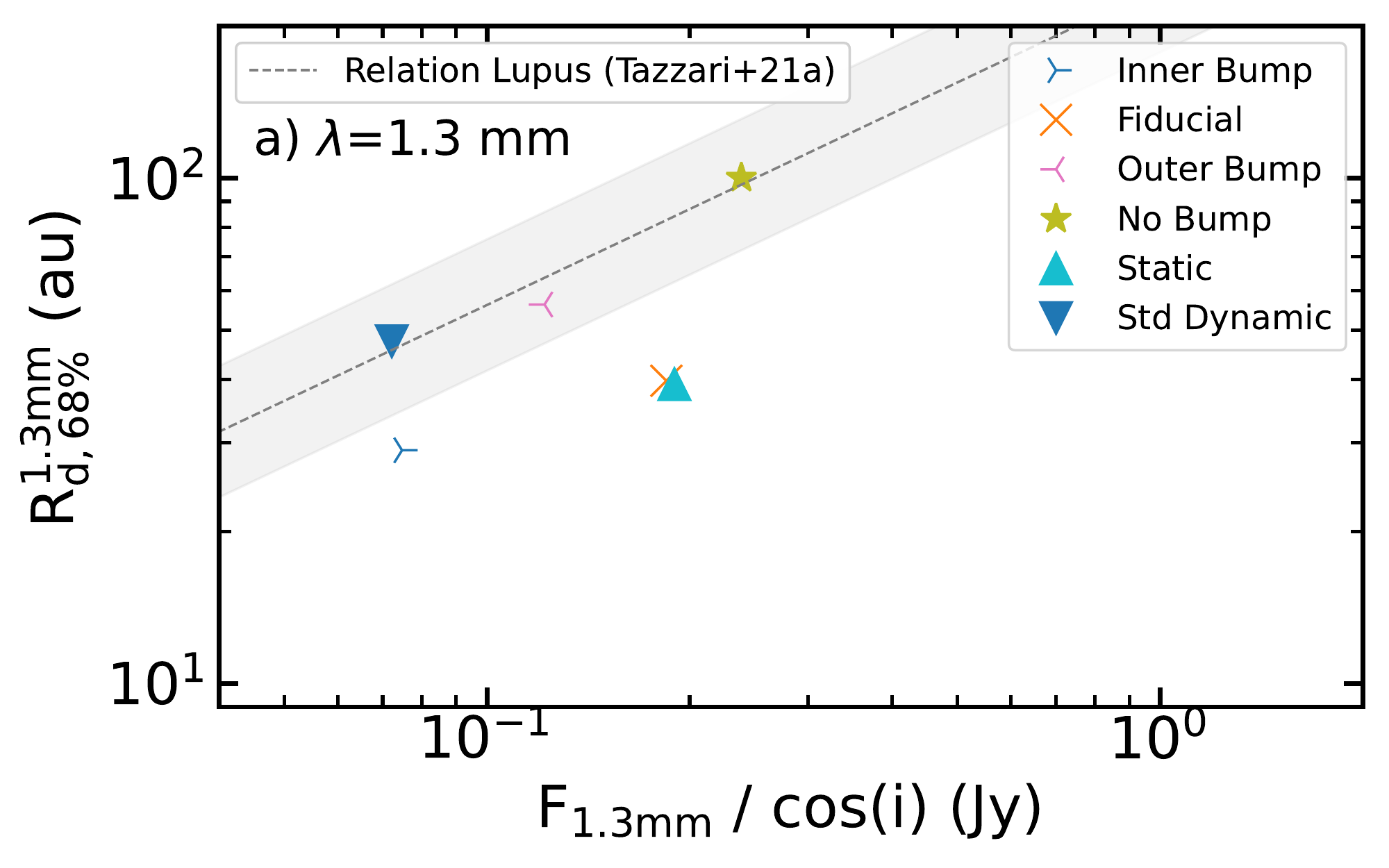}
\includegraphics[width=\linewidth]{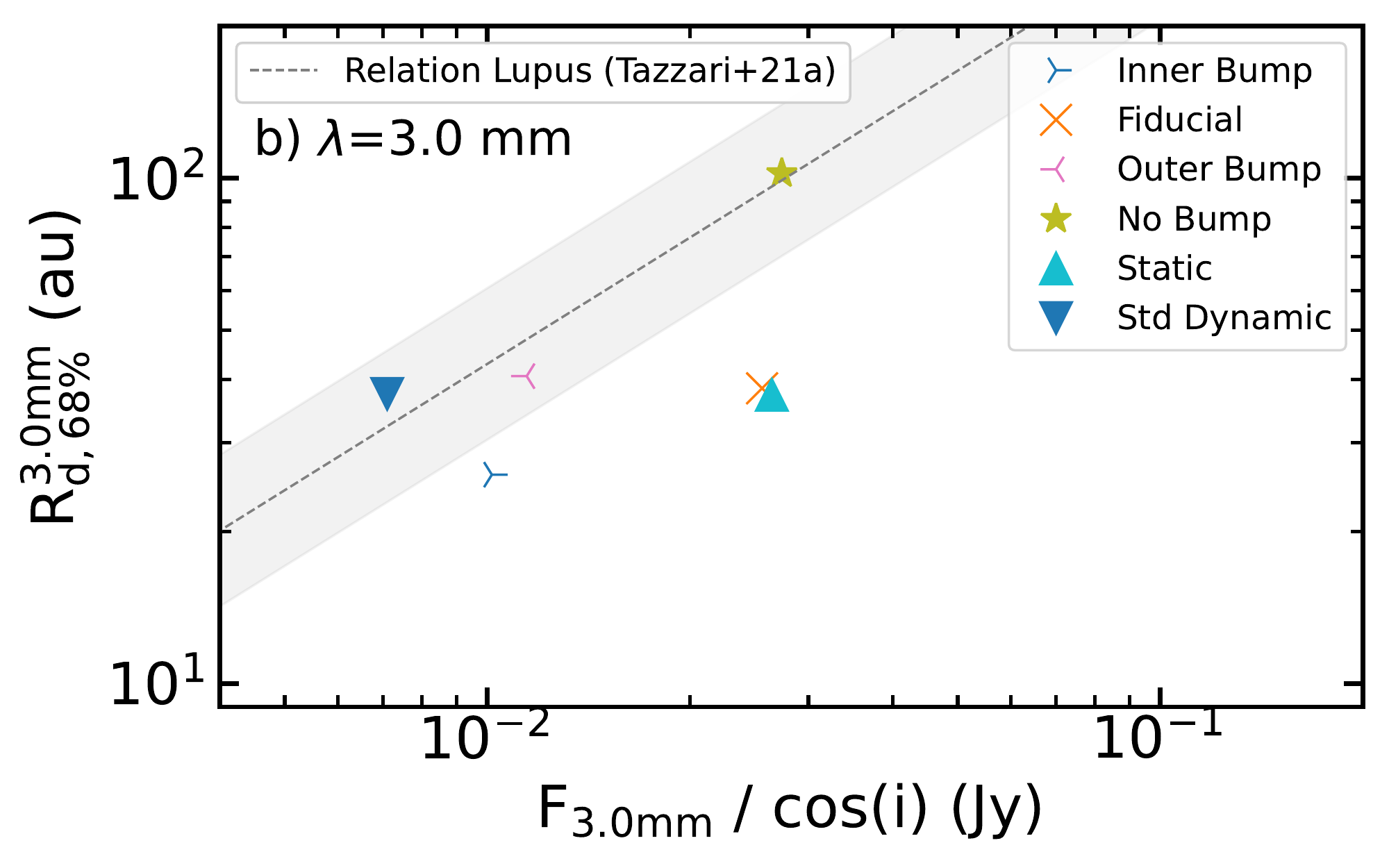}
\includegraphics[width=\linewidth]{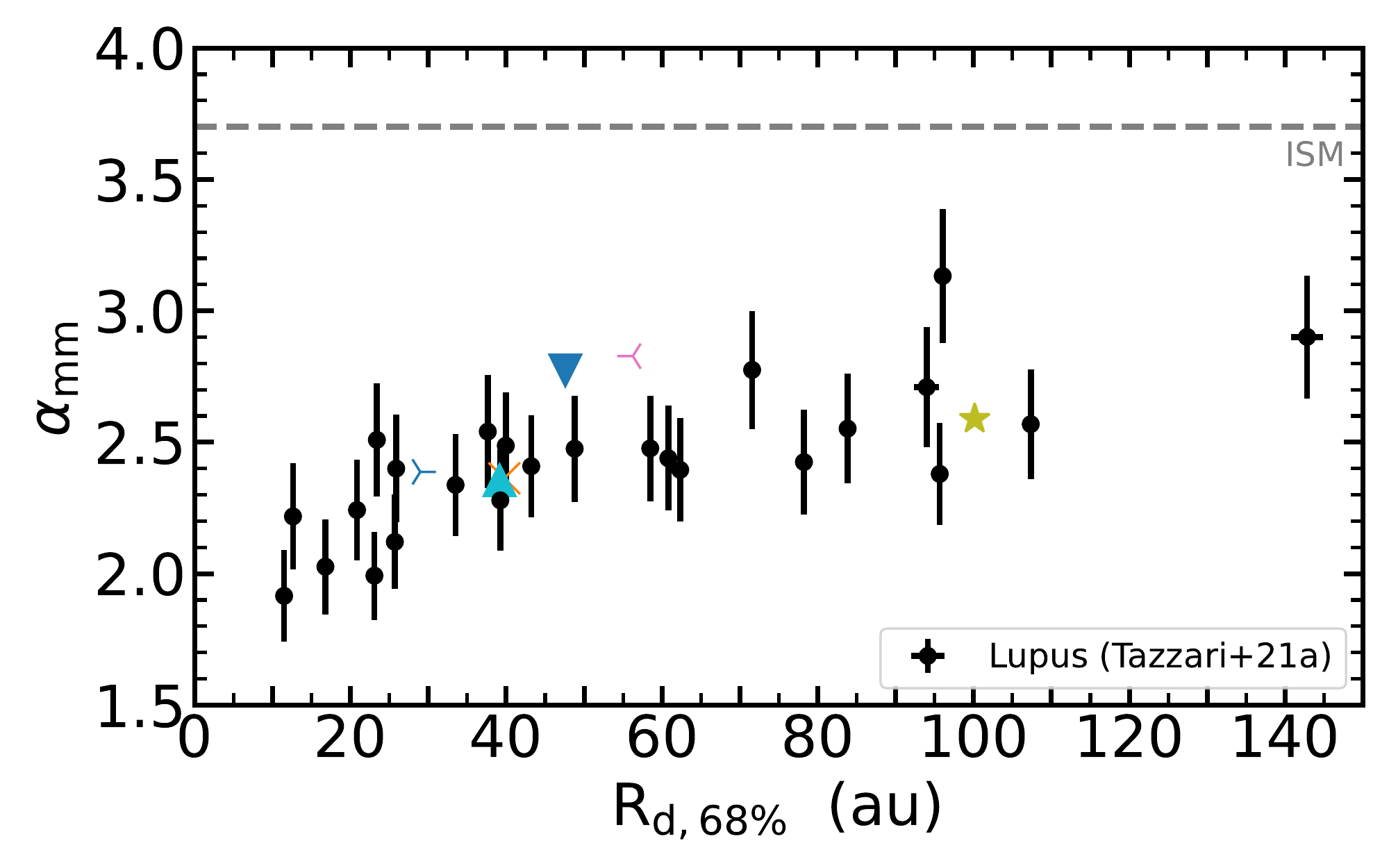}
\caption[Size-luminosity relations and spectral indices dependent on disk sizes]{Size-luminosity relations and spectral indices dependent on disk sizes. As Fig.~\ref{fig:Tazzari_relations} but using the \cite{Ricci2010} opacities.}
\label{fig:Tazzari_relations_Ricci}
\vspace{-2.5mm}
\end{figure}

 \begin{table*}[t]
    \centering
    \small
    \begin{threeparttable}
    \caption{Observational disk properties calculated using the opacities by \cite{Ricci2010}.}
    \label{table:obs_proper_rici}
    \begin{tabular}{l | c c c c c c c c c}
     \hline \hline
     Name & $M_\mathrm{d}^\mathrm{1Myr}$ & $M_\mathrm{d, obs}$ &  $F_\mathrm{1.3mm}$ & $F_\mathrm{3.0mm}$ & $\alpha_\mathrm{mm}$ &  $R_\mathrm{d,68\%}$ & $R_\mathrm{d,90\%}$ & $\tau_\mathrm{abs}^\mathrm{peak}$ &
     $\tau_\mathrm{obs}^\mathrm{peak}$ \\
    
     & [$\mathrm{M_\oplus}$] & [$\mathrm{M_\oplus}$] &  [mJy] & [mJy]  & & [au] & [au] & & \\
    
    (1) & (5) & (8) & (9) &(10)& (11)&(12)&(13)& (14)& (15) \\
    \hline
    Inner Bump & 115 &  39.0 & 70.2 & 9.5 & 2.4 & 13 & 40 & 337 & 1.3 \\
    \hline
    Fiducial   & 87.3 & 96.3 & 174 & 24.1 & 2.4 & 38 & 47 & 36 & -- \\
    \hline
    Outer Bump  & 38.9 & 124 & 224 & 25.7 & 2.6 & 104 & 112 & 2.8 & 1.9 \\
    \hline
    No Bump  & 6.8 &  63.5  & 114  & 10.8  & 2.8 & 55 & 77 & 0.15  & 0.19 \\
    \hline
    Std Dynamic & 7.4 & 37.6 & 67.9 & 6.7 & 2.8 & 37 & 86 & 0.14 & 0.15 \\
    \hline
    Static  & 126 & 98.9  & 178 & 24.9 & 2.4 & 32 & 42 & 140 & 1.2 \\
    \hline
    Planet. Form.  & 42.3 & 91.0  & 164 & 21.5 & 2.4 & 38 & 49 & 14 & 2.5 \\
    \hline
    \end{tabular}
    \begin{tablenotes}
      \small
      \item NOTE - All observational properties have columns similarly labelled as columns in Table \ref{table:disk_properties}. For the \textit{Fiducial} model $\tau_\mathrm{obs}^\mathrm{peak}$ could not be computed, as its peak intensity was larger than the blackbody emission.
    \end{tablenotes}
    \end{threeparttable}
\end{table*}

Laboratory experiments, that have revisited the sticking properties of dust aggregates in protoplanetary disks even support this idea of a lower fragmentation velocity \citep{Gundlach2018,Musiolik2019,Steinpilz2019}. Recently, \cite{pinilla21} studied disk models with such a fragmentation velocity of 1 m/s, combined with different values of the turbulence parameters in disks. Even though, they have a similar disk setup and also conduct radiative transfer calculations with scattering, they obtain disk millimetre fluxes that are nearly an order of magnitude higher than the values we found. However, they use amorphous grains with a different opacity prescription, namely the one stated in \cite{Ricci2010}, whose generally higher opacity values for mm-sized grains translate into their larger millimetre fluxes. This stresses the fact that the opacities of grains are still the largest uncertainty we face in modelling the disks millimetre appearances as discussed in the next subsection. 

\subsection{The effect of the dust opacity}
\label{sec:disc_dust_opacity}

The composition of interstellar dust grains is still not well known. We initially chose the DSHARP opacities (which considers compact grains), since most of the studies we compare our results to, used this prescription as well \citep[as][]{dullemond18, Stammler2019, Zhu2019, Franceschi2021}. Nevertheless, it is important to discuss the effect a different dust opacity model has, when modelling the observational properties of protoplanetary disks. 

For this subsection, we calculated the opacity of the grains following the widely used prescription of \cite{Ricci2010}. In contrast to the DSHARP opacity model by \cite{birnstiel18}, the grains are assumed to be porous spheres with a vacuum volume fraction of 40 \%, composed of astronomical silicates \citep[][]{Draine2003}, carbonaceous material \citep[][]{Zubko1996} and water ice \citep[][]{warren&brandt08}. First, this leads to a decreased dust material density of $\rho_\mathrm{s,Ric}=0.96\,\mathrm{g/cm^3}$ versus $\rho_\mathrm{s,Dsharp}=1.68\,\mathrm{g/cm^3}$. However, we are not rerunning our dust evolution models with this lower density, since we only investigate the change of opacity here. 

The resulting millimetre absorption and scattering opacities for the two prescriptions can be compared in Fig.~\ref{fig:Opacities} of the Appendix. Notably, the Ricci absorption opacities are an order of magnitude higher than the ones of \cite{birnstiel18}. Additionally, they also display no major difference between the two mm-wavelengths for larger than millimetre grains, whereas the 3.0\,mm DSHARP opacities are lower than the 1.3\,mm ones. These differences have a major impact on the calculated mm-fluxes and spectral indices of our modelled disks. The resulting observational properties for some chosen models, calculated using the Ricci opacities are shown in Table \ref{table:obs_proper_rici}.

As it can be seen in the table, all models increase their millimetre fluxes by at least a factor of three, in the case of the \textit{No Bump} and \textit{Std Dynamic} models even by over an order of magnitude. Through the increases of $F_\mathrm{1.3mm}$, we also obtain larger observed dust masses which are in most cases now even higher than the actual model masses at 1 Myr. The highest increase is experienced for $F_\mathrm{3.0mm}$, as the Ricci absorption opacities are over an order of magnitude larger than the DSHARP prescription. This has a direct consequence for the calculated spectral index that is for all chosen models now $\alpha_\mathrm{mm} \leq 2.8$. In turn, $\alpha_\mathrm{mm}$ just mirrors the difference in the opacity values between 1.3\,mm and 3.0\,mm, as in nearly all models we grow a sufficient amount of pebbles which in both opacity prescriptions have high opacity values.

Fig.~\ref{fig:Tazzari_relations_Ricci} shows how the changed opacity prescription affects the models' size-luminosity relations and spectral indices when compared to the Lupus sample of \cite{Tazzari2021_sizes}. As discussed above, the mm-luminosities for all models increase, however, their disk sizes stay roughly the same. This leads to the point that (smooth) models, which did not fit the relation in Fig.~\ref{fig:Tazzari_relations}, fit it now the best, while the \textit{Fiducial} and \textit{Static Bumps} model, display now too high luminosities compared to their sizes. \cite{Zormpas2022} also finds that whether disks lie on the size-luminosity relation heavily depends on the chosen opacity model and that for DSHARP opacities preferably disks with sub-structures populate the relation. Looking at the spectral indices, we now find all models to lie within the range of the Lupus sample, even the ones which before did not.

The above discussed substantial differences in the observational properties highlight the need for more studies investigating the opacities of protoplanetary disks, in order to weigh up the usage of the various opacity prescriptions when calculating the disks' observables. For instance, dust particles may grow more fluffy than assumed in this paper, until they grow and compact to planetesimals \citep{Akimasa2013}. The absorption and scattering mass opacity are highly influenced by the filling factor of the dust particles, which observationally can be distinguished by deep and high angular resolution at multiple wavelengths and polarisation observations of protoplanetary disks \citep{Akimasa2014}. However, if the dust particles are very fluffy, it is expected that they do not drift towards regions of high pressure in contraction to the observed substructures.

\subsection{Optical depths and dust masses}
\label{sec:disc_dynamic_bumps}

\begin{figure*}[t]
\centering
\includegraphics[width=\linewidth]{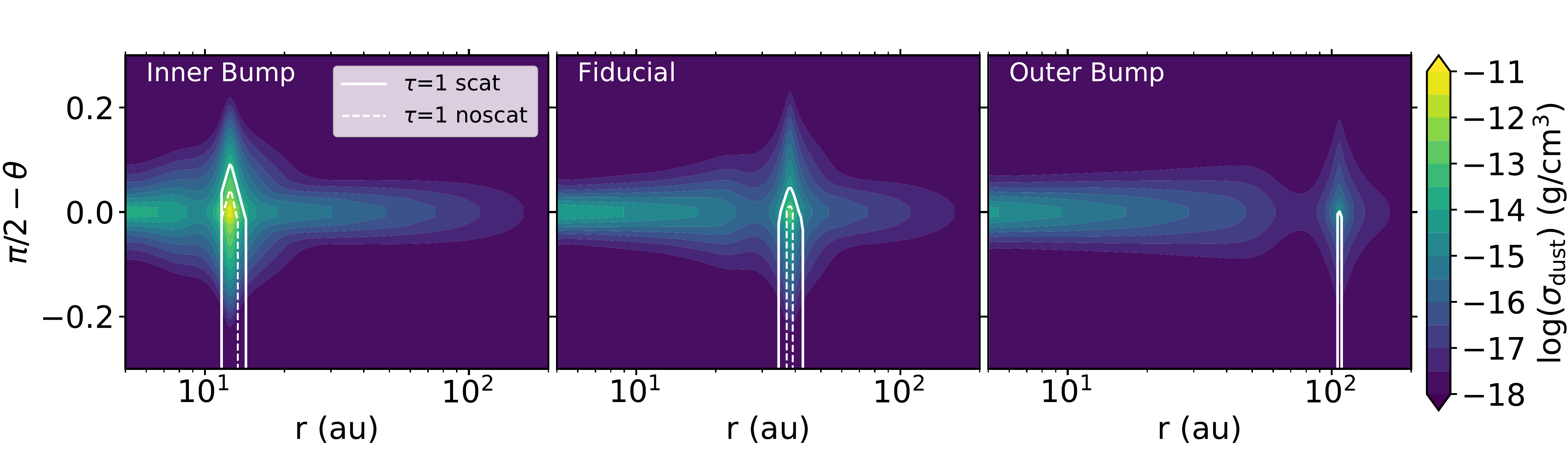}
\caption[2D dust density distribution of models with varying gap locations at 1 Myr]{Vertical and radial dust density distribution. Shown are the models with varying gap locations after 1 Myr of evolution, over-plotted by the contours of the optical depth emission surface corresponding to $\tau=1$ with either treatment of full (solid) or no (dashed) scattering.}
\label{fig:Tausurf_DDensity}
\end{figure*}

The discrepancy between the observed and modelled dust masses is directly related to the optical depth at the peak location of our dust rings. This has already been seen by comparing the according variables in Fig.~\ref{fig:Mdust_tau_effect} that show the same dependencies on the bump model parameters. If the dust rings have $\tau_\mathrm{abs} \geq 1$, the optically thin approximation breaks down and we find that the bulk of the dust mass, which is located in the rings, cannot accurately be traced anymore. This is shown in Fig.~\ref{fig:Tausurf_DDensity}, where we display the particle density distribution for our models, over-plotted by the emission surface corresponding to an optical depth of $\tau=1$. Above this surface radiation or photons can escape the rings unimpeded, without being absorbed or scattered. Therefore, the dust mass that lies below the white contours is hidden from our view and can thus not be traced by observations anymore. 

It is important to note that the height of the emission surface changes whether we include scattering in our radiative transfer calculations or not. Ignoring dust scattering leads to an underestimation $\tau$, especially in the inner regions of the disks ($< 50$ au) \citep{Zhu2019}. Similar to their results, we find that dust scattering reduces the emission from optically thick regions, as shown in Fig.~\ref{fig:RadFlux_Scat}, but increases it for optically thin ones. The latter is the case for the \textit{Outer Bump} model where it leads to a further decrease in its optical depth, hence the model without scattering not even shows an emission surface of $\tau=1$.
On the contrary, the rings of the \textit{Inner Bump} and \textit{Fiducial} model raise their \textit{observed} optical depths to $\tau_\mathrm{obs}=0.58$ and $\tau_\mathrm{obs}=1.02$, respectively, when scattering is not included. This is an increase of over 40\% for both models compared to the values including scattering, and is due to their increased emission at the peak location of the ring.

The general case is that we do not \textit{observe} the dust rings with $\tau_\mathrm{abs} > 1$ to be optically thick, as all of our $\tau_\mathrm{obs} < 1$. However, the way of calculating the observed optical depth (through Eq. \eqref{eq:opt_dept_obs}) is already assuming the thermal emission of the dust to be optically thin \citep{dullemond18}, so it is not useful to measure the optical depth of optically thick rings with this formula. The argument of \cite{dullemond18}, to use the optically thin assumption for their DSHARP sample is that they do not observe the source's radial ring profiles to be flat-topped, which should be the case for highly optically thick rings. Yet, we do not observe a flat top of our radial ring emission profiles for our (highly) optically thick models in Fig.~\ref{fig:RadFlux_Scat}. This is likely due to the fact that such a flat-topped region is very narrow, such that it cannot be resolved by observations. The inclusion of scattering further reduces and broadens the surface, as well as the convolution with the Gaussian beam which additionally adds to the smoothing of the peaked emission profile and thus also to a decreased observed optical depth. In the end, it remains challenging to thoroughly determine the optical depths of observed disks with substructures.

To conclude the discussion on the optical depths of the rings, we want to point out that there are still large sources of uncertainty in calculating these values. Not only the temperature prescription for $\tau_\mathrm{obs}$, but also the grain composition and the thereof derived opacities used to calculate $\tau_\mathrm{abs}$ are still not well known. Furthermore, we assume $\tau_\mathrm{abs}$ to be the 'actual' optical depth of our models. This is not entirely correct, since we ignore the scattering opacity $\kappa_\lambda^{\mathrm{scat}}$ of the grains which would significantly contribute to the then summed up extinction optical depth $\tau_\mathrm{ext}=\tau_\mathrm{abs}+\tau_\mathrm{sca}$. However, as scattering itself is angle dependent, this summation is not straightforward to do, so in first order it is reasonable to compare $\tau_\mathrm{obs}$ to $\tau_\mathrm{abs}$, instead of $\tau_\mathrm{ext}$. \\

About dust disk masses, we first want to
point out that we confirm the findings of \cite{Binkert2021}, which also find the dust masses of optically thick inner bumps are underestimated by up to an order of magnitude. While we treat dust evolution in our one dimensional disk models, they carried out three dimensional two-fluid (gas + dust) hydrodynamic simulations to obtain their results. 

Regarding the effects the initial conditions have on the obtained dust masses, the most important parameters are of course the initial disk mass and dust-to-gas ratio. Doubling the initial disk's mass of the \textit{Fiducial} model, we also found the total dust mass at 1 Myr to double. However, due to optical thickness of the dust ring, where the bulk of the mass is located, this leads only to an increase in the mm-fluxes (and $M_\mathrm{d, obs}$) of about 35\%.

Another very important parameter, we have not yet discussed, is the turbulent viscosity $\alpha$. Since we induce our Gaussian bumps onto the $\alpha$-profile, the moderate value of $\alpha_0=$1e-3 makes sure that we do not experience too long growth times of our pressure bumps within the disks. The key parameter is the viscous timescale $\tau_\mathrm{visc} \propto (\alpha_0 \Omega_\mathrm{k})^{-1}$ which sets the lower limit on how fast the gas can viscously adapt to changes. If for example, we take very low $\alpha$ values, it may take up to 1 Myr for a bump in the outer region of the disk to reach its final amplitude. An increased turbulence value on the other hand, would lead to less dust trapping within the rings, as the higher dust diffusivity would smooth out the density gradients, eventually decreasing the overall dust mass.

Another initial bump model parameter that effects the disk masses is the time, after which we initially induce the bumps into our viscosity profile, namely at $T_\mathrm{ini}=1$ kyr. Most of studies are already including their pressure bumps at the beginning of the simulation  and then let the viscous timescale set their according growth times \citep[e.g.][]{dullemond18,pinilla21}. Recently, \cite{Kuznetsova2022} also argued for a early formation of pressure bumps in disks, which in their work are generated through anisotropic streams of infalling material. As we are particularly interested in the growth times of pressure bumps, we ran two models, similar to our \textit{Fiducial} setup with a bump at 30 au, but where we start to grow the bump later than 1 kyr. For the first model, we induced the bump only at ${T_\mathrm{ini}=\,0.1\,\mathrm{Myr}}$ of evolution which led to a decrease in total dust mass of 35\% compared to the \textit{Fiducial} model. In a second setup, we even increased $T_\mathrm{ini}$ to 0.5 Myr and further let the bump grow over the same period of time, in comparison to the fiducial value of $T_\mathrm{grow}=$\,0.1 Myr at 30 au. For this model, the according dust masses at 1 and 3 Myr of evolution, even decreased by more than a factor of eight. Both models show significantly less dust mass, than the \textit{Fiducial} setup that would lead to even lower observed mm-fluxes and higher spectral indices for these models.

Regarding the formation time of zonal flows in global MHD-simulations, they can develop as fast as a few tens to hundreds of orbits, depending on the magnetisation and ionisation of the gas. They then remain stable in the disk for several hundreds of orbits. In a recent study by \cite{Riols2020} their zonal flow structures even remains steady over their entire simulation time ($\sim$1000 orbits). We ran another simulation taking this value for $T_\mathrm{life}$ together with the formation time of their zonal flows ($\sim$200 orbits). It turned out that the increased lifetime does not help in halting the dust loss (see Fig.~\ref{fig:dynamic_bumps}). The important point is, that the inner-most zone passes $\sim$25 bump life cycles till 1\,Myr in (each 38\,kyr at 10\,au), which sets the bottleneck of how much dust the disk looses towards the star. Therefore, we only obtain a total dust mass of 12$M_\oplus$ for the above model, still an order of magnitude less than in the \textit{Static} one.

To conclude, the observed dust masses of disks are still hard to measure, especially as we cannot be sure whether observed substructures are optically thick or not, therewith hiding a bulk of the dust mass contained within disks. However, optically thick substructures might ease the discrepancy between the observed dust masses in mature class II disks, compared to the solid material actually needed to reproduce observed exoplanetary systems \citep{Tychoniec2020}. Another possibility is that part of the dust mass could be contained within larger boulders of metre to kilometre sized objects.

\subsection{Planetesimal formation}
\label{sec:disc_pltes_form}
In this last subsection, we shortly want to discuss the impact planetesimal formation has on our \textit{Fiducial} model, where it might reduce the dust mass stored within the ring, converting it into `invisible' planetesimals and therewith reducing its optical depth.

Dust rings may be ideal places to form metre to kilometre sized objects, as the local dust-to-gas ratio $\epsilon$ is strongly enhanced here. This might trigger the so-called `streaming instability' \citep[e.g.][]{Youdin+Goodman2005} that can lead to the process of gravoturbulent planetesimal formation \citep{Johansen2006}. We refer to section 7.3 of \cite{dullemond18} for a more detailed discussion about the mechanisms that can lead to the streaming instability in dust rings.

To this end, we simply follow the setup of \cite{Stammler2019} that builds on the planetesimal model of \cite{Schoonenberg2018}. Here, planetesimal formation is triggered as soon as a characteristic value $\epsilon_\mathrm{crit}=1.0$ of the dust-to-gas ratio at the midplane is reached. If this is the case, a fraction $\zeta=0.1$ of the dust surface density per settling time scale of each dust mass bin will be converted into the surface density of planetesimals. Similarly, we do not further evolve the planetesimals' surface density, even though planetesimal-planetesimal collisions could lead to dust reproduction and hence a lower planetesimal mass.

\begin{figure}[t]
\centering
\includegraphics[width=\linewidth]{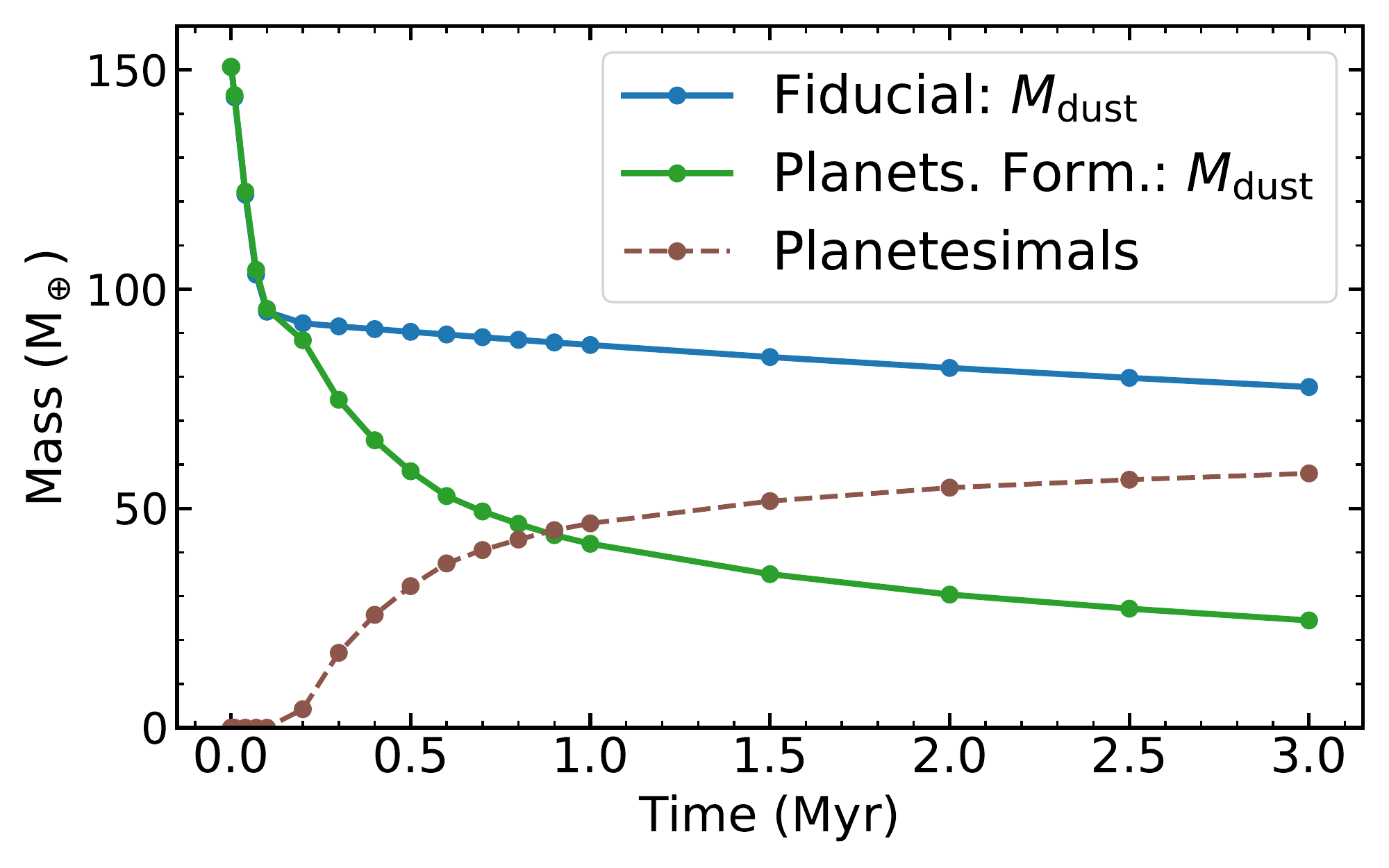}\\
\caption[Dust mass evolution of the model with planetesimal formation]{Dust mass evolution of the model with planetesimal formation. Total mass of the fiducial and planetesimal formation model in dust (blue \& green) and in  planetesimals (brown) over the disk lifetime.} 
\label{fig:planetis}
\vspace{-2.0mm}
\end{figure}

We take our \textit{Fiducial} model, with the bump at 30 au and now add the formation of planetesimals to the model. The dust size distribution of the according model, can be found in Fig.~\ref{fig:DustSizeDens_Static} of the Appendix. As expected, we find the dust density to be decreased only at the location of the ring (by a factor of $\Sigma_\mathrm{d}^\mathrm{fid}/\Sigma_\mathrm{d}^\mathrm{plt}\sim3$), compared to the size distribution of the \textit{Fiducial} model. Apart from the density at the ring location, there are no differences in the dust size distribution between the models.

In Fig.~\ref{fig:planetis}, we plotted the total masses of the models stored in dust and planetesimals over time. First, we see that the difference in dust mass between the models does not increase further after about 1.5 Myr of evolution. Already at 1 Myr, we find over half of the total dust mass of the \textit{Fiducial} model to be converted into planetesimals, namely $M_\mathrm{pltes}=47 M_\oplus$, even more than what is stored in dust $M_\mathrm{d}^\mathrm{1 Myr}= 42 M_\oplus$. This leads to decreased fluxes at 1.3 and 3.0\,mm of 15\% and 35\%, respectively. As $F_\mathrm{3.0mm}$ drops stronger in emission than $F_\mathrm{1.3mm}$, this directly impacts the spectral index which increased to $\alpha_\mathrm{mm}=3.0$.

But how about the optical depths of the model? Checking again the absorption optical depth at the peak location of the dust rings, we find $\tau_\mathrm{abs}^\mathrm{plts}=1.4$ versus $\tau_\mathrm{abs}^\mathrm{fid}=3.5$, a decrease of 60\%. However, calculating the optical depth from the intensity peak of the synthetic observation, we only find $\tau_\mathrm{obs}^\mathrm{plts}=0.55$ versus ${\tau_\mathrm{obs}^\mathrm{fid}=0.70}$, a small reduction of $\sim20$\%. 

The same trend can be found using the higher \cite{Ricci2010} opacities (see Table \ref{table:obs_proper_rici}) for obs. prop.), here we find both rings highly optically thick $\tau_\mathrm{abs}^\mathrm{plts}=14$ versus $\tau_\mathrm{abs}^\mathrm{fid}=36$, still the planetesimal model with a decrease of over 60\%. For the observed optical depths the picture is harder to interpret, as for our \textit{Fiducial} model, we were not able to determine ${\tau_\mathrm{obs}^\mathrm{fid}}$ anymore, since the peak emission exceeds the prescribed blackbody emission. Therefore, we increase the blackbody temperature prescription of Eq. \eqref{eq:temp_profile} by 5 K to 165 K at 1 au, to obtain $\tau_\mathrm{obs}^\mathrm{plts}=2.2$ versus ${\tau_\mathrm{obs}^\mathrm{fid}=4.2}$. On contrary to the DSHARP-opacities, which were also used by \cite{Stammler2019}, we now find both rings to be optically thick, even though planetesimal formation lowered $\tau_\mathrm{obs}$ by nearly a factor of two. Still, the problem persists that $\tau_\mathrm{obs}$ underestimates the `actual' absorption optical depth $\tau_\mathrm{abs}$ of the rings, for the Ricci-prescription even greater than for the DSHARP-opacities. 

\cite{Stammler2019} proposed that planetesimal formation provides a mechanism to explain the low observed optical depths of dust rings in the DSHARP-sample \citep{dullemond18}, as it naturally regulates the dust density in the rings.  In our adaption, we similarly find the absorption optical depth of our dust ring to decrease, but only at the cost of lower millimetre fluxes and more importantly an increased spectral index of the disk (similarly using the DSHARP-opacities), which is not supported by observations (though if we use the high \cite{Ricci2010} opacities the $\alpha_\mathrm{mm}$ remains unchanged). Additionally, \cite{Stammler2019} calculated their intensity profile using equation \eqref{eq:opt_dept_obs} without the treatment of scattering. We on the contrary, performed full radiative transfer calculations with an-isotropic scattering to obtain synthetic observations. From the convolved intensity profiles, we further obtain the observed optical depths of the ring(s) similar to \cite{dullemond18} and find them in both cases to be optically thin. 

To conclude this section, we first note that it is challenging to obtain correct optical depth values from observations, as we measure them already assuming optical thinness and additionally, as the peak of the emission and thus derived optical depth is also depending on the beam-size, decreasing with larger beams. Secondly, we should be cautious as to compare these observed values to the ones of models without scattering, as scattering can significantly reduce the emission and therewith observed optical depths, at least for those of optically thick dust rings in inner disk regions ($\leq$ 40 au), as we have shown in the above sections. These points make it hard to conclude whether planetesimal formation is actually at work in dust rings, regulating their optical depths, or if the fine tuned values of $\tau$ are just a result of the error-prone determination of the observed optical depths. A recent study by \cite{Carrera2022} also presents evidence that the streaming instability is indeed unlikely to form planetesimals from mm-grains inside axissymmetric pressure bumps.

\section{Conclusions} \label{sec:conclusions}

In this work, we investigated how dynamic pressure bumps affect the observational properties of protoplanetary disks. Furthermore, we aimed to differentiate between the planetary- versus zonal flow-origin of pressure bumps. To this end, we performed gas and dust evolution simulations, setting up models with varying pressure bump features, including their amplitude and location, growth time and number of bumps. We subsequently ran radiative transfer calculations, including the full treatment of anisotropic scattering, to create synthetic images from which we inferred the disk's observed properties. Finally, we compared our results to multi-wavelength observations of protoplanetary disks in the millimetre dust continuum.

The main conclusions of this paper are summarised as follows:

\begin{enumerate} \label{sec:conlcusions}
\item The location of the farthest out pressure bump determines the dust disk effective size, calculated through its emission radius enclosing (68\% or) 90\% of the flux. This delivers a possible answer to why disks in low-resolution surveys appear to have the same sizes across different wavelengths \citep{Tazzari2021_sizes}. 

\item Dust traps need to form early (< 0.1 Myr), fast (on viscous timescales) and must be long lived (> Myr) in order to obtain low spectral indices of disks. This favours the planetary-origin of pressure bumps and supports the idea that planet formation may already occur in the class 0-I phases of circumstellar disks \citep[e.g.][]{Tychoniec2020} and that their migration is type II during most of the disk's lifetime.

\item Unlike the planetary bumps, our dynamic bump model with dis- and re-appearing pressure bumps of finite lifetimes can only reproduce the disk's high fluxes and low spectral indices observed with the highest prescription of opacities that we investigated, thus disfavouring the zonal flow-origin of most of the observed disk substructures. 

\item The wavelength-independent sizes of unresolved small disks ($R_\mathrm{dust}\leq$ beam-size) with low spectral indices may indicate that they contain unresolved dust traps. By resolving them, we could measure the disks to be significantly smaller in size. This would lead to a steepening of the slope for the size-luminosity relation found in disks, since the determination of the luminosity (in general) and the sizes of larger disks, are less affected by the chosen resolution.

\item We find a large discrepancy between the observed and modelled optical depths of dust rings within our disks, especially for those in inner regions ($<40$ au), independently of the opacity model. For the DSHARP opacities, the observed values all lie between ${\tau_\mathrm{obs} \approx 0.2-0.7}$ for the models with bumps, even though some of them are optically thick, reaching optical depths in the models of ${\tau_\mathrm{abs} \approx 2-35}$. This difference is because the determination of $\tau_\mathrm{obs}$ implicitly assumes the emission to be optically thin. Additionally, the non-treatment of scattering underestimates the optical depth of dust rings, particularly in the inner part of the disks.

\item The observed dust masses of disks with optical thick inner bumps ($\leq40$ au) are underestimated by up to an order of magnitude. This might ease the discrepancy between observed dust masses of mature class II disks and the solid material actually needed to reproduce the observed exoplanetary systems \citep{Tychoniec2020}. However, these results are particularly affected by the chosen opacity model.

\item Planetesimal formation within dust rings can help to reduce the modelled optical depths of rings, but only at the cost of lower millimetre luminosities and high spectral indices, when using the DSHARP-opacities. The prevalent determination of observed optical depths (assuming optical thinness), however, makes it hard to conclude whether planetisimal formation is actually at work in dust rings, or if the observed fine-tuned values of $\tau$ are just a result of the procedure determining them in the first place.

\item Most of the inferred observational disk properties are sensitive to the chosen opacity prescription \citep{birnstiel18}. Using another prescription \citep{Ricci2010} leads to higher fluxes, dust masses and optical depths, as well as lower spectral indices ($\alpha_\mathrm{mm}\leq 2.8$). The dust disk effective sizes for models with long-lived pressure bumps are less affected by the choice of opacity.

 \end{enumerate}
 
In summary, our findings favour the planetary- over the zonal flow-origin of ring and gap substructures in the dust continuum of protoplanetary disks. Furthermore, this supports the idea that planet formation may already occur in early Class 0-I stages of circumstellar disks. It remains an open question if planetesimals can be formed in our zonal flow model, who then can deliver the initial seeds to form planetary cores in early times, depending on their formation timescales.

\begin{acknowledgements}
 We would like to thank referee Clément Baruteau for his very constructive feedback, as well as Mario Flock for the fruitful discussion regarding zonal flows. The authors acknowledge funding from the Alexander von Humboldt Foundation in the framework of the Sofja Kovalevskaja Award endowed by the Federal Ministry of Education and Research, from the European Research Council (ERC) under the European Union’s Horizon 2020 research and innovation programme under grant agreement No 714769 and funding by the Deutsche Forschungsgemeinschaft (DFG, German Research Foundation) under grants 361140270, 325594231, and Germany's Excellence Strategy - EXC-2094 - 390783311.\\
\textbf{Software:}
 DustPy \citep{Stammler2022}, Matplotlib \citep{Hunter_mpl}, Numpy \citep{vanderWalt_np}, OpTool \citep{Dominik2021}, RADMC-3D \citep{Dullemond2012}, Scipy \citep{Virtanen_scipy}.
 \end{acknowledgements}

\bibpunct{(}{)}{;}{a}{}{,} 
\bibliographystyle{aa} 
\bibliography{TheBib.bib} 

\begin{appendix}
\onecolumn
\section{Supporting figures}
\renewcommand\thefigure{A.\arabic{figure}}   
\begin{figure*}[ht]
\centering
\includegraphics[width=0.9\linewidth]{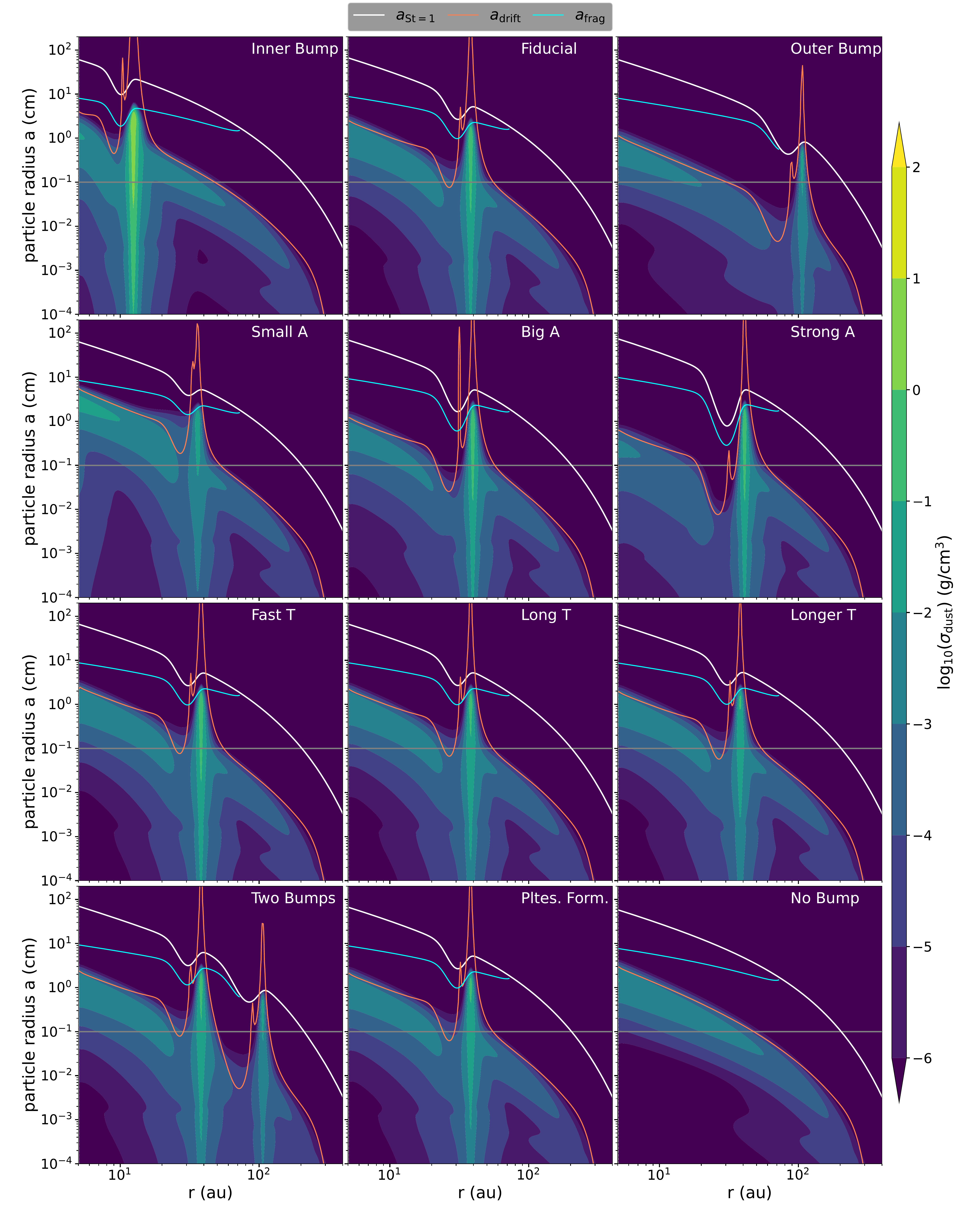}
\captionsetup{width=\linewidth}
\caption[Dust size distributions planetary bump models]{Dust size distributions for the `planetary' bump models at 1 Myr. The contours show the growth limits for drift (orange) and fragmentation (cyan), as well as the the particle sizes corresponding to St = 1 (white) and 1 mm (grey). The parameters for the displayed models can be found in Table \ref{table:disk_properties}.}
\label{fig:DustSizeDens_Static}
\end{figure*}

\begin{figure*}[ht]
\centering
\includegraphics[width=\linewidth]{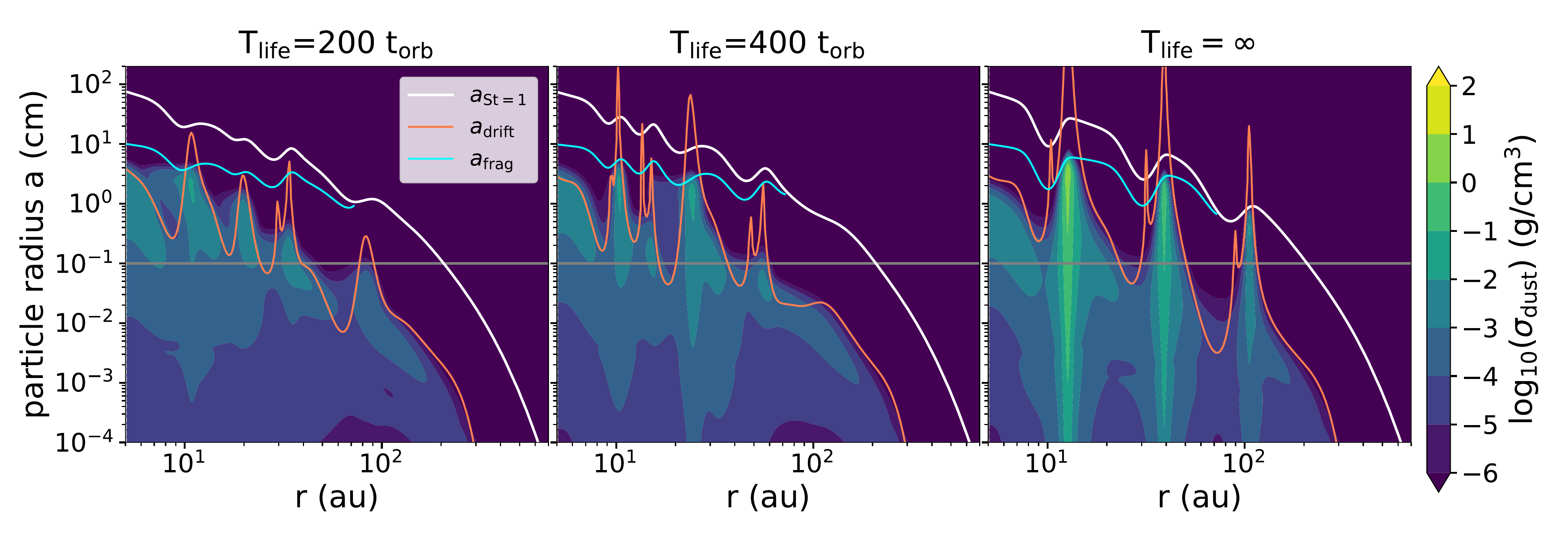}
\captionsetup{width=\linewidth}
\caption[Dust size distributions dynamic bump models]{Dust size distributions dynamic bump models. As Fig.~\ref{fig:DustSizeDens_Static} but for the dynamic bumps models.}
\label{fig:DustSizeDens_Dynamic}
\end{figure*}

\begin{figure*}[h]
\centering
\includegraphics[width=150mm]{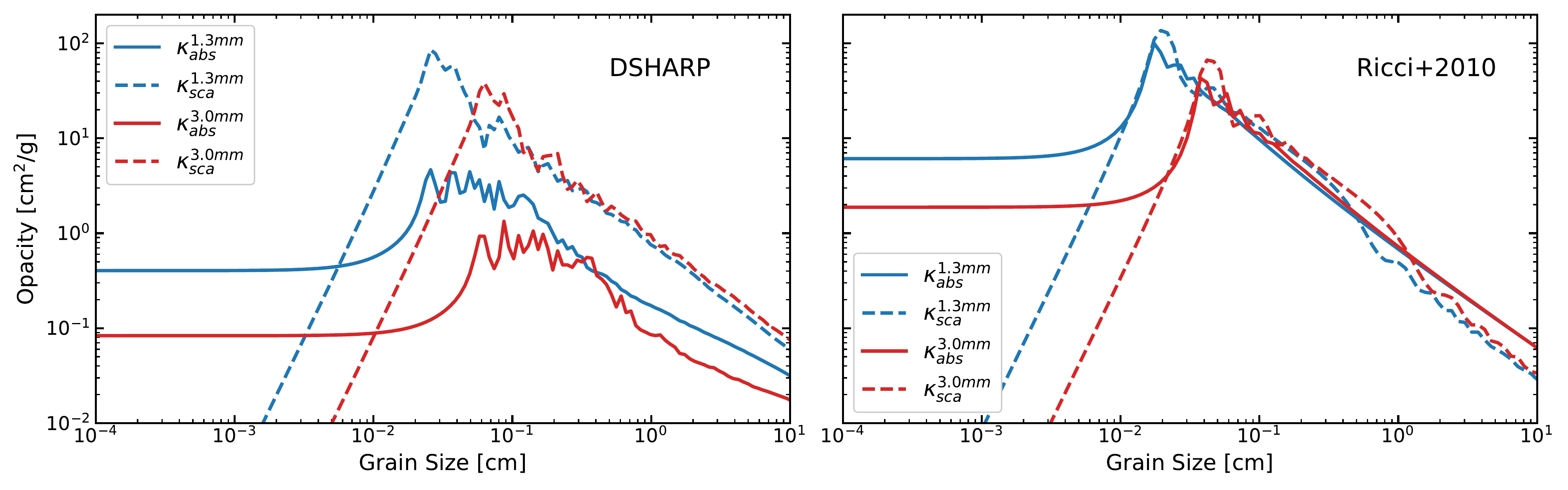}
\captionsetup{width=\linewidth}
\caption[Millimetre opacities over dust grainsize]{Millimetre opacities over dust grainsize. Absorption and scattering opacities over dust grainsize for the observation wavelengths of $\lambda=1.3$ (blue) \& 3.0\,mm (red). The composition of the grains leading to the displayed opacities are adopted from the DSHARP paper of \cite{birnstiel18} and the paper of \cite{Ricci2010}}
\label{fig:Opacities}
\end{figure*}

\begin{figure*}[ht]
\centering
\includegraphics[width=0.8\linewidth]{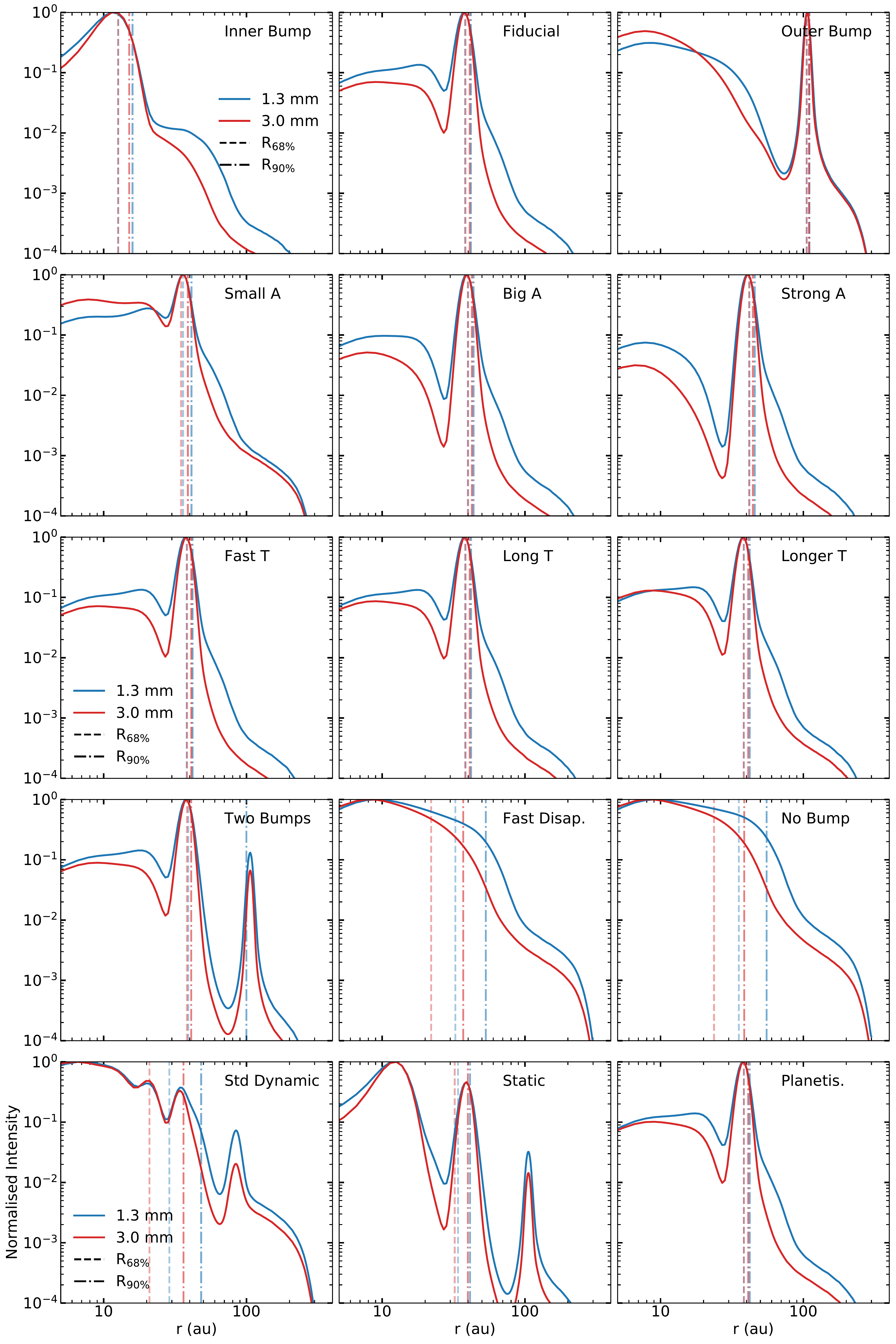}
\captionsetup{width=\linewidth}
\caption[Radial millimetre intensity profiles]{Radial millimetre intensity profiles. Profiles for synthetic images at 1.3\,mm (blue) and 3.0\,mm (red), both convolved with a Gaussian beam with a FWHM of 0.04''x0.04''. The dashed and dotted-dashed lines show the 68\% and 90\% emission disk radii of the models, respectively.}
\label{fig:NormIntenstiy_Profiles}
\end{figure*}

\end{appendix}

\end{document}